\documentclass[12pt]{article}
\usepackage{latexsym}
\usepackage{amsthm}
\usepackage{amssymb,amsbsy,amsfonts,amsmath}
\usepackage{bm,bbm}
\usepackage{epsfig}
\usepackage{graphicx,rotating,lscape}

\usepackage[hypertexnames=false]{hyperref}
\usepackage{verbatim}
\usepackage{natbib}
\usepackage{multirow,color}
\usepackage{colortbl}
\usepackage{geometry}
\usepackage{setspace}
\usepackage{subfigure}
\usepackage[flushleft]{threeparttable}
\usepackage{enumitem}
\usepackage{algorithm}                
\usepackage{algpseudocode}  

\newtheorem{assum}{{\bf Assumption}}

\newtheorem{Pro}{Proposition}
\newtheorem{Lem}{\underline{\bf Lemma}}

\newtheorem{Th}{\underline{\bf Theorem}}

\newtheorem{Rem}{\underline{\bf Remark}}
\newtheorem{Cor}{\underline{\bf Corollary}}
\newtheorem{Ex}{Example}

\newcommand{\norm}[1]{\left\lVert #1 \right\rVert}

\def\sign{\mbox{sign}}

\def\wh{\widehat}
\def\wt{\widetilde}
\def\n{\nonumber}   
\def\var{\mbox{var}}
\def\cov{\mbox{cov}}

\def\argmin{\mbox{argmin}}

\def\pr{\mbox{pr}}
\def\log{\mbox{log}}
\def\trans{^{\top}}

\def\sumi{\sum_{i=1}^n}
\def\sumia{\sum_{i=1}^{n_1}}

\def\0{{\bf 0}}
\def\1{{\bf 1}}

\def\O{{\bf O}}
\def\o{{\bf o}}

\def\V{{\bf V}}

\def\s{{\bf s}}
\def\S{{\bf S}}

\def\u{{\bf u}}

\def\v{{\bf v}}
\def\W{{\bf W}}

\def\w{{\bf w}}
\def\X{{\bf X}}
\def\b0{{\bf 0}}

\def\x{{\bf x}}
\def\Y{{\bf Y}}

\def\Z{{\bf Z}}

\def\calT{\mathcal{T}}
\def\calF{\mathcal{F}}

\def\mR{\mathbb{R}}

\def\bg{{\boldsymbol\gamma}}

\def\bSigma{{\boldsymbol \Sigma}}

\def\bse{\begin{eqnarray*}}
\def\ese{\end{eqnarray*}}
\def\be{\begin{eqnarray}}
\def\ee{\end{eqnarray}}
\def\bsq{\begin{equation*}}
\def\esq{\end{equation*}}
\def\bq{\begin{equation}}
\def\eq{\end{equation}}

\def\squarebox#1{\hbox to #1{\hfill\vbox to #1{\vfill}}}
\def\boxit#1{\vbox{\hrule\hbox{\vrule\kern6pt\vbox{\kern6pt#1\kern6pt}\kern6pt\vrule}\hrule}}

\begin{document}

\title{\LARGE Is It Worth Collecting Additional Covariates?}

\date{}

\author{ Wen Zhang\\
  Institute of Statistics and Big Data, Renmin University of China
  \\
  \texttt{zhangwen.stat@gmail.com}
  \and
  Yanyuan Ma\\
  Department of Statistics, The Pennsylvania State University
  \\
  \texttt{yanyuanma@gmail.com}
  \and
  Liping Zhu\\
  Centre for Applied Statistics and the Institute of Statistics and Big Data,\\
  Renmin University of China
  \\
  \texttt{zhu.liping@ruc.edu.cn}}

\maketitle
\begin{abstract}
{In many studies, basic covariates are routinely recorded at
  negligible cost, whereas acquiring additional covariates requires
  expensive or time-consuming procedures.  
Thus, a natural question is whether it is worth to collecting these
additional covariates. Using 
the reduction in prediction interval width as criterion, we evaluate
and test the gain brought by the additional covariates 
and hence answer this question. We consider the gain both marginally
and 
conditionally on the observable characteristics.
The conditional formulation captures subpopulation heterogeneity and
supports individualized data acquisition decisions.
To avoid restrictive parametric assumptions, we adopt a model-free
framework that accommodates flexible nonparametric or machine learning
methods. We propose an efficient influence function based method, 
with proven theoretical properties to achieve our purpose 
under mild nuisance convergence rates. Interestingly, we find
the natural conformal prediction based method cannot achieve the same
purpose: It fails in the marginal case and imposes much more
stringent or even infeasible requirement on the nuisance parameter
estimation procedures.
We demonstrate the superior finite-sample performance
of the proposed procedures through simulations and real data
applications. }
\end{abstract}

{\it Keywords:} Conformal prediction; Efficient influence function; Prediction interval; Semiparametric inference; Uncertainty quantification.

\vfill

\newpage
\section{Introduction}

In many scientific, medical and engineering applications, basic
covariates are 
routinely recorded at negligible cost, whereas some additional
covariates may require laboratory assays, genetic testing, or specialized equipment. 
Although these supplementary covariates may improve prediction, collecting them can be costly and time-consuming.
For example, in clinical dosing of warfarin, routine demographic and clinical characteristics can be supplemented with pharmacogenetic variants, which require genotyping assays \citep{iwpc2009}.
In diabetes screening, standard physical examination data can be augmented with blood biomarkers such as fasting plasma glucose or glycated hemoglobin to predict post-load glucose levels, at the expense of venipuncture and laboratory analysis.
Similarly, in civil engineering, ultrasonic pulse velocity measurements can supplement simple rebound hammer tests for predicting concrete compressive strength \citep{shih2015improving}, and in environmental monitoring, visible and near-infrared reflectance spectra can complement geographic and climatic variables for predicting soil organic carbon, provided that field sampling and laboratory scanning are undertaken \citep{orgiazzi2018lucas,jones2020lucas}.
Across these settings, a fundamental practical question arises: \textit{do the gains in predictive performance justify the additional cost and effort of acquiring the extra covariates?}

This question can be cast as a variable importance problem that evaluates the incremental contribution of candidate covariates relative to a baseline set of routinely available information.
An intuitive approach is to quantify the statistical relevance of candidate predictors, a problem extensively studied in the feature screening literature.
Moving beyond sure independence screening based on marginal linear correlations \citep{fan2008sure}, model-free screening procedures assess covariate relevance through conditional distribution functions \citep{zhu2011model}, distance correlation \citep{li2012feature}, or the fused Kolmogorov filter \citep{mai2015fused}.
However, a strong statistical dependence does not necessarily imply better predictive performance \citep{pepe2004limitations}, and such screening methods typically rank observed covariates by their marginal dependence with the outcome to reduce dimensionality prior to modeling.
In contrast, we evaluate the targeted gain from augmenting a fixed baseline set of inexpensive covariates with a prespecified set of costly measurements.
This perspective aligns closely with cost-sensitive prediction, which weighs predictive improvements against the resources required for data acquisition \citep{nan2017adaptive}.

Most existing variable importance measures, however, evaluate improvements in point prediction accuracy, such as reductions in mean squared error \citep{breiman2001random,fisher2019all,williamson2021nonparametric,hooker2021unrestricted}.
Yet statistical association need not imply predictive utility \citep{pepe2004limitations}, and improvements in point prediction accuracy do not necessarily translate into narrower uncertainty bands for an individual outcome.
To directly capture uncertainty reduction, we quantify incremental predictive value by the reduction in prediction interval width required to maintain a prespecified coverage probability.
Comparing interval widths at a fixed coverage level adheres to the
classical principle of maximizing sharpness subject to calibration
\citep{gneiting2007probabilistic} and directly measures how much more
precisely a future outcome can be predicted once the additional
covariates become available.
Specifically, let $\X$ denote the readily available covariates, $\Z$ the costly additional covariates, and $\W=(\X\trans,\Z\trans)\trans$ the augmented covariate vector.
Following the semiparametric prediction interval formulation \citep{jiang2024prediction}, we consider symmetric prediction intervals centered at the conditional means $E(Y\mid \X)$ and $E(Y\mid \W)$, whose half-widths $\zeta_{\x}$ and $\zeta_{\w}$ are determined by the $(1-\alpha)$-quantiles of the corresponding absolute residuals.
We define the marginal incremental predictive value of $\Z$ as the half-width reduction $\Delta=\zeta_{\x}-\zeta_{\w}$.
Because a smaller mean squared error does not automatically guarantee a smaller $(1-\alpha)$-quantile of the absolute residuals, both the sign and magnitude of $\Delta$ must be formally assessed through statistical inference.

While the marginal contrast summarizes the population wide gain in predictive precision, it can obscure substantial heterogeneity across subpopulations.
An additional measurement may markedly narrow prediction intervals for
certain individuals while offering negligible improvement for others. 
In metabolic screening, for instance, blood assays may be far more
informative for older or overweight individuals than for young adults
of normal weight. 
Such heterogeneity motivates dynamic and active feature acquisition,
in which the value of further testing depends on an individual's
observed profile \citep{shim2018joint,ma2019eddi,covert2023dynamic}. 
To capture this variation, for a prespecified subset $\S\subseteq\X$
of readily available covariates, we define
$\Delta(\s_0)=\zeta_{\x}(\s_0)-\zeta_{\w}(\s_0)$ as the half-width
reduction required to attain conditional coverage $1-\alpha$ given
$\S=\s_0$. 
Crucially, because $\S$ is observed prior to collecting $\Z$,
evaluating $\Delta(\s_0)$ enables practitioners to tailor acquisition
decisions to observable characteristics. 
The central question is therefore not merely whether the additional
covariates are useful on average, but for which observable profiles
their predictive contribution is large enough to warrant collection. 

Because our criterion compares interval half-widths at a fixed
coverage level, it is naturally connected to conformal prediction,
which constructs prediction sets with finite sample marginal coverage
guarantees under exchangeability
\citep{lei2014bands,lei2018distribution,romano2019conformalized} and
has also been used for leave-one-covariate-out importance assessments
\citep{lei2018distribution}. 
While exact finite sample conditional coverage is impossible for
continuous covariates without distributional assumptions
\citep{barber2021limits}, recent advances achieve approximate local or
group conditional coverage via localization and weighting
\citep{guan2023localized,gibbs2025conditional}.
However, an essential distinction must be drawn between
\textit{predictive coverage} for a future outcome and
\textit{statistical inference} on the population half-width contrast $\Delta$ or $\Delta(\s)$. 
Split conformal calibration estimates $\zeta_{\x}$ and $\zeta_{\w}$ using the empirical quantiles of fitted residuals. Although this plug-in approach guarantees valid marginal coverage for
a future observation, its calibrated half-width $\wh\Delta$ or $\wh\Delta(\s)$ is contaminated by first-order estimation errors in the fitted
regression functions.
Consequently, valid predictive coverage does
not by itself yield root-$n$ consistent estimators, valid confidence
intervals, or well-calibrated hypothesis tests for the population
predictive gain $\Delta$ or $\Delta(\s)$. 

To avoid restrictive parametric assumptions that could distort the residual distribution under model misspecification, we allow the conditional mean functions to be estimated via flexible nonparametric or machine learning methods.
Conducting valid inference in this flexible regime requires overcoming three intertwined challenges.
First, because residuals depend on unknown conditional means, plug-in empirical quantiles of fitted residuals inherit first-order bias from regression estimation errors \citep{akritas2001residual,muller2009error}.
We resolve this sensitivity by deriving the efficient influence
function (EIF) for the residual quantile and constructing a locally
robust estimating equation that is first-order insensitive to nuisance
perturbations \citep{chernozhukov2022locally}. 
Second, evaluating estimating equations on the same sample used to fit
flexible nuisance models induces overfitting dependence; we employ
cross-fitting to decouple nuisance estimation from target evaluation
\citep{chernozhukov2018double}. 
Third, targeting the conditional gain requires local estimation around
the fixed location, which reduces the effective sample size and
introduces smoothing bias. 
We address this by integrating the orthogonalized estimating equation
with kernel localization \citep{fan2022estimation} and employing
undersmoothing so that localization bias becomes asymptotically
negligible relative to sampling variability
\citep{calonico2018effect}. 

In this paper, we develop a unified semiparametric framework for
estimating and conducting inference on both the marginal and
conditional predictive gain. Our main methodological and theoretical
contributions are threefold:
\begin{itemize}[leftmargin=*, itemsep=2pt, topsep=4pt]
      \item \textbf{Uncertainty-based incremental importance and
          individualized decision framework.} Moving beyond point
        prediction accuracy, we formulate the value of costly
        covariates through the reduction in prediction interval
        half-width at both the population and conditional levels. We
        pair this measure with one-sided hypothesis tests against a
        prespecified nonnegative threshold $\tau$: setting
        $\tau=0$ tests for a positive gain, whereas setting
        $\tau>0$ tests whether the reduction is large enough to
        justify the acquisition cost. 
    \item \textbf{Locally robust semiparametric estimation and asymptotic theory.} Building on the efficient influence function, we construct cross-fitted estimators for $\Delta$ and kernel-localized estimators for $\Delta(\s_0)$, accompanied by consistent variance estimators. We prove that the marginal estimator achieves $\sqrt{n}$-asymptotic normality when the regression and conditional residual density estimators converge at the rate of $o_p(n^{-1/4})$, and the conditional estimator achieves $(nh^{d_s})^{1/2}$-asymptotic normality under an analogous local rate of $o_p\{(nh^{d_s})^{-1/4}\}$.
    \item \textbf{Theoretical comparison with conformal calibration.} We rigorously characterize the asymptotic behavior of plug-in split-conformal residual quantile estimators for $\Delta$ and $\Delta(\s_0)$. We demonstrate that, because plug-in calibration fails to remove first-order regression estimation error, it generally suffers from slower convergence rates or inflated asymptotic variance.
\end{itemize}
Extensive simulations confirm that our semiparametric procedures maintain nominal coverage and type~I error control across diverse error distributions, including skewed and heavy-tailed settings, whereas plug-in conformal inference is often severely distorted.
We further demonstrate the practical utility of our framework through an analysis of oral glucose tolerance testing in NHANES, showing how the value of blood assays varies across age and body mass index, alongside applications to concrete strength, soil organic carbon, and warfarin dosing. 

The remainder of the paper is organized as follows. Section~\ref{sec:formulation} formulates the prediction-interval comparison problem and hypothesis testing framework. Sections~\ref{sec:marginal} and~\ref{sec:4} develop the marginal and conditional estimators respectively, and establish their asymptotic properties alongside theoretical comparisons with conformal calibration. Section~\ref{sec:delta} presents variance estimation, confidence intervals, and hypothesis tests for the predictive gain. Section~\ref{sec:sim} evaluates finite sample performance through simulation studies, and Section~\ref{sec:real} demonstrates the methodology in real data applications.

\section{Problem formulation}\label{sec:formulation}
Consider a regression setting with outcome $Y$ and covariate vector
$\W=(\X\trans, \Z\trans)\trans$, where $\X$ denotes easily available
baseline covariates and $\Z$ denotes additional covariates that are
costly or difficult to acquire. 
The researcher would like to decide whether it is worth making extra
effort to obtain $\Z$.  
We characterize the additional gain from
$\Z$ by the improvement of the prediction interval length.
Let $\S\subseteq\X$ be a prespecified subset of baseline covariates.
Specifically, 
for a given coverage level $1-\alpha\in(0,1)$,
the conditional prediction
interval of $Y$ based on $\X$ is formulated as 
\bse
\pr\{\left| Y- m_\x(\X) \right| \leq
\zeta_\x(\s_0)\mid\S=\s_0\} = 1-\alpha,
\ese
and the conditional prediction interval of $Y$ based on  $\W$ is 
\bse
\pr\{\left| Y- m_\w(\W) \right| \leq \zeta_\w(\s_0)\mid\S=\s_0\} =
1-\alpha.
\ese
Here, $m_\x(\cdot)$ and $m_\w(\cdot)$ are the centers of the
prediction intervals, and can be chosen as, for example, the
conditional mean, conditional
median, or conditional mode. We take the conditional mean as a
specific choice to provide details below. In this case, 
\bse
Y&=&m_\x(\X)+ \epsilon_\x,\\ 
Y&=&m_\w(\W)+ \epsilon_\w,
\ese
where $m_\w(\W)=E(Y \mid \W)$ and $m_\x(\X)=E(Y \mid \X)$.
The notation $\zeta_\w(\s_0), \zeta_\x(\s_0)$ 
reflects the fact that the prediction interval lengths depend on $\s_0$.
The special case of setting
$\S=\emptyset$ corresponds to marginal prediction probability, while the
special case of setting $\S=\X$ corresponds to conditional prediction
probability given the entire easily accessible $\x_0$.
Because incorporating additional covariates is generally expected to
sharpen prediction, one typically anticipates $\zeta_\w(\s_0)\le
\zeta_\x(\s_0)$. We quantify the incremental predictive value of $\Z$
at profile $\s_0$ by the half-width reduction 
\begin{equation}
\Delta(\s_0)=\zeta_\x(\s_0)-\zeta_\w(\s_0).
\end{equation}
When $\S=\emptyset$, we drop the argument $\s_0$ and write $\zeta_\x$,
$\zeta_\w$, and $\Delta=\zeta_\x-\zeta_\w$. 

 In practice, $\Delta(\s_0)$ can be used directly to evaluate the uncertainty reduction against the acquisition cost
  of $\Z$, or formally tested against a prespecified nonnegative threshold $\tau(\s_0)$, which we write as
  $\tau$ when $\S=\emptyset$. Specifically, if $\Delta(\s_0) >
  \tau(\s_0)$, then the gain is considered
substantial and  including $\Z$ in the model is recommended for more
precise 
prediction. In contrast, if $\Delta(\s_0) \le \tau(\s_0)$, then
 the additional gain  from engaging $\Z$ is considered not large enough to
justify the additional cost,  and hence one should be content with using
solely $\X$.
This motivates the one-sided hypothesis testing problem
\bse
H_0: \Delta(\s_0) \le \tau(\s_0), \mbox{ versus } H_1: \Delta(\s_0) > \tau(\s_0).
\ese
Our goal in the subsequent sections is to develop valid estimation and
inference procedures for $\zeta_\x(\s_0)$, $\zeta_\w(\s_0)$, and
$\Delta(\s_0)$. 

\section{Marginal prediction interval estimation}\label{sec:marginal}
We begin with the marginal setting $\S=\emptyset$, which highlights the core methodological ideas. We focus on estimating $\zeta_\x$, as the same procedure applies directly to $\zeta_\w$. Here, the problem reduces to finding $\zeta_\x$ such that
\bse
\pr\{\left| Y_0- m_\x(\X_0) \right| \leq
\zeta_\x\} = 1-\alpha.
\ese
The formulation directly puts the problem in the conformal
  prediction setting, and very naturally one would adopt a split
  conformal estimator, where one part of the data can be used to
  estimate $m_\x(\cdot)$ and the other part performs conformal
  prediction based on the residual quantiles. Interestingly, we find
  that this natural approaches fails, as will explained in detail
  later. Thus, instead, we view $\zeta_\x$ as a parameter of interest
  and cast the problem of estimating $\zeta_\x$ as a semiparameric
  problem, and subsequent develop a semiparametric estimator based on
the efficient influence function.

\subsection{Semiparametric estimation}\label{sec:marginal-semi}
 The estimation of $\zeta_\x$ can be viewed as a semiparametric
   problem because 
$\zeta_\x$ is a finite-dimensional target parameter, while the
regression function 
$m_\x$ and the conditional residual density $f_{\epsilon_\x\mid\x}$
are infinite-dimensional nuisance parameters.
In Section \ref{sec:phix}, we show that the efficient influence
function for estimating $\zeta_\x$ is
\bse
\phi_{\x,\rm{eff}}(\O,\zeta_\x,m_\x,f_{\epsilon_\x\mid\x})=\frac{\phi_\x(\O,\zeta_\x,m_\x,f_{\epsilon_\x\mid\x})}{E[\delta\{\zeta_\x-r(\O,m_\x)\}]},
\ese
where
\bse
\phi_\x(\O,\zeta_\x, m_\x,f_{\epsilon_\x\mid\x})=(1-\alpha)-I\{
  r(\O,m_\x)<\zeta_\x\}-\epsilon_\x
  E[\sign(\epsilon_\x)\delta\{\zeta_\x-r(\O,m_\x)\}\mid\X],
\ese
and  $\epsilon_\x=Y-m_\x(\X)$,
$r(\O,m_\x)\equiv|\epsilon_\x|$, $\O=(\X_i\trans, Y_i)\trans$, the dependence on $f_{\epsilon_\x\mid\x}$ is via $E[\delta\{\zeta-r(\O,m_\x)\}\mid
\X]$. Since the denominator of the efficient influence function is a nonzero
normalizing constant and therefore does not change the root of the estimating
equation,
we subsequently estimate $\zeta_\x$ via solving 
\bse \sumi \phi_\x(\O_i,\zeta_\x, m_\x, f_{\epsilon_\x\mid\x})=0.\ese
We denote the resulting estimator  by $\wh\zeta_\x$.

In implementation, the nuisance functions are replaced by their estimates.
These include $m_\x(\cdot)$ and
$E[\sign(\epsilon_\x)\delta\{\zeta_\x-|Y_i-m_\x(\x_i)|\}\mid
\x_i]$, which can be estimated using flexible nonparametric or machine
 learning methods. To avoid using the same observations both to estimate
 the nuisance functions and to solve the target estimating equation, we
 use two-fold cross-fitting.
Specifically, we partition the observations into two
approximately equal folds,
$D_j=\{\O_i:i\in\mathcal I_j\}$, $j=1,2$, with $|\mathcal I_1|=\lfloor n/2\rfloor$.  
We write $D_{-j}$ for the fold complementary to $D_j$,  use the
superscript $(-j)$ for estimators trained on $D_{-j}$, and use
  superscript $j$ for estimators trained on $D_j$.
On each fold $D_j$, we estimate
$m_\x$ and $f_{\epsilon_\x\mid\x}$,
obtaining $\wh m_\x^{(j)}$ and
$\wh f_{\epsilon_\x\mid\x}^{(j)}$, respectively.  To estimate the
target on $D_1$, we evaluate the estimating equation using only the
nuisance estimators trained on $D_2$, which yields
$\wh\zeta_\x^{(1)}$.  We then reverse the roles of the two folds to
obtain $\wh\zeta_\x^{(2)}$.  The final two-fold cross-fitted estimator
is
\[
\wh\zeta_\x=\frac{\wh\zeta_\x^{(1)}+\wh\zeta_\x^{(2)}}{2}.
\]
Thus, each observation is used once to evaluate the target estimating
equation and once to train the nuisance estimators, but it is never
used for both purposes within the same fold-specific estimating
equation. Further details are given in Algorithm~\ref{alg:crossfit-semi}.

\begin{algorithm}[h]
\caption{Two-fold cross-fitted semiparametric estimation of $\zeta_\x$}
\label{alg:crossfit-semi}
\begin{algorithmic}[1]
\Require Data $\{\O_i=(\X_i\trans,Y_i)\trans\}_{i=1}^n$ and
miscoverage level $\alpha$
\Ensure Cross-fitted estimator $\wh\zeta_\x$
\State Partition $\{1,\ldots,n\}$ into two disjoint,
approximately equal sets $\mathcal I_1$ and $\mathcal I_2$.
\For{$j\in\{1,2\}$}
  \State Use $\{\O_i:i\in\mathcal I_j\}$ to obtain
  $\wh m_\x^{(j)}$.
  \State Compute the training residuals
  $\wh\epsilon_{\x,i}^{(j)}=Y_i-\wh m_\x^{(j)}(\X_i)$,
  $i\in\mathcal I_j$, and estimate
  $\wh f_{\epsilon_\x\mid\x}^{(j)}$.
  \State Define
  $\wh A_\x^{(j)}(\zeta,\x)
  =\wh f_{\epsilon_\x\mid\x}^{(j)}(\zeta,\x)
  -\wh f_{\epsilon_\x\mid\x}^{(j)}(-\zeta,\x)$.
\EndFor
\For{$j\in\{1,2\}$}
For each $i\in\mathcal I_j$, compute
  $\wh\epsilon_{\x,i}^{(-j)}
  =Y_i-\wh m_\x^{( -j)}(\X_i)$.
  \State Obtain $\wh\zeta_\x^{(j)}$ by solving
  \Statex \hspace{\algorithmicindent}
  $\displaystyle
  \sum_{i\in\mathcal I_j}
  \left[
  (1-\alpha)-I\{|\wh\epsilon_{\x,i}^{(-j)}|<\zeta\}
  -\wh\epsilon_{\x,i}^{(-j)}
   \wh A_\x^{(-j)}(\zeta,\X_i)
  \right]=0.$
\EndFor
\State \Return
$\displaystyle
\wh\zeta_\x=(\wh\zeta_\x^{(1)}+\wh\zeta_\x^{(2)})/2$.
\end{algorithmic}
\end{algorithm}

Next, we establish the asymptotic properties of the semiparametric
estimator under some standard regularity conditions.

\begin{assum}
      \label{ass:eps0}
      For $\v\in\{\x,\w\}$, $E(\epsilon_\v^2 ) < \infty$.
    \end{assum}
          
    \begin{assum}
      \label{ass:lip0}
      For $\v\in\{\x,\w\}$, the conditional density $f_{\epsilon_\v\mid \v}$ exists and is Lipschitz 
      continuous in neighborhoods of $\zeta_\v$ and $-\zeta_\v$, i.e., there exist constants
      $L>0$, $C_2>0$, and $c_2>0$ such that for all
      $s,t\in[\zeta_\v-c_2,\zeta_\v+c_2]$ or
      $s,t\in[-\zeta_\v-c_2,-\zeta_\v+c_2]$,
      \bse
      |f_{\epsilon_\v\mid \v}(s,\v)-f_{\epsilon_\v\mid \v}(t,\v)| \le L|s-t|\quad\mbox{and}\quad
        \sup_\x
        |f_{\epsilon_\v\mid \v}(\zeta_\v,\v)|\vee
        |f_{\epsilon_\v\mid \v}(-\zeta_\v,\v)| \le C_2.
        \ese
        Moreover, $\inf_\v f_{\epsilon_\v\mid\v}\{\zeta_\v,\v\}>0$.
        \end{assum}

        Assumption~\ref{ass:eps0} is a mild moment condition that guarantees the stability of the influence-function based expansion. Assumption~\ref{ass:lip0} imposes local smoothness of the conditional residual density at the two boundary points $\pm\zeta_\v$, where the prediction interval changes coverage. The lower bound rules out a flat quantile, ensuring that small changes in the coverage probability correspond to well-behaved changes in the interval half-length. 
        \begin{Th}\label{th:semi0}
            For $\v\in\{\x,\w\}$, under
            Assumptions~\ref{ass:eps0}--\ref{ass:lip0}, assume that, 
            $\sup_\v|\wh m_\v^{(j)}(\v)-m_\v(\v)|=o_p(n^{-1/4})$ and $\sup_{\zeta}\sup_\v
            |\wh f_{\epsilon_\v\mid\v}^{(j)}(\zeta,\v)
            -f_{\epsilon_\v\mid\v}(\zeta,\v)|
            =o_p(n^{-1/4})$,
          for $j\in\{1,2\}$. Then $\wh\zeta_\v$ satisfies
            \be
            \label{equ:theo1-linear0}
            n^{1/2}(\wh\zeta_\v-\zeta_\v)
            =-n^{-1/2}g_{\v}^{-1}\sumi
            \phi_\v(\O_i,\zeta_\v,m_\v,f_{\epsilon_\v\mid\v})+o_p(1).
            \ee
            Furthermore,
            \bse
            n^{1/2}(\wh\zeta_\v-\zeta_\v)
            \xrightarrow{d} N\big(0, \sigma_\v\big),
            \ese
            where
            $\sigma_\v=
            g_{\v}^{-2}
            E\{\phi_\v^2(\O_i, \zeta_\v, m_\v,f_{\epsilon_\v\mid\v})\}$, and
            $g_{\v}=f_{\epsilon_\v}(\zeta_\v)+f_{\epsilon_\v}(-\zeta_\v)$.
          \end{Th}
          
Define $\Delta\equiv\zeta_\x-\zeta_\w$, together with $\wh\Delta\equiv\wh\zeta_\x-\wh\zeta_\w$. Applying Theorem~\ref{th:semi0} jointly to $\v=\x$ and $\v=\w$ yields the following corollary.
          \begin{Cor}[Marginal, Semiparametric]\label{th:semi-delta-uncondi}
          Under the conditions of Theorem~\ref{th:semi0}, 
          \bse
          n^{1/2}(\wh\Delta-\Delta)\xrightarrow{d}N(0,\sigma_{\x\x}^0+\sigma_{\w\w}^0-2\sigma_{\x\w}^0),
          \ese
          where $\v_1,\v_2 \in\{\x,\w\}$, 
          \bse \sigma_{\v_1\v_2}^0\equiv E[\phi_{\v_1}\{\O_i, \zeta_{\v_1},m_{\v_1},f_{\epsilon_{\v_1}\mid\v_1}\}\phi_{\v_2}\{\O_i, \zeta_{\v_2},m_{\v_2},f_{\epsilon_{\v_2}\mid\v_2}\}]/(g_{\v_1}g_{\v_2}).\ese
          \end{Cor}
          
Theorem~\ref{th:semi0} shows that the semiparametric estimator admits an asymptotically linear representation with influence function $\phi_\v$. A key feature is the robustness with respect to the nuisance estimates: the regression function and the conditional density only need to converge at the rate $o_p(n^{-1/4})$. This rate is compatible with many modern nonparametric and machine learning procedures under standard smoothness or complexity assumptions. Consequently, $\wh\zeta_\v$ can retain the root-$n$ convergence rate even when the nuisance functions are estimated flexibly.

\subsection{Conformal prediction approach}
We consider the split data version of conformal prediction. Based on observations $\O_{n_1+1}, \dots,\O_n$, we obtain the same estimator $\wh m_\x(\x)$ as in Section \ref{sec:marginal-semi}. The subsequent estimation of $\zeta_\x$ is an empirical quantile problem for the absolute residuals. Equivalently, it can be estimated by minimizing
\bse
\sumia \rho_{1-\alpha}\{\left|y_i-\wh m_\x(\x_i)\right|-\zeta_\x\},
\ese
where $\rho_{1-\alpha}(t)\equiv t\{1-\alpha-I(t<0)\}$ is the check function. We denote the resulting estimator  by $\wt\zeta_\x$. Its statistical behavior, however, is affected directly by the quality of $\wh m_\x$, because the residual quantile is computed after plugging in the estimated regression function.  We similarly use two-fold cross-fitting. For $j=1,2$, let $\wt\zeta_\x^{(j)}$ be estimated on $D_j$ using $\wh m_\x^{(-j)}$, and set $\wt\zeta_\x=(\wt\zeta_\x^{(1)}+\wt\zeta_\x^{(2)})/2$. 

To establish the asymptotic properties of the conformal estimator, we
impose the following marginal density condition, which is analogous to
Assumption~\ref{ass:lip0}.

    \begin{assum}
        \label{ass:lip20}
        For $\v\in\{\x,\w\}$, the density of $\epsilon_\v$, denoted by
        $f_{\epsilon_\v}(t)$, exists.
        There exist constants $1>c_3>0$ and $L'>0$ such that for all
        $t\in[\zeta_\v-c_3,\zeta_\v+c_3]\cup[-\zeta_\v-c_3,-\zeta_\v+c_3]$,
        \bse
        0<f_{\epsilon_\v}(t)<L'.
        \ese
    \end{assum}

\begin{Th}\label{th:conformal0}
 For $\v\in\{\x,\w\}$, under Assumption~\ref{ass:lip20}, let
 $a_n^{(j)}\equiv\sup_\v|\wh m_\v^{(j)}-m_\v|=o_p(1)$, for $j\in\{1,2\}$.  
 Then $\wt\zeta_\v$ satisfies
 \be
 \label{equ:theo2-linear0}
  n^{1/2}(\wt\zeta_\v-\zeta_\v)
 =-g_\v^{-1}n^{1/2}\overline B_\v
 +g_\v^{-1}n^{-1/2}\sumi
 \psi_{1-\alpha}(|\epsilon_{\v,i}|-\zeta_\v)
 +O_p(\overline a_n^2 n^{1/2})+o_p(1),
 \ee
 where
 \[
 \overline B_\v=\frac{B_\v^{(1)}+B_\v^{(2)}}{2},\qquad
 \overline a_n^2=\frac{(a_n^{(1)})^2+(a_n^{(2)})^2}{2},
 \]
 \[
 B_\v^{(j)}=E[\{f_{\epsilon_\v\mid\v}(\zeta_\v,\V)
 -f_{\epsilon_\v\mid\v}(-\zeta_\v,\V)\}
 \{\wh m_\v^{(j)}(\V)-m_\v(\V)\}]=O_p(a_n^{(j)}),
 \]
 $g_\v=f_{\epsilon_\v}(\zeta_\v)+f_{\epsilon_\v}(-\zeta_\v)$,
 and $\psi_{1-\alpha}(t)=1-\alpha-I(t<0)$.
\end{Th}

Combining the two fold-specific expansions in
Theorem~\ref{th:conformal0} for $\v=\x$ and $\v=\w$ yields the
corresponding expansion for $\wt\Delta
\equiv\wt\zeta_\x-\wt\zeta_\w$.
\begin{Cor}[Marginal, Conformal]\label{th:conf-delta-uncondi} 
Under the conditions of Theorem~\ref{th:conformal0}, assume 
$\sup_\v|\wh m_\v^{(j)}(\v)-m_\v(\v)|=o_p(n^{-1/4})$
 for $\v\in\{\x,\w\}$ and $j\in\{1,2\}$. Then
\begin{align}
n^{1/2}(\wt\Delta-\Delta)
=&n^{-1/2}\sumi\left\{
g_\x^{-1}\psi_{1-\alpha}(|\epsilon_{\x,i}|-\zeta_\x)
-g_\w^{-1}\psi_{1-\alpha}(|\epsilon_{\w,i}|-\zeta_\w)
\right\}\notag\\
&-n^{1/2}(g_\x^{-1}\overline B_\x-g_\w^{-1}\overline B_\w)+O_p(n^{1/2}\overline a_n^2)+o_p(1),
\label{equ:theo-conf-delta-linear0}
\end{align}
where $\overline B_\v$, $B_\v^{(j)}$, $\overline a_n^2$, $g_\v$, and $\psi_{1-\alpha}$
are as in Theorem~\ref{th:conformal0}.
\end{Cor}
Because the nuisance estimation effect of $\wh m_\v$ does not generally vanish in the conformal prediction method, this corollary shows that in general $\wt\Delta$ is not a root-$n$ consistent estimator of $\Delta$.

The expansion in Theorem~\ref{th:conformal0} separates two sources of uncertainty. The first term is the usual empirical quantile fluctuation. The third term, of order $O_p\{n^{1/2}(a_n^{(1)}+a_n^{(2)})\}$, records the effect of replacing $m_\v$ by the two fold-specific estimators $\wh m_\v^{(1)}$ and $\wh m_\v^{(2)}$.  This term is not ignorable unless both estimators achieve the rate $o_p(n^{-1/2})$, which is generally infeasible. Even in the situation when $\wh m_\v$ is parametrically estimated and has the $n^{1/2}$ rate, it will still affect the estimation variability of $\wt\zeta_\v$. 

\begin{Rem}[Comparison]
When nonparametric or machine learning methods are used to estimate $\wh m_\x$, the convergence rate is often slower than $n^{-1/2}$, for example, it can only reach $o_p(n^{-1/4})$. In such cases, the semiparametric method is superior to the conformal prediction method. Specifically, Theorem~\ref{th:semi0} shows that the semiparametric method can achieve the parametric root-$n$ convergence rate in estimating $\zeta_\v$. Conversely, Theorem~\ref{th:conformal0} shows that in such cases, $\wt\zeta_{\v}-\zeta_{\v}=O_p(a_n)$, where $a_n$ is the convergence rate of the fitted regression function $\wh m_\v$. Hence, the convergence rate of $\wt\zeta_\v$ is slower than root-$n$. Furthermore, even if a parametric method is used so that $\wh m_\x$ converges  at the $n^{-1/2}$ rate, it still leads to an inflated asymptotic variance for $\wt\zeta_\v$ compared to the semiparametric estimator $\wh\zeta_\v$.
\end{Rem}

\begin{Rem}[Symmetry]
  \label{rem:sym}
  For the conformal prediction estimator $\wt\zeta_\v$, the leading effect of estimating the nuisance regression function is proportional to $E\{f_{\epsilon_\v\mid\V}(\zeta_\v,\V)-f_{\epsilon_\v\mid\V}(-\zeta_\v,\V)\}$. Hence, when the distribution of $\epsilon_\v\mid \V$ is symmetric, this coefficient vanishes and the first-order impact of the regression error in $\wh m_\v$ disappears. In this case,
  \bse
   n^{1/2}(\wt\zeta_\v-\zeta_\v)
  &=&n^{-1/2}g_{\v}^{-1}\sumi\psi_{1-\alpha}(| \epsilon_{\v,i}|-\zeta_\v)+O_p(\overline a_n^2 n^{1/2})+o_p(1),
  \ese
and the root-$n$ convergence rate of $\wt\zeta_\v$ only requires $\sup_\v|\wh m_\v^{(j)}-m_\v|=o_p(n^{-1/4})$ for $j\in\{1,2\}$.
 
This can be corroborated from another perspective. Comparing $\phi_\v$ with $\psi_\v$, we see that $\phi_\v$ attains orthogonality by explicitly subtracting the first-order nuisance effect through the additional term
\[
-\epsilon_\v E[\sign(\epsilon_\v)\delta\{\zeta_\v-r(\O,m_\v)\}\mid \V]
=-\epsilon_\v \{f_{\epsilon_\v\mid\v}(\zeta_\v,\V)
-f_{\epsilon_\v\mid\v}(-\zeta_\v,\V)\}.
\]
When the distribution of $\epsilon_\v\mid\V$ is symmetric, this correction term also vanishes and the two estimators are asymptotically equivalent.
\end{Rem}

\section{Conditional prediction interval estimation}\label{sec:4}
We now turn to the conditional setting with a nonempty $\s_0$, focusing again on $\zeta_\x(\s_0)$ as the same procedures apply to $\zeta_\w(\s_0)$.
\subsection{Semiparametric estimation}
Relative to the marginal case, the main additional difficulty is that the target prediction interval is local to a stratum of $\S$. We therefore combine the semiparametric estimating equation with kernel localization and estimate $\zeta_\x(\s_0)$ via solving
\bse
\sumi \phi_\x\{\O_i,\zeta_\x(\s_0),m_\x, f_{\epsilon_\x\mid\x}\}K_h(\S_i-\s_0)=0,
\ese
where $K_h(\cdot)=h^{-d_s}K(\cdot/h)$, $K(\cdot)$ is a kernel function, and $d_s$ is the dimension of $\S$. The bandwidth $h$ controls the trade-off between bias and variance: smaller $h$ uses more local data hence reduces estimation bias while larger $h$ uses more observations and reduces variance.
We denote the resulting estimator  by $\wh\zeta_\x(\s_0)$. Next, we
establish the asymptotic properties of $\wh\zeta_\v(\s_0)$ under the
following regularity conditions.

\begin{assum}
      \label{ass:eps}
    For $\v\in\{\x,\w\}$, there exist constants $C_1>0$ and $c_1>0$ such that
      $\sup_{\s\in\mathcal N(\s_0,c_1)} E(\epsilon_\v^2 \mid \S=\s) \le C_1$,
      where $\mathcal N(\s_0,c_1)$ denotes an open neighborhood of $\s_0$.
    \end{assum}
    \begin{assum}
      \label{ass:lip}
      For $\v\in\{\x,\w\}$, the conditional density $f_{\epsilon_\v\mid \v}$ exists and is Lipschitz 
      continuous in neighborhoods of $\zeta_\v(\s_0)$ and $-\zeta_\v(\s_0)$, i.e., there exist constants
      $L>0$, $C_2>0$, and $c_2>0$ such that for all
      $s,t\in[\zeta_\v(\s_0)-c_2,\zeta_\v(\s_0)+c_2]$ or
      $s,t\in[-\zeta_\v(\s_0)-c_2,-\zeta_\v(\s_0)+c_2]$,
      \bse
      |f_{\epsilon_\v\mid \v}(s,\v)-f_{\epsilon_\v\mid \v}(t,\v)| \le L|s-t|\quad\mbox{and}\quad
        \sup_\x
        |f_{\epsilon_\v\mid \v}\{\zeta_\v(\s_0),\v\}|\vee
        |f_{\epsilon_\v\mid \v}\{-\zeta_\v(\s_0),\v\}| \le C_2.
        \ese
        Moreover, $\inf_\v f_{\epsilon_\v\mid\v}\{\zeta_\v(\s_0),\v\}>0$.
        \end{assum}
        
        \begin{assum}
          \label{ass:kernel}
          $K(\cdot)$ is a $t$th order symmetric kernel and $\int K^2(\u)d\u<\infty$.
        \end{assum}
        Assumption~\ref{ass:eps} is a local moment condition around $\s_0$; it prevents observations near $\s_0$ from having unstable residual variation. Assumption~\ref{ass:lip} is used for the semiparametric expansion because the estimating function depends on the conditional density of the residual given the covariates. Assumption~\ref{ass:kernel} is a standard assumption and determines the order of the kernel estimation bias.

        \begin{Th}\label{th:semi}
            For $\v\in\{\x,\w\}$, under
            Assumptions~\ref{ass:eps}--\ref{ass:kernel}, assume
            $nh^{d_s}\to\infty$, $nh^{4t+d_s}\to0$, and
            $\sup_\v|\wh m_\v^{(j)}(\v)-m_\v(\v)|
            =o_p\{(nh^{d_s})^{-1/4}\}$, 
            $\sup_{\zeta}\sup_\v|\wh
            f_{\epsilon_\v\mid\v}^{(j)}(\zeta,\v)
            -f_{\epsilon_\v\mid\v}(\zeta,\v)|
            =o_p\{(nh^{d_s})^{-1/4}\}$, 
            for $j\in\{1,2\}$.
            Then $\wh\zeta_\v(\s_0)$ satisfies
            \be
            \label{equ:theo1-linear}
            &&(nh^{d_s})^{1/2}\{\wh\zeta_\v(\s_0)-\zeta_\v(\s_0)-b_\v^{\rm semi}\}\notag\\
            &=&-\frac{(nh^{d_s})^{-1/2}\sumi\bigl\{\phi_\v(\O_i,\zeta_\v(\s_0),m_\v,f_{\epsilon_\v\mid\v})K_h(\S_i-\s_0)-E(\phi_\v K_h)\bigr\}}{g_{\v,0}f_\S(\s_0)}+o_p(1),
            \ee
            where $E(\phi_\v K_h)=E\{\phi_\v(\O_i,\zeta_\v(\s_0),m_\v,f_{\epsilon_\v\mid\v})K_h(\S_i-\s_0)\}$ and
            \bse
            b_{\v}^{\rm semi}\equiv h^{t}
            \left\{g_{\v,0}f_\S(\s_0)\right\}^{-1}
            \frac{\partial^t  E[\phi_\v\{\O_i,\zeta_\v(\s),f_{\epsilon_\v\mid\v}\}\mid\s_0]f_{\S}(\s_0) }{\partial \s_0^t}\int \u^tK(\u)d\u+o(h^t).
            \ese
            Further it satisfies
            \bse
            (nh^{d_s})^{1/2}\{\wh\zeta_\v(\s_0)-\zeta_\v(\s_0)-b_{\v}^{\rm semi}\}
            \to N\big(0, \frac{\int K^2(\u)d\u}{f_\S(\s_0)}\sigma_\v\big)
            \ese
            in distribution as $n\to\infty$, where
            \bse
            \sigma_\v=g_{\v,0}^{-2}
            E[\phi_\v^2\{\O_i, \zeta_\v(\s_0),f_{\epsilon_\v\mid\v}\}\mid\s_0],
            \ese
            where $\phi_\v=(1-\alpha)-I\{
            r(\O,m_\v)<\zeta\}-\epsilon_\v E[\sign(\epsilon_\v)\delta\{\zeta-r(\O,m_\v)\}\mid\V]$
            represents the numerator of $\phi_\v$ and
             $g_{\v,0}=f_{\epsilon_\v\mid\s}\{\zeta_\v(\s_0),\s_0\}
             +f_{\epsilon_\v\mid\s}\{-\zeta_\v(\s_0),\s_0\}$.
            \end{Th}
            
            Define $\Delta(\s_0)\equiv\zeta_\x(\s_0)-\zeta_\w(\s_0)$ and the estimators $\wh\Delta(\s_0)\equiv\wh\zeta_\x(\s_0)-\wh\zeta_\w(\s_0)$ and $\wt\Delta(\s_0)\equiv\wt\zeta_\x(\s_0)-\wt\zeta_\w(\s_0)$. The joint application of Theorem~\ref{th:semi} to $\v=\x$ and $\v=\w$ yields the following corollary.

            \begin{Cor}[Conditional, Semiparametric]\label{th:semi-delta}
            Under the conditions of Theorem~\ref{th:semi},  we have
            \bse
            (nh^{d_s})^{1/2}\{\wh\Delta(\s_0)-\Delta(\s_0)-(b_\x^{\rm semi}-b_\w^{\rm semi})\}\xrightarrow{d}N(0,v_\Delta^{\rm semi}),
            \ese
            with \bse v_\Delta^{\rm semi}=(\sigma_{\x\x}+\sigma_{\w\w}-2\sigma_{\x\w})\int K^2(\u)d\u/f_\S(\s_0),\ese
            where $b_\v^{\rm semi}$, $g_{\v,0}$ and $\phi_\v$ are as in
            Theorem~\ref{th:semi}, and $\v_1, \v_2 \in\{\x,\w\}$,
            \bse \sigma_{\v_1\v_2}\equiv E[\phi_{\v_1}\{\O_i, \zeta_{\v_1}(\s_0),m_{\v_1},f_{\epsilon_{\v_1}\mid\v_1}\}\phi_{\v_2}\{\O_i, \zeta_{\v_2}(\s_0),m_{\v_2},f_{\epsilon_{\v_2}\mid\v_2}\}\mid\s_0]/(g_{\v_1,0}g_{\v_2,0}).\ese
            \end{Cor}
            
Theorem~\ref{th:semi} is an analogue of Theorem~\ref{th:semi0}. The effective sample size is $nh^{d_s}$, which reflects the number of observations  in the neighborhood of $\s_0$. The term $b_\v^{\rm semi}$ is the bias induced by kernel localization and is of order $O(h^t)$, which is $o\{(nh^{d_s})^{-1/2}\}$ under undersmoothing condition $nh^{2t+d_s}\to0$. Importantly, only a convergence rate of $o_p\{(nh^{d_s})^{-1/4}\}$ is needed for the nuisance component estimation, reflecting the benefit of the semiparametric estimating equation.

\subsection{Conformal prediction approach}
The subsequent estimation of $\zeta_\x(\s_0)$ can also be formulated as a localized quantile problem for the absolute residuals. We estimate it by minimizing
\bse
\sumia \rho_{1-\alpha}\{\left|y_i-\wh m_\x(\x_i)\right|-\zeta_\x(\s_0)\}K_h(\s_i-\s_0)
\ese
to obtain $\wt\zeta_\x(\s_0)$, where $\rho_{1-\alpha}(t)\equiv t\{1-\alpha-I(t<0)\}$ is the check function. This estimator is a kernel-weighted version of split conformal calibration. 
To establish the asymptotic properties of the localized conformal
estimator $\wt\zeta_\v(\s_0)$, we impose the following condition on the conditional density.

\begin{assum}
  \label{ass:lip2}
  For $\v\in\{\x,\w\}$, the conditional density of $\epsilon_\v$ given $\S=\s$, denoted by
$f_{\epsilon_\v\mid\S}(t\mid\s)$, exists.
There exist constants $1>c_3>0$ and $L'>0$ such that for all
$t\in[\zeta_\v(\s_0)-c_3,\zeta_\v(\s_0)+c_3]\cup[-\zeta_\v(\s_0)-c_3,-\zeta_\v(\s_0)+c_3]$,
\bse
0<f_{\epsilon_\v\mid \s}(t,\s_0)\quad\mbox{and}\quad \sup_\s f_{\epsilon_\v\mid \s}(t,\s)<L'.
\ese
Moreover, $f_{\epsilon_\v\mid\s}\{\zeta_\v(\s_0),\s\}$ and
$f_{\epsilon_\v\mid\s}\{-\zeta_\v(\s_0),\s\}$ are continuous at $\s_0$.
\end{assum}

Assumption~\ref{ass:lip2} is weaker in the sense that it only concerns the conditional density of the residual given $\S=\s$ near the  boundary; it is sufficient for the localized conformal quantile expansion.

\begin{Th}\label{th:conformal}
 For $\v\in\{\x,\w\}$, under Assumptions~\ref{ass:kernel}
 and~\ref{ass:lip2}, assume $nh^{d_s}\to\infty$ and
 $\sup_\v|\wh m_\v^{(j)}-m_\v|
 =o_p\{(nh^{d_s})^{-1/2}\}$ for
 $j\in\{1,2\}$.
 Then $\wt\zeta_\v(\s_0)$ satisfies
 \be
 \label{equ:theo2-linear}
&&(nh^{d_s})^{1/2}\{\wt\zeta_\v(\s_0)-\zeta_\v(\s_0)-b_\v^{\rm conf}\}\notag\\
&=&\frac{(nh^d_s)^{-1/2}\sumi\bigl[\psi_{1-\alpha}\{|\epsilon_{\v,i}|-\zeta_\v(\s_0)\}K_h(\S_i-\s_0)-E(\psi_{\v,1-\alpha}K_h)\bigr]}{g_{\v,0}f_\S(\s_0)}+o_p(1),
\ee
where $E(\psi_{\v,1-\alpha}K_h)=E\bigl[\psi_{1-\alpha}\{|\epsilon_{\v,i}|-\zeta_\v(\s_0)\}K_h(\S_i-\s_0)\bigr]$,
$\epsilon_{\v,i}=Y_i-m_\v(\V_i)$ and
\bse
b_{\v}^{\rm conf}=\{g_{\v,0}f_\S(\s_0)\}^{-1}h^t
\frac{\partial^t\bigl[E\{\psi_{1-\alpha}(|\epsilon_\v|-\zeta_\v(\s_0))\mid\s\}f_\S(\s)\bigr]}{\partial\s^t}\bigg|_{\s=\s_0}
\int\u^tK(\u)d\u+o(h^t).
\ese
Further, it satisfies, as $n\to\infty$,
\bse
(nh^{d_s})^{1/2}\{\wt\zeta_\v(\s_0)-\zeta_\v(\s_0)-b_\v^{\rm conf}\}
\xrightarrow{d}N\left(0,\frac{\int K^2(\u)d\u}{f_\S(\s_0)}g_{\v,0}^{-2}\alpha(1-\alpha)\right),
\ese
where $\psi_{1-\alpha}(t)=1-\alpha-I(t<0)$ and
 $g_{\v,0}=f_{\epsilon_\v\mid\s}\{\zeta_\v(\s_0),\s_0\}
 +f_{\epsilon_\v\mid\s}\{-\zeta_\v(\s_0),\s_0\}$.
\end{Th}

 Applying Theorem~\ref{th:conformal} jointly to $\v=\x$ and $\v=\w$ yields the corresponding result for $\wt\Delta(\s_0)$.
\begin{Cor}[Conditional, Conformal]\label{th:conformal-delta}
Under the conditions of Theorem~\ref{th:conformal}, we have
\bse
(nh^{d_s})^{1/2}\{\wt\Delta(\s_0)-\Delta(\s_0)-(b_\x^{\rm conf}-b_\w^{\rm conf})\}\xrightarrow{d}N(0,v_\Delta^{\rm conf}),
\ese
with
\bse
v_\Delta^{\rm conf}=\frac{\int K^2(\u)d\u}{f_\S(\s_0)}\!\left\{\frac{\alpha(1-\alpha)}{g_{\x,0}^{2}}+\frac{\alpha(1-\alpha)}{g_{\w,0}^{2}}-\frac{2\rho_0}{g_{\x,0}g_{\w,0}}\right\},
\ese
where $b_\v^{\rm conf}$ and $g_{\v,0}$ are as in
Theorem~\ref{th:conformal}, and
\bse
\rho_0\equiv\cov\bigl(I\{|\epsilon_\x|<\zeta_\x(\s_0)\},\;I\{|\epsilon_\w|<\zeta_\w(\s_0)\}\bigm|\s_0\bigr).
\ese
\end{Cor}
 
\begin{Rem}\label{rem:condition}
Theorem~\ref{th:conformal} is an analogue of Theorem~\ref{th:conformal0}. Theorem~\ref{th:conformal} implies that the effective sample size is $nh^{d_s}$, and $b_\v^{\rm conf}$ is the bias induced by kernel localization. Unlike the semiparametric estimator, the conformal estimator does not eliminate the first-order effect in estimating $m_\v$. Consequently, it requires a faster convergence rate $o_p\{(nh^{d_s})^{-1/2}\}$ for the nuisance parameter estimation, in contrast to $o_p\{(nh^{d_s})^{-1/4}\}$ as required in Theorem~\ref{th:semi}. For many nonparametric or machine learning methods, the convergence rate of $o_p\{(nh^{d_s})^{-1/4}\}$ required in Theorem~\ref{th:semi} is often considered attainable and mild. In contrast, $o_p\{(nh^{d_s})^{-1/2}\}$ required in Theorem~\ref{th:conformal} is much harder to achieve.
\end{Rem}

\begin{Ex}
We use kernel estimation to illustrate the convergence rates required for the nuisance parameter. Suppose that $\wh m_\v$ is estimated using a kernel method with bandwidth $b$ based on a sample of size $n_2$, and that $m_\v$ is $\beta$-smooth. Assume $n_1\asymp n_2\asymp n$. Then, $\sup_\v |\wh m_\v-m_\v|=O_p\left\{b^\beta+(nb^{d_v})^{-1/2}\right\}$, where $d_v$ denotes the dimension of $\v$. Choosing the optimal bandwidth $b\asymp n^{-1/(2\beta+d_v)}$ yields
\[
\sup_\v |\wh m_\v-m_\v|=O_p\left\{n^{-\beta/(2\beta+d_v)}\right\}.
\]
To satisfy the $o_p\{(nh^{d_s})^{-1/2}\}$ condition in Theorem~\ref{th:conformal}, we require a very small bandwidth $h$ that satisfies $h^{d_s}=o(n^{-d_v/(2\beta+d_v)})$. However, to satisfy the $o_p\{(nh^{d_s})^{-1/4}\}$ condition in Theorem~\ref{th:semi}, we only require $h^{d_s}=o(n^{(2\beta-d_v)/(2\beta+d_v)})$, which is much larger.
\end{Ex}

\section{Inference on $\Delta(\s_0)$}\label{sec:delta} 

We now construct consistent estimators of the asymptotic variances for both procedures and use them to conduct inference on $\Delta(\s_0)$.

\subsection{Variance estimation}\label{sec:variance}
To apply Corollaries~\ref{th:semi-delta-uncondi}--\ref{th:conformal-delta}, we provide the corresponding two-fold cross-fitted plug-in variance estimators. For the marginal setting, the formulas below remain valid by taking $K_h(\s_i-\s_0)=1$ throughout and omitting $\s_0$. Let $n_j=|\mathcal I_j|$, $j=1,2$, and, for $i\in\mathcal I_j$, define
\[
\wh\epsilon_{\v,i}^{(-j)}=y_i-\wh m_\v^{(-j)}(\v_i),\qquad \v\in\{\x,\w\}.
\]
Let
\bse
\wh f_\S(\s_0)&\equiv\frac12\sum_{j=1}^2\frac1{n_j}
\sum_{i\in\mathcal I_j}K_h(\s_i-\s_0).
\ese
 Take $c_n=o(1)$  and, for $j=1,2$, define
\bse
\wt\zeta_\v^{(j),+}&\equiv&\argmin_{\zeta\in\mR}
\sum_{i\in\mathcal I_j}\rho_{1-\alpha+c_n}
\{|y_i-\wh m_\v^{(-j)}(\v_i)|-\zeta\}K_h(\s_i-\s_0);\\
\wt\zeta_\v^{(j),-}&\equiv&\argmin_{\zeta\in\mR}
\sum_{i\in\mathcal I_j}\rho_{1-\alpha-c_n}
\{|y_i-\wh m_\v^{(-j)}(\v_i)|-\zeta\}K_h(\s_i-\s_0).
\ese
Then we estimate $g_{\v,0}$ by
\bse
(\wh g_{\v,0})^{-1}=\frac12\sum_{j=1}^2
\frac{\wt\zeta_\v^{(j),+}-\wt\zeta_\v^{(j),-}}{2c_n}.
\ese
\begin{Rem}
  Because \bse g_{\v,0}\equiv f_{\epsilon_\v\mid\s}\{\zeta_\v(\s_0),\s_0\}+f_{\epsilon_\v\mid\s}\{-\zeta_\v(\s_0),\s_0\}=\frac{\partial \pr\{|\epsilon_\v|<\zeta\mid\S=\s_0\}}{\partial \zeta}\mid_{\zeta=\zeta_\v(\s_0)}.\ese
Define the conditional quantile function $Q(p,\s_0)=\inf\{\zeta\in\mR: \pr\{|\epsilon_\v|<\zeta\mid\S=\s_0\}\ge p\}$, which is the inverse function of the conditional CDF $\pr\{|\epsilon_\v|<\zeta\mid\S=\s_0\}$.
Thus, 
\bse
g_{\v,0}^{-1}=\frac{\partial Q(p,\s_0)}{\partial p}\mid_{p=1-\alpha}=\lim_{c_n\to 0}\frac{Q(1-\alpha+c_n,\s_0)-Q(1-\alpha-c_n,\s_0)}{2c_n}=\lim_{c_n\to 0}\frac{\zeta_\v^{+}-\zeta_\v^{-}}{2c_n}.
\ese
The advantage of this approach is that it avoids introducing another kernel to estimate the conditional density.
\end{Rem}

\medskip\noindent\textbf{(a) Semiparametric variance.}
For $\v_1,\v_2\in\{\x,\w\}$,
\bse
\wh\sigma_{\v_1\v_2}\equiv
\frac{\displaystyle\frac12\sum_{j=1}^2\frac1{n_j}
\sum_{i\in\mathcal I_j}\wh\phi_{\v_1,i}^{(-j)}
\wh\phi_{\v_2,i}^{(-j)}K_h(\s_i-\s_0)}
{\wh f_\S(\s_0)\wh g_{\v_1,0}\wh g_{\v_2,0}},
\ese
where, for $i\in\mathcal I_j$,
$\wh\phi_{\v,i}^{(-j)}=\phi_\v(\o_i,\wh\zeta_\v^{(j)}(\s_0),
\wh m_\v^{(-j)},\wh f_{\epsilon_\v\mid\v}^{(-j)})$.
Then the plug-in variance estimator is
\be
\label{var-est:semi}
\wh v_\Delta^{\rm semi}\equiv\frac{\int K^2(\u)d\u}{\wh f_\S(\s_0)}\bigl(\wh\sigma_{\x\x}+\wh\sigma_{\w\w}-2\wh\sigma_{\x\w}\bigr).
\ee

\medskip\noindent\textbf{(b) Conformal variance.}
Let
\bse
\wh\rho_0&\equiv
\frac{1}{2\wh f_\S(\s_0)}\sum_{j=1}^2\frac1{n_j}
\sum_{i\in\mathcal I_j}
I\{|\wh\epsilon_{\x,i}^{(-j)}|<\wt\zeta_\x^{(j)}(\s_0)\}
I\{|\wh\epsilon_{\w,i}^{(-j)}|<\wt\zeta_\w^{(j)}(\s_0)\}
K_h(\s_i-\s_0)-(1-\alpha)^2,
\ese
Then the plug-in variance estimator is
\be
\label{var-est:conf}
\wh v_\Delta^{\rm conf}\equiv\frac{\int K^2(\u)d\u}{\wh f_\S(\s_0)}\!\left\{\frac{\alpha(1-\alpha)}{\wh g_{\x,0}^{2}}+\frac{\alpha(1-\alpha)}{\wh g_{\w,0}^{2}}-\frac{2\wh\rho_0}{\wh g_{\x,0}\wh g_{\w,0}}\right\}.
\ee

\begin{Pro}\label{pro:var-cons}
Under the assumptions of  Corollaries~\ref{th:semi-delta}
and~\ref{th:conformal-delta} and  $c_n\to0$, assume
$nh^{d_s}c_n^2\to\infty$ and $h^tc_n^{-1}\to0$.
\bse
\wh v_\Delta^{\rm semi}\xrightarrow{p} v_\Delta^{\rm semi},\qquad
\wh v_\Delta^{\rm conf}\xrightarrow{p} v_\Delta^{\rm conf}.
\ese
 For the marginal setting, under the assumptions of
Corollary~\ref{th:semi-delta-uncondi}, let $c_n\to0$ and
$nc_n^2\to\infty$. Then the first result reads
$\wh v_\Delta^{\rm semi}\xrightarrow{p}
\sigma_{\x\x}^0+\sigma_{\w\w}^0-2\sigma_{\x\w}^0$ by taking $K_h=1$.
\end{Pro}

\subsection{Hypothesis test}\label{sec:test}
We now construct a test of
\bse
H_0:\Delta(\s_0)\le\tau(\s_0)\quad\mbox{versus}\quad H_1:\Delta(\s_0)>\tau(\s_0),
\ese
where $\tau(\s_0)$  is a pre-determined threshold for evaluating the gain. Let the nominal level of the test be $\alpha'$.

For the two methods, define the test statistics
\be
&&T_n^{\rm semi}\equiv(nh^{d_s}_{\rm semi})^{1/2}\{\wh\Delta(\s_0)-\tau(\s_0)\}/(\wh v_\Delta^{\rm semi})^{1/2},\label{eq:teststat}\\
&&T_n^{\rm conf}\equiv(nh ^{d_s}_{\rm conf})^{1/2}\{\wt\Delta(\s_0)-\tau(\s_0)\}/(\wh v_\Delta^{\rm conf})^{1/2}.\label{eq:teststat2}
\ee
Note that we stress the different bandwidths used in the two methods because they satisfy different requirements and are indeed two different values. Specifically $h_{\rm semi}$ is much larger than $h_{\rm conf}$.
Accordingly, we rejection $H_0$ when
\be\label{eq:inference}
T_n^{\rm semi}>z_{1-\alpha'}, \mbox{ or }  T_n^{\rm conf}>z_{1-\alpha'}
\ee
depending on if we perform the test based on semiparametric method or based on the conformal prediction method. Here $z_{1-\alpha'}$ is the standard normal $1-\alpha'$ quantile.

When $nh^{2t+d_s}_{\rm semi}\to 0$ and $h$ satisfies conditions required in Theorem \ref{th:semi}, the bias terms will vanish. Then by Corollary~\ref{th:semi-delta},
\bse
T_n^{\rm semi}-(nh^{d_s}_{\rm semi})^{1/2}\{\Delta(\s_0)-\tau(\s_0)\}/(\wh v_{\Delta}^{\rm semi})^{1/2}\xrightarrow{d}N(0,1).
\ese
Moreover,
\bse
\pr(T_n^{\rm semi}>z_{1-\alpha'})\to 1-\Phi\!\left(z_{1-\alpha'}-(nh^{d_s}_{\rm semi})^{1/2}\{\Delta(\s_0)-\tau(\s_0)\}/(\wh v_{\Delta}^{\rm semi})^{1/2}\right).
\ese
Thus, under $H_0$, $\pr(T_n^{\rm semi}>z_{1-\alpha'})\le\alpha'$; under $H_1$, if there exists a constant $c>0$ such that $\Delta(\s_0)-\tau(\s_0)>c$, then $\pr(T_n^{\rm semi}>z_{1-\alpha'})\to 1$. We have the  parallel results for $T_n^{\rm conf}$.

Using the same normal approximation, when $nh^{2t+d_s}_{\rm conf}\to 0$ and $h_{\rm conf}$ satisfies Conditions required in Theorem~\ref{th:conformal}, a two-sided $1-\alpha'$ confidence interval for $\Delta(\s_0)$ is given by
\be
\label{eq:interval}
\wh\Delta(\s_0)\;\pm\;z_{1-\alpha'/2}\sqrt{\wh v_\Delta^{\rm semi}/(nh^{d_s}_{\rm semi})},\qquad
\wt\Delta(\s_0)\;\pm\;z_{1-\alpha'/2}\sqrt{\wh v_\Delta^{\rm conf}/(nh^{d_s}_{\rm conf})}.
\ee

In the marginal case where $\S=\emptyset$, the test statistic in \eqref{eq:teststat} for the semiparametric method reduces to $T_n^{\rm semi}=n^{1/2}(\wh\Delta-\tau)/(\wh v_\Delta^{\rm semi})^{1/2}$, and the confidence interval in \eqref{eq:interval} reduces to $\wh\Delta\pm z_{1-\alpha'/2}\sqrt{\wh v_\Delta^{\rm semi}/n}$. However, for the conformal method, the nonvanishing estimation bias generally prevents standard inference based on $\wt\Delta$.

\section{Numerical experiments}\label{sec:sim}

In this section, we evaluate the finite-sample performance of the proposed
estimators of $\zeta_\x$ and $\zeta_\w$ and the corresponding inference
procedures for the contrast $\Delta=\zeta_\x-\zeta_\w$, which
quantifies the value of the additional covariate $Z$ in terms of prediction.

\subsection{Marginal Prediction Interval}
We first consider the marginal target where $\S=\emptyset$. Let
$X\sim\operatorname{Unif}(-1,1)$ and
$Z\sim\operatorname{Unif}(-\sqrt{3},\sqrt{3})$ be independent. Let
$G\sim N(0,1)$ and $C\sim\operatorname{Bernoulli}(p)$, with $p=0.2$,
where $G$, $C$, $X$ and $Z$ are mutually independent. We
consider three for cases for the standardized error
  written as $\epsilon_0$: (a) symmetric errors, with $\epsilon_0=G$, (b) right-skewed errors, with
$\epsilon_0=\{G+5(C-p)\}/
\sqrt{1+25p(1-p)}$, and (c) left-skewed errors, with $\epsilon_0=
-\{G+5(C-p)\}/
\sqrt{1+25p(1-p)}$.
These errors have mean zero and variance one. 
The outcome is generated according to
\bse
Y=m_0(X)+\beta Z+\sigma(X)\epsilon_0,
\ese
where
\bse
m_0(x)=x+2x^2+\frac{1}{2}\sin(\pi x),\qquad
\sigma(x)=1+0.4x.
\ese
The regression functions for
$X$ and $\W=(X,Z)\trans$ are
$m_\x(x)=m_0(x)$ and $m_\w(x,z)=m_0(x)+\beta z$, respectively. The
corresponding residuals are
$\epsilon_\x=\beta Z+\sigma(X)\epsilon_0$ and $\epsilon_\w=\sigma(X) \epsilon_0$.
The coefficient $\beta$ controls the incremental contribution of
$Z$. We set the nominal miscoverage level of the prediction interval to
$\alpha=0.1$.
All estimators are implemented with the two-fold cross-fitting procedure in
Algorithm~\ref{alg:crossfit-semi}.
Nuisances are estimated by kernel regression. The covariate smoother
uses an 
Epanechnikov kernel, 
the conditional
density 
estimator uses a Gaussian kernel. Bandwidths are selected by
cross-validation. We use $n\in\{1000,2000,4000,8000,16000\}$ and
$R=1000$ replications. We report the left- and right-skewed
  designs in the main text and relegate the symmetric error results to
Appendix~\ref{sec:sim-sym}.

\begin{figure}[H]
  \centering
\includegraphics[width=0.98\textwidth]{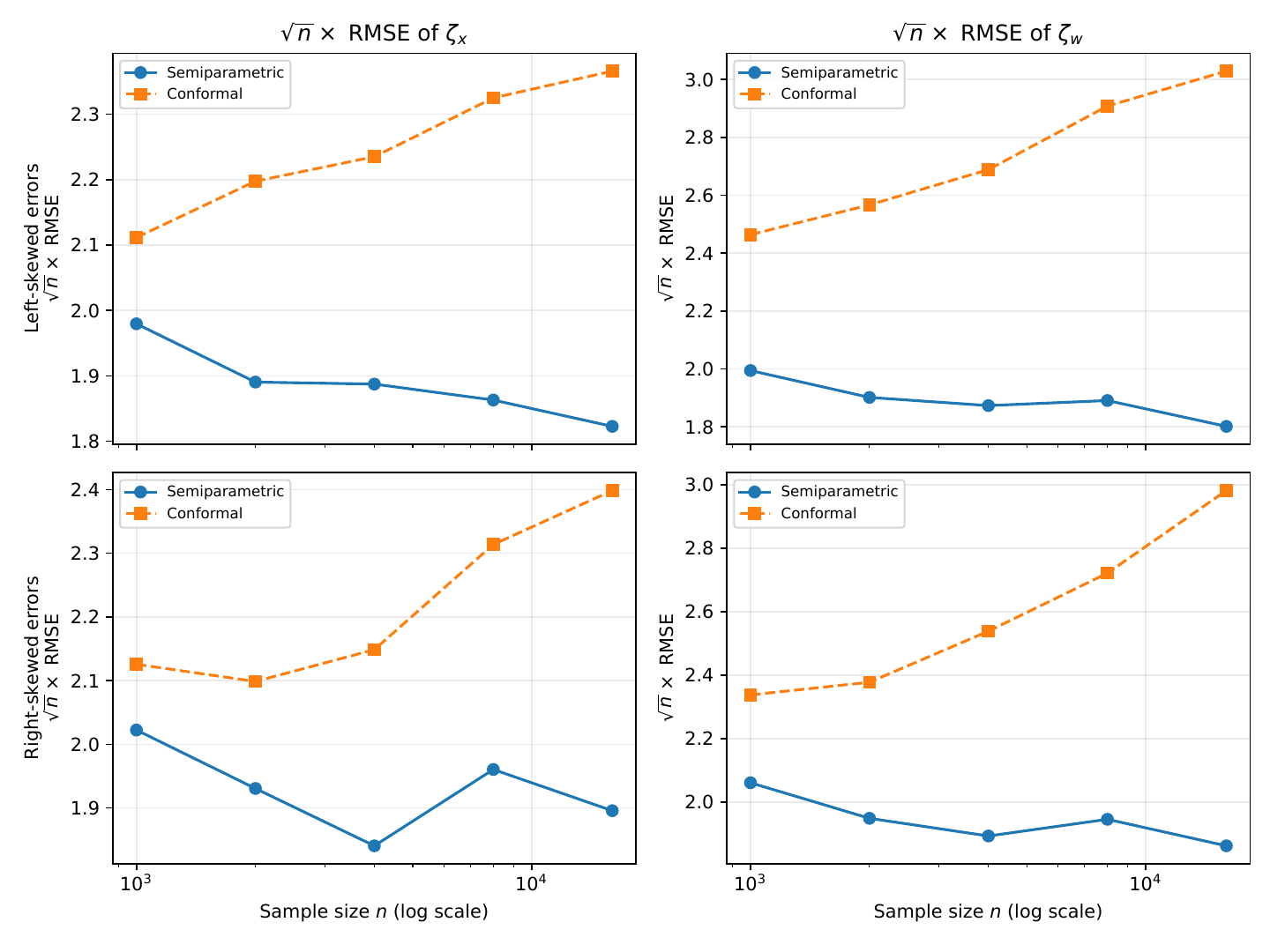}
  \caption{
  Monte Carlo $\sqrt n$ times RMSE of the semiparametric and conformal
  estimators of $\zeta_\x$ and $\zeta_\w$ under left- and right-skewed errors.
  The horizontal axis is logarithmic. Rows correspond to the error direction and
  columns to the target.
  }
  \label{fig:learned-cv-zeta-rmse}
\end{figure}

\medskip\noindent\textbf{Estimation of $\zeta_\v$.}
We compare the two methods for estimating $\zeta_\x$ and $\zeta_\w$
under right-skewed and left-skewed error distributions at
$\beta=0.2$. 
Table~\ref{tab:learned-cv-zeta} reports the bias, SD, RMSE, and
 $95\%$ confidence interval coverage for the
left- and right-skewed error distributions.
Figure~\ref{fig:learned-cv-zeta-rmse} reports $\sqrt n$ times RMSE against $n$,
for the
left- and right-skewed error distributions.
\begin{table}[H]
\centering
\caption{Bias, SD, SE, RMSE, and $95\%$ confidence interval coverage (Cvrg)
  for the estimators of $\zeta_\x$ and 
$\zeta_\w$ under left-skewed (Left) and right-skewed (Right) errors.  For each sample size,
Semi and Conf denote the semiparametric and conformal estimators,
respectively. All entries are reported in units of
$0.01$.}
\label{tab:learned-cv-zeta}
\setlength{\tabcolsep}{3.5pt}
\renewcommand{\arraystretch}{1.1}
\begin{tabular}{lrlrrrrrrrrrrr}
\hline
& & & \multicolumn{5}{c}{$\zeta_\x$} & \multicolumn{5}{c}{$\zeta_\w$} \\
\cline{4-8}\cline{9-13}
Error & $n$ & Method & Bias & SD & SE & RMSE & Cvrg & Bias & SD & SE & RMSE & Cvrg \\
\hline
      & 1000  & Semi & 0.37 & 6.25 & 6.34 & 6.26 & 94.9 & 1.03 & 6.23 & 6.22 & 6.31 & 93.3 \\
\rowcolor[gray]{0.93}
\cellcolor[gray]{1} & & Conf & 3.07 & 5.93 & 6.64 & 6.68 & 93.6 & 5.20 & 5.81 & 6.57 & 7.79 & 89.7 \\[2pt]
      & 2000  & Semi & 0.47 & 4.20 & 4.38 & 4.23 & 94.8 & 0.72 & 4.19 & 4.26 & 4.25 & 94.6 \\
\rowcolor[gray]{0.93}
\cellcolor[gray]{1} & & Conf & 2.71 & 4.10 & 4.68 & 4.91 & 92.7 & 4.04 & 4.08 & 4.62 & 5.74 & 87.9 \\[2pt]
      & 4000  & Semi & 0.21 & 2.98 & 3.04 & 2.98 & 95.4 & 0.40 & 2.94 & 2.95 & 2.96 & 93.7 \\
\rowcolor[gray]{0.93}
\cellcolor[gray]{1} & & Conf & 1.97 & 2.93 & 3.30 & 3.53 & 92.9 & 3.11 & 2.90 & 3.26 & 4.25 & 86.0 \\[2pt]
      & 8000  & Semi & 0.23 & 2.07 & 2.12 & 2.08 & 96.4 & 0.34 & 2.09 & 2.06 & 2.11 & 93.4 \\
\rowcolor[gray]{0.93}
\cellcolor[gray]{1} & & Conf & 1.57 & 2.07 & 2.33 & 2.60 & 91.7 & 2.51 & 2.07 & 2.30 & 3.25 & 84.1 \\[2pt]
      & 16000 & Semi & 0.03 & 1.44 & 1.49 & 1.44 & 95.0 & 0.07 & 1.42 & 1.45 & 1.42 & 95.3 \\
\rowcolor[gray]{0.93}
\cellcolor[gray]{1}\multirow{-10}{*}[4pt]{Left}
      & & Conf & 1.13 & 1.49 & 1.64 & 1.87 & 91.5 & 1.87 & 1.49 & 1.64 & 2.39 & 79.3 \\[2pt]
\hline
      & 1000  & Semi & -1.22 & 6.28 & 6.38 & 6.40 & 94.7 & -1.46 & 6.35 & 6.40 & 6.52 & 94.8 \\
\rowcolor[gray]{0.93}
\cellcolor[gray]{1} & & Conf & -2.92 & 6.06 & 6.69 & 6.72 & 95.0 & -4.15 & 6.12 & 6.79 & 7.39 & 91.9 \\[2pt]
      & 2000  & Semi & -0.43 & 4.30 & 4.39 & 4.32 & 95.3 & -0.52 & 4.33 & 4.33 & 4.36 & 94.3 \\
\rowcolor[gray]{0.93}
\cellcolor[gray]{1} & & Conf & -2.17 & 4.16 & 4.73 & 4.69 & 94.2 & -3.14 & 4.29 & 4.75 & 5.32 & 91.8 \\[2pt]
      & 4000  & Semi & -0.22 & 2.90 & 3.02 & 2.91 & 95.0 & -0.31 & 2.98 & 3.00 & 2.99 & 95.2 \\
\rowcolor[gray]{0.93}
\cellcolor[gray]{1} & & Conf & -1.74 & 2.92 & 3.30 & 3.40 & 93.9 & -2.58 & 3.08 & 3.35 & 4.01 & 90.1 \\[2pt]
      & 8000  & Semi & -0.02 & 2.19 & 2.11 & 2.19 & 94.0 & -0.11 & 2.17 & 2.09 & 2.17 & 94.9 \\
\rowcolor[gray]{0.93}
\cellcolor[gray]{1} & & Conf & -1.26 & 2.26 & 2.34 & 2.59 & 92.3 & -2.02 & 2.28 & 2.37 & 3.04 & 87.2 \\[2pt]
      & 16000 & Semi & -0.07 & 1.50 & 1.49 & 1.50 & 95.1 & -0.12 & 1.47 & 1.47 & 1.47 & 94.6 \\
\rowcolor[gray]{0.93}
\cellcolor[gray]{1}\multirow{-10}{*}[4pt]{Right}
      & & Conf & -1.09 & 1.55 & 1.65 & 1.90 & 91.6 & -1.76 & 1.57 & 1.67 & 2.36 & 82.3 \\
\hline
\end{tabular}
\end{table}
Under the two asymmetric error distributions, between the two methods, the
advantage of the 
semiparametric estimator is clearly manifested in the estimation of
$\zeta_\x$ and $\zeta_\w$: its bias and RMSE are substantially smaller
than those of the 
conformal estimator. 
Moreover, for a given sample size, the semiparametric methods yield
similar RMSEs for estimating $\zeta_\x$ and $\zeta_\w$. 
In contrast, the conformal method consistently yields a larger RMSE
for estimating $\zeta_\w$ than for estimating $\zeta_\x$.
 The coverage column shows the same separation.
The semiparametric intervals stay near the nominal $95\%$ level for both targets.
The conformal coverage for $\zeta_\w$ declines with $n$, reaching
$79.3\%$ under left-skewed errors and $82.3\%$ under right-skewed errors
at $n=16000$.
These patterns can be explained by the different
levels of difficulty in estimating $\zeta_\w$ and $\zeta_\x$, since the
former involves the higher-dimensional nuisance regression function
$m_\w$.
More specifically, $m_\x$ and $m_\w$ are estimated using a
second-order kernel smoother. Because $m_\x$ depends on a one-dimensional
covariate and $m_\w$ depends on a two-dimensional covariate, their
sup-norm convergence rates are, respectively,
\[
\|\wh m_\x-m_\x\|_\infty=O_p(n^{-2/5}),
\qquad
\|\wh m_\w-m_\w\|_\infty=O_p(n^{-1/3}).
\]
$\widehat m_\x$ converges faster and has a smaller estimation
error, whereas $\widehat m_\w$ converges more slowly. The resulting
nuisance error is therefore more pronounced in the estimation of
$\zeta_\w$. This dimensionality effect is consistent with the observed
differences between the two targets: the semiparametric correction can
reduce the first-order impact of the slower nuisance estimate for
$\zeta_\w$, whereas the conformal estimator retains more of that impact.

Figure~\ref{fig:learned-cv-zeta-rmse} confirms that the curves the RMSE of the
semiparametric estimator for $\zeta_\w$ decreases faster than that of the
conformal estimator.  
The curves for the semiparametric estimator remain approximately flat, consistent with the theoretical $n^{-1/2}$ convergence rate.
In contrast, the scaled RMSE, $\sqrt{n}\times\mathrm{RMSE}$, for the
conformal estimator increases with $n$.
This finite-sample pattern suggests the presence of a remainder term
that decays more slowly than $n^{-1/2}$. 
This pattern matches the theoretical analysis: the slower nonparametric convergence rates affect the conformal
estimator,  
while the semiparametric correction removes its first-order effect.

\medskip\noindent\textbf{Inference based on $\Delta$.}
For inference on $\Delta$, we the null hypothesis 
$H_0: \Delta\le\tau=0.20$ and set the test level to be
$\alpha'=0.05$. We adjust $\beta$ so that the true values
  of
$\Delta(\beta)-\tau$ are
-0.05,0,0.05,0.10,0.20.
The corresponding
oracle-calibrated values of $\beta$ are $\beta=0.53,0.61,0.68,0.74,0.85$.
These values apply to both reflected asymmetric error distributions.
The rejection regions, and confidence intervals used below
are given in \eqref{eq:inference} and
\eqref{eq:interval}, respectively.
The
plug-in
variance estimators $\wh v_\Delta^{\rm semi}$ and
$\wh v_\Delta^{\rm conf}$ include the covariance terms described in
Section~\ref{sec:variance}. 
We use $\mathrm{SD}(T)$ to denote the Monte Carlo standard
deviation of the corresponding test statistic defined in \eqref{eq:teststat} or \eqref{eq:teststat2}.

\begin{table}[H] 
\centering
\caption{Finite-sample inference on $\Delta$ under $H_0:
  \Delta(\beta)=\tau=0.20$ when error distribution is left-skewed
  (Left) or right-skewed (Right).
Size is computed using the standard-normal critical value. The test level is $0.05$, and the confidence level is $0.95$.  For each sample size,
Semi and Conf denote the semiparametric and conformal estimators,
respectively. All entries are reported in units of
$0.01$ except for SD($T$).}
\label{tab:delta_calib}
\setlength{\tabcolsep}{6pt}
\renewcommand{\arraystretch}{1.1}
\begin{tabular}{lrlrrrrr}
\hline
Error & $n$ & Method & Size & Bias &  Cvrg & CI length & SD($T$) \\
\hline
      & 1000  & Semi & 3.9 & -0.4 & 94.1 & 24.2 & 1.001 \\
\rowcolor[gray]{0.93}
\cellcolor[gray]{1} & & Conf & 1.6 & -3.1 & 91.9 & 23.9 & 1.000 \\[2pt]
      & 2000  & Semi & 4.9 & -0.1 & 95.4 & 16.7 & 0.997 \\
\rowcolor[gray]{0.93}
\cellcolor[gray]{1} & & Conf & 1.0 & -2.3 & 91.4 & 16.6 & 0.990 \\[2pt]
      & 4000  & Semi & 5.1 & 0.0 & 95.3 & 11.6 & 0.961 \\
\rowcolor[gray]{0.93}
\cellcolor[gray]{1} & & Conf & 0.8 & -1.9 & 91.4 & 11.6 & 0.987 \\[2pt]
      & 8000  & Semi & 4.6 & 0.0 & 94.8 & 8.1 & 0.988 \\
\rowcolor[gray]{0.93}
\cellcolor[gray]{1} & & Conf & 0.9 & -1.5 & 89.1 & 8.2 & 1.002 \\[2pt]
      & 16000 & Semi & 5.4 & 0.0 & 95.3 & 5.7 & 1.015 \\
\rowcolor[gray]{0.93}
\cellcolor[gray]{1}\multirow{-10}{*}[4pt]{Left}
      & & Conf & 0.6 & -1.3 & 84.8 & 5.8 & 1.029 \\[2pt]
\hline
      & 1000  & Semi & 6.8 & 0.9 & 94.4 & 23.6 & 1.041 \\
\rowcolor[gray]{0.93}
\cellcolor[gray]{1} & & Conf & 13.8 & 3.4 & 91.7 & 23.5 & 0.976 \\[2pt]
      & 2000  & Semi & 6.2 & 0.0 & 93.6 & 16.3 & 1.072 \\
\rowcolor[gray]{0.93}
\cellcolor[gray]{1} & & Conf & 14.2 & 2.3 & 91.5 & 16.4 & 1.006 \\[2pt]
      & 4000  & Semi & 5.0 & 0.1 & 93.6 & 11.4 & 1.034 \\
\rowcolor[gray]{0.93}
\cellcolor[gray]{1} & & Conf & 16.3 & 1.9 & 90.6 & 11.5 & 0.989 \\[2pt]
      & 8000  & Semi & 6.6 & 0.1 & 94.9 & 8.0 & 1.014 \\
\rowcolor[gray]{0.93}
\cellcolor[gray]{1} & & Conf & 19.8 & 1.6 & 87.4 & 8.2 & 1.021 \\[2pt]
      & 16000 & Semi & 5.1 & 0.1 & 96.1 & 5.6 & 0.967 \\
\rowcolor[gray]{0.93}
\cellcolor[gray]{1}\multirow{-10}{*}[4pt]{Right}
      & & Conf & 22.0 & 1.3 & 86.4 & 5.8 & 0.993 \\
\hline
\end{tabular}
\end{table}
Table~\ref{tab:delta_calib} reveals that the
conformal procedure is highly sensitive to the direction of error
skewness, whereas the semiparametric procedure has substantially more
stable size behavior.
Under left-skewed errors, the conformal procedure has nearly zero
rejection probability and severe undercoverage. Under right-skewed errors,
the conformal rejection probability increases from $13.8\%$ to $22.0\%$ as
$n$ grows, indicating severe size distortion. In contrast, the
semiparametric rejection probability decreases toward the nominal level
of $0.05$.

The coverage results show the same pattern. The semiparametric confidence
interval achieves coverage close to the nominal level under
left-skewed errors and improves with the sample size under right-skewed
errors. The conformal confidence interval is substantially
undercovering under both asymmetric error distributions. 
Moreover, the empirical
standard deviations of the standardized statistics are generally close
to one, suggesting that the main issue is finite-sample centering and
nuisance-estimation bias rather than a failure of variance estimation.

\begin{figure}[H]
  \centering
 \includegraphics[width=0.98\textwidth]{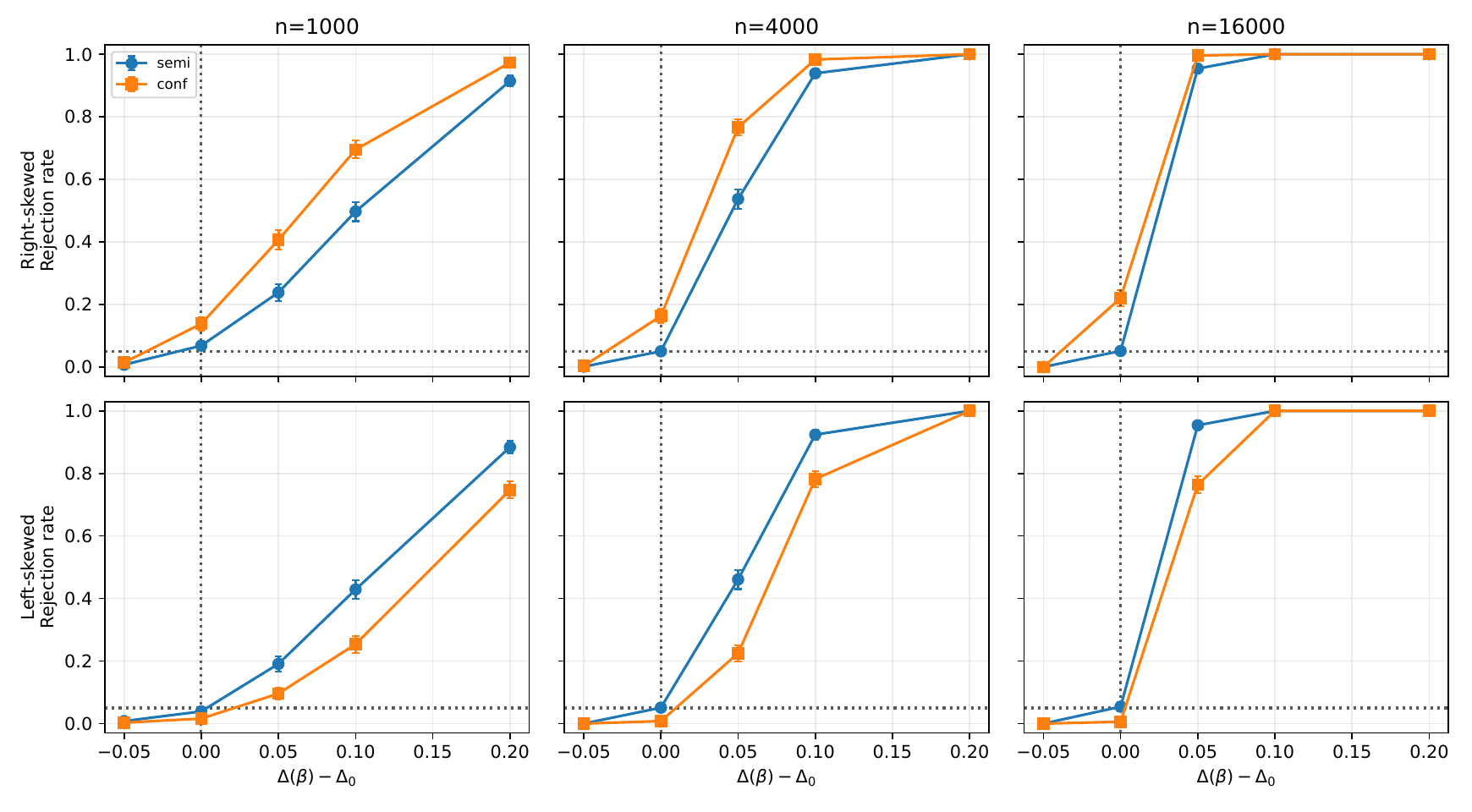}
  \caption{ Rejection probabilities for testing
  $H_0:\Delta\leq\tau$ against $H_1:\Delta>\tau$ with
  $\tau=0.20$. Rows correspond
  to right- and left-skewed errors, columns to $n\in\{1000,4000,16000\}$. The horizontal dotted line denotes the nominal level $0.05$, the
  vertical dotted line marks the boundary gap $\Delta(\beta)-\tau=0$,
  and error bars are pointwise 95\% Monte Carlo confidence intervals.}
  \label{fig:delta_power}
\end{figure}
Figure~\ref{fig:delta_power} shows that the rejection probability increases
with both the contrast gap and the sample size. At the boundary, the
conformal procedure has large type one error under right-skewed
errors and too conservative
under left-skewed errors, whereas the semiparametric procedure is more
stable and approaches the nominal level as $n$ increases. Under
  the alternative, the power of both procedures becomes substantial at $n=4000$ and
$n=16000$, and is close to one when the gap equals $0.20$.

\subsection{Conditional Prediction Interval}
\label{sec:sim-conditional-design}

We consider a conditional experiment with $S=X$, retaining the same covariate
distribution, regression function, scale function, and standardized
error distributions as in the marginal experiment. We set
$\alpha=0.1$ and consider the prespecified
interior points $s_0\in\{-0.5,0,0.5\}$. The sample sizes are
$n\in\{1000,2000,4000,8000,16000\}$, with $R=1000$ replications per
data-generating configuration. 
Writing $F_0$ for the distribution function of $\epsilon_0$, and
$q_0$ for the $0.9$ quantile of $|\epsilon_0|$, we have
$\zeta_\w(s)=\sigma(s)q_0$. The value $\zeta_\x(s;\beta)$ is the
positive solution $q$ of
\[
\frac{1}{2\sqrt{3}}\int_{-\sqrt{3}}^{\sqrt{3}}
\left[
F_0\left\{\frac{q-\beta z}{\sigma(s)}\right\}
-F_0\left\{\frac{-q-\beta z}{\sigma(s)}\right\}
\right]dz=0.9.
\]
For each fixed $(\beta,s)$ pair, these specifications determine the corresponding
$\Delta(s;\beta)=\zeta_\x(s;\beta)-\zeta_\w(s)$.  

Both procedures use the same two folds and the same nuisance estimation procedure as in the marginal experiment.
The fold-specific estimating equations and residual
quantiles are localized using
$K_h(u)=h^{-1}K(u/h)$, where
$K(u)=3(1-u^2)I(|u|\le1)/4$.
The final estimator averages the two fold-specific estimates.
The primary comparison uses
\[
h_{\rm semi}=1.5 n^{-1/4},\qquad
h_{\rm conf}=1.5 n^{-2/5},
\]
where $n$ denotes the total sample size. Both choices satisfy
$nh\to\infty$ and $nh^5\to0$, the undersmoothing condition for a
second-order kernel in one localization dimension.
Regarding the convergence rate requirements for the regression
estimators in Theorems~\ref{th:semi}--\ref{th:conformal}, 
the slower, two-dimensional regression rate $n^{-1/3}$ is sufficient for
the semiparametric requirement
$\sup_{\w}|\wh m_\w-m_\w|=o_p\{(nh_{\rm semi})^{-1/4}\}$
and the conformal requirement
$\sup_{\w}|\wh m_\w-m_\w|=o_p\{(nh_{\rm conf})^{-1/2}\}$.
We use the variance estimators in
\eqref{var-est:semi} and \eqref{var-est:conf} and set $c_n=\min\{0.05, 0.1(nh/2)^{-1/5}\}$, where $h=h_{\rm semi}$ for the
semiparametric method and $h=h_{\rm conf}$ for the conformal method.

\begin{figure}[H]
  \centering
  \includegraphics[width=0.98\textwidth]{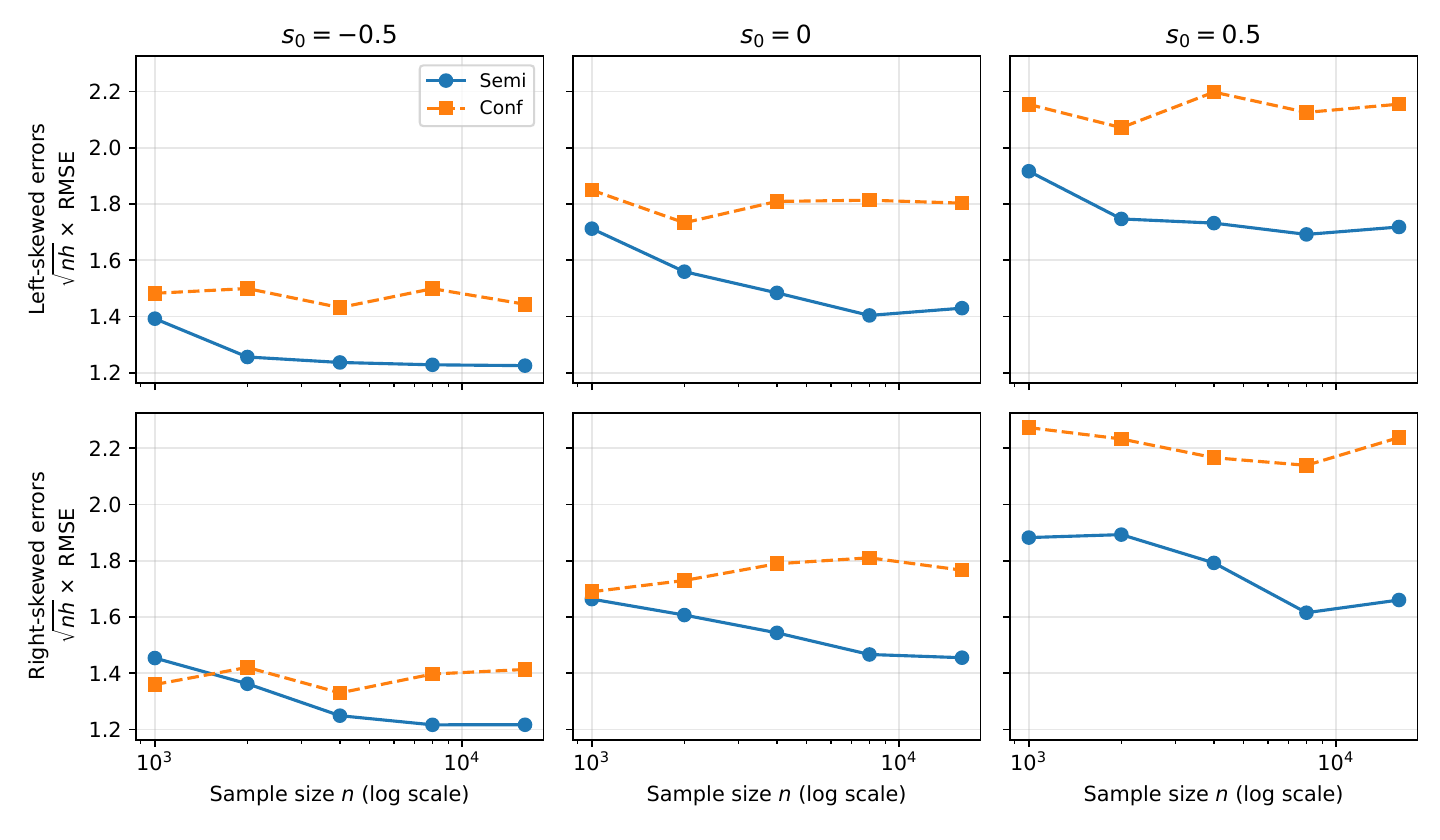}
  \caption{Monte Carlo $\sqrt{nh}$ times RMSE of the semiparametric
    ($h=1.5n^{-1/4}$) and
  conformal ($h=1.5n^{-2/5}$) estimators of $\zeta_\x(s_0)$. The horizontal
  axis is logarithmic.  Rows correspond
  to right- and left-skewed errors and columns to
  $s_0\in\{-0.5,0,0.5\}$.}
  \label{fig:cond-zeta-rmse}
\end{figure}

\medskip\noindent\textbf{Estimation of $\zeta_\v(s_0)$.}
We fix $\beta=0.2$. Tables~\ref{tab:cond-zeta-m05}--\ref{tab:cond-zeta-p05} report bias, SD, SE,
RMSE, and $95\%$ confidence interval coverage at $s_0\in\{-0.5,0,0.5\}$.
Figure~\ref{fig:cond-zeta-rmse} reports $\sqrt{nh}$ times RMSE against $n$.

\begin{table}[H]
\centering
\caption{Bias, SD, SE, RMSE, and $95\%$ confidence interval coverage
  (Cvrg) for the estimators of $\zeta_\x(s_0)$ and 
$\zeta_\w(s_0)$ at $s_0=-0.5$ under left-skewed (Left) and
right-skewed (Right) errors. For each sample size, Semi and Conf
denote the semiparametric and conformal estimators. All entries are reported
in units of $0.01$.}
\label{tab:cond-zeta-m05}
\setlength{\tabcolsep}{3.5pt}
\renewcommand{\arraystretch}{1.1}
\begin{tabular}{lrlrrrrrrrrrrr}
\hline
& & & \multicolumn{5}{c}{$\zeta_\x(s_0)$} & \multicolumn{5}{c}{$\zeta_\w(s_0)$} \\
\cline{4-8}\cline{9-13}
Error & $n$ & Method & Bias & SD & SE & RMSE & Cvrg & Bias & SD & SE & RMSE & Cvrg \\
\hline
      & 1000  & Semi & -0.19 & 8.53 & 8.53 & 8.52 & 94.0 & 0.14 & 8.29 & 7.86 & 8.29 & 93.1 \\
\rowcolor[gray]{0.93}
\cellcolor[gray]{1} & & Conf & 1.45 & 15.18 & 16.26 & 15.24 & 93.8 & -1.15 & 13.84 & 14.67 & 13.88 & 94.0 \\[2pt]
      & 2000  & Semi & 0.54 & 5.91 & 6.25 & 5.93 & 93.9 & 0.87 & 5.79 & 5.63 & 5.85 & 92.7 \\
\rowcolor[gray]{0.93}
\cellcolor[gray]{1} & & Conf & 1.58 & 12.43 & 13.34 & 12.52 & 94.5 & -0.29 & 10.46 & 12.49 & 10.46 & 96.9 \\[2pt]
      & 4000  & Semi & 0.20 & 4.50 & 4.70 & 4.50 & 95.3 & 0.26 & 4.22 & 4.10 & 4.23 & 94.4 \\
\rowcolor[gray]{0.93}
\cellcolor[gray]{1} & & Conf & 1.29 & 9.63 & 10.79 & 9.71 & 96.3 & -0.82 & 8.31 & 10.08 & 8.34 & 97.0 \\[2pt]
      & 8000  & Semi & 0.07 & 3.45 & 3.52 & 3.45 & 94.8 & 0.03 & 3.07 & 3.07 & 3.07 & 95.2 \\
\rowcolor[gray]{0.93}
\cellcolor[gray]{1} & & Conf & 0.98 & 8.20 & 8.71 & 8.26 & 95.3 & -0.51 & 7.10 & 8.06 & 7.12 & 96.8 \\[2pt]
      & 16000 & Semi & 0.06 & 2.65 & 2.68 & 2.65 & 96.0 & 0.23 & 2.36 & 2.31 & 2.37 & 93.3 \\
\rowcolor[gray]{0.93}
\cellcolor[gray]{1}\multirow{-10}{*}[4pt]{Left}
      & & Conf & 1.12 & 6.36 & 7.04 & 6.46 & 95.4 & 0.22 & 5.50 & 6.46 & 5.50 & 97.1 \\[2pt]
\hline
      & 1000  & Semi & -1.75 & 8.73 & 7.65 & 8.90 & 90.6 & 0.10 & 8.36 & 7.53 & 8.36 & 91.2 \\
\rowcolor[gray]{0.93}
\cellcolor[gray]{1} & & Conf & -4.26 & 13.32 & 14.19 & 13.98 & 91.2 & -2.58 & 13.24 & 14.59 & 13.48 & 95.2 \\[2pt]
      & 2000  & Semi & -0.65 & 6.40 & 5.79 & 6.43 & 92.4 & 0.21 & 5.89 & 5.48 & 5.89 & 92.0 \\
\rowcolor[gray]{0.93}
\cellcolor[gray]{1} & & Conf & -4.04 & 11.16 & 12.04 & 11.86 & 92.7 & -2.54 & 10.58 & 12.03 & 10.88 & 95.6 \\[2pt]
      & 4000  & Semi & -0.59 & 4.51 & 4.37 & 4.55 & 93.6 & 0.10 & 4.29 & 4.04 & 4.29 & 93.0 \\
\rowcolor[gray]{0.93}
\cellcolor[gray]{1} & & Conf & -2.80 & 8.58 & 10.03 & 9.02 & 95.8 & -1.53 & 7.90 & 9.76 & 8.04 & 97.5 \\[2pt]
      & 8000  & Semi & -0.41 & 3.39 & 3.32 & 3.42 & 94.3 & -0.20 & 3.22 & 3.03 & 3.23 & 92.5 \\
\rowcolor[gray]{0.93}
\cellcolor[gray]{1} & & Conf & -2.16 & 7.39 & 8.20 & 7.70 & 95.9 & -1.45 & 6.91 & 7.97 & 7.06 & 96.6 \\[2pt]
      & 16000 & Semi & -0.16 & 2.63 & 2.57 & 2.63 & 94.7 & 0.01 & 2.30 & 2.32 & 2.30 & 94.8 \\
\rowcolor[gray]{0.93}
\cellcolor[gray]{1}\multirow{-10}{*}[4pt]{Right}
      & & Conf & -1.55 & 6.13 & 6.85 & 6.32 & 96.1 & -0.75 & 5.49 & 6.53 & 5.54 & 97.5 \\[2pt]
\hline
\end{tabular}
\end{table}

\begin{table}[H]
      \centering
      \caption{Bias, SD, SE, RMSE, and $95\%$ confidence interval coverage (Cvrg) for the estimators of $\zeta_\x(s_0)$ and
      $\zeta_\w(s_0)$ at $s_0=0$ under left-skewed (Left) and
      right-skewed (Right) errors. For each sample size, Semi and Conf
      denote the semiparametric and conformal estimators. All entries are reported
      in units of $0.01$.}
      \label{tab:cond-zeta-0}
      \setlength{\tabcolsep}{3.5pt}
      \renewcommand{\arraystretch}{1.1}
      \begin{tabular}{lrlrrrrrrrrrrr}
      \hline
      & & & \multicolumn{5}{c}{$\zeta_\x(s_0)$} & \multicolumn{5}{c}{$\zeta_\w(s_0)$} \\
      \cline{4-8}\cline{9-13}
      Error & $n$ & Method & Bias & SD & SE & RMSE & Cvrg & Bias & SD & SE & RMSE & Cvrg \\
      \hline
            & 1000  & Semi & 0.15 & 10.49 & 9.90 & 10.48 & 91.7 & 0.35 & 10.15 & 9.31 & 10.15 & 91.7 \\
      \rowcolor[gray]{0.93}
      \cellcolor[gray]{1} & & Conf & 0.82 & 19.00 & 18.74 & 19.01 & 91.7 & -0.50 & 18.26 & 17.95 & 18.26 & 91.2 \\[2pt]
            & 2000  & Semi & 0.12 & 7.37 & 7.24 & 7.36 & 94.1 & 0.41 & 7.22 & 6.82 & 7.23 & 92.4 \\
      \rowcolor[gray]{0.93}
      \cellcolor[gray]{1} & & Conf & 1.54 & 14.40 & 15.87 & 14.47 & 95.0 & 0.29 & 14.04 & 15.20 & 14.04 & 95.7 \\[2pt]
            & 4000  & Semi & 0.03 & 5.41 & 5.48 & 5.40 & 94.4 & 0.15 & 5.12 & 5.10 & 5.12 & 94.4 \\
      \rowcolor[gray]{0.93}
      \cellcolor[gray]{1} & & Conf & 0.45 & 12.27 & 13.15 & 12.27 & 94.4 & -0.13 & 11.88 & 12.70 & 11.88 & 95.8 \\[2pt]
            & 8000  & Semi & 0.07 & 3.94 & 4.14 & 3.94 & 95.1 & 0.30 & 3.75 & 3.81 & 3.76 & 95.2 \\
      \rowcolor[gray]{0.93}
      \cellcolor[gray]{1} & & Conf & 1.22 & 9.92 & 10.61 & 9.99 & 94.9 & 0.50 & 9.39 & 10.15 & 9.39 & 94.9 \\[2pt]
            & 16000 & Semi & 0.08 & 3.10 & 3.17 & 3.09 & 94.7 & 0.02 & 2.87 & 2.90 & 2.87 & 95.1 \\
      \rowcolor[gray]{0.93}
      \cellcolor[gray]{1}\multirow{-10}{*}[4pt]{Left}
            & & Conf & 0.91 & 8.02 & 8.67 & 8.07 & 94.7 & 0.35 & 7.61 & 8.25 & 7.61 & 96.3 \\[2pt]
      \hline
            & 1000  & Semi & -1.30 & 10.11 & 9.09 & 10.19 & 91.3 & -0.61 & 10.52 & 8.92 & 10.53 & 90.5 \\
      \rowcolor[gray]{0.93}
      \cellcolor[gray]{1} & & Conf & -4.35 & 16.82 & 17.49 & 17.37 & 92.1 & -4.93 & 16.36 & 16.91 & 17.08 & 92.8 \\[2pt]
            & 2000  & Semi & -0.80 & 7.55 & 6.87 & 7.59 & 90.1 & -0.17 & 7.09 & 6.59 & 7.09 & 91.7 \\
      \rowcolor[gray]{0.93}
      \cellcolor[gray]{1} & & Conf & -3.91 & 13.91 & 15.04 & 14.44 & 94.6 & -4.15 & 13.62 & 14.64 & 14.23 & 94.1 \\[2pt]
            & 4000  & Semi & -0.40 & 5.61 & 5.19 & 5.62 & 92.6 & -0.03 & 5.20 & 4.90 & 5.20 & 93.1 \\
      \rowcolor[gray]{0.93}
      \cellcolor[gray]{1} & & Conf & -3.26 & 11.69 & 12.31 & 12.13 & 94.1 & -3.50 & 11.27 & 11.94 & 11.80 & 94.1 \\[2pt]
            & 8000  & Semi & -0.27 & 4.11 & 4.00 & 4.12 & 94.4 & 0.05 & 3.90 & 3.73 & 3.90 & 94.2 \\
      \rowcolor[gray]{0.93}
      \cellcolor[gray]{1} & & Conf & -2.86 & 9.56 & 10.19 & 9.97 & 94.7 & -2.72 & 9.00 & 9.97 & 9.40 & 96.4 \\[2pt]
            & 16000 & Semi & -0.25 & 3.14 & 3.06 & 3.15 & 94.1 & 0.00 & 2.84 & 2.83 & 2.84 & 95.5 \\
      \rowcolor[gray]{0.93}
      \cellcolor[gray]{1}\multirow{-10}{*}[4pt]{Right}
            & & Conf & -2.35 & 7.55 & 8.32 & 7.90 & 95.4 & -1.80 & 7.11 & 8.01 & 7.33 & 95.4 \\[2pt]
      \hline
      \end{tabular}
\end{table}

\begin{table}[H]
\centering
\caption{Bias, SD, SE, RMSE, and $95\%$ confidence interval coverage (Cvrg) for the estimators of $\zeta_\x(s_0)$ and
$\zeta_\w(s_0)$ at $s_0=0.5$ under left-skewed (Left) and
right-skewed (Right) errors. For each sample size, Semi and Conf
denote the semiparametric and conformal estimators. All entries are reported
in units of $0.01$.}
\label{tab:cond-zeta-p05}
\setlength{\tabcolsep}{3.5pt}
\renewcommand{\arraystretch}{1.1}
\begin{tabular}{lrlrrrrrrrrrrr}
\hline
& & & \multicolumn{5}{c}{$\zeta_\x(s_0)$} & \multicolumn{5}{c}{$\zeta_\w(s_0)$} \\
\cline{4-8}\cline{9-13}
Error & $n$ & Method & Bias & SD & SE & RMSE & Cvrg & Bias & SD & SE & RMSE & Cvrg \\
\hline
      & 1000  & Semi & -1.29 & 11.67 & 11.15 & 11.73 & 92.7 & -0.81 & 11.80 & 10.97 & 11.82 & 93.3 \\
\rowcolor[gray]{0.93}
\cellcolor[gray]{1} & & Conf & -3.77 & 21.83 & 23.73 & 22.14 & 93.5 & -4.25 & 21.82 & 22.96 & 22.22 & 92.2 \\[2pt]
      & 2000  & Semi & 0.04 & 8.25 & 8.22 & 8.25 & 94.1 & -0.06 & 8.20 & 7.93 & 8.19 & 93.4 \\
\rowcolor[gray]{0.93}
\cellcolor[gray]{1} & & Conf & -1.92 & 17.20 & 18.62 & 17.30 & 94.0 & -2.77 & 17.19 & 18.11 & 17.41 & 94.0 \\[2pt]
      & 4000  & Semi & -0.19 & 6.31 & 6.21 & 6.31 & 94.4 & -0.07 & 6.19 & 5.92 & 6.19 & 93.0 \\
\rowcolor[gray]{0.93}
\cellcolor[gray]{1} & & Conf & -3.22 & 14.56 & 15.00 & 14.91 & 93.8 & -3.73 & 14.72 & 14.64 & 15.18 & 93.0 \\[2pt]
      & 8000  & Semi & -0.14 & 4.75 & 4.75 & 4.75 & 94.6 & 0.25 & 4.59 & 4.52 & 4.60 & 94.5 \\
\rowcolor[gray]{0.93}
\cellcolor[gray]{1} & & Conf & -1.42 & 11.63 & 12.31 & 11.71 & 95.0 & -1.25 & 11.53 & 11.93 & 11.60 & 94.6 \\[2pt]
      & 16000 & Semi & 0.00 & 3.72 & 3.58 & 3.72 & 94.1 & -0.08 & 3.44 & 3.42 & 3.44 & 95.4 \\
\rowcolor[gray]{0.93}
\cellcolor[gray]{1}\multirow{-10}{*}[4pt]{Left}
      & & Conf & -1.59 & 9.51 & 9.90 & 9.64 & 94.7 & -1.65 & 9.47 & 9.65 & 9.61 & 94.2 \\[2pt]
\hline
      & 1000  & Semi & -1.58 & 11.42 & 11.26 & 11.53 & 95.5 & -1.28 & 11.66 & 11.00 & 11.73 & 93.6 \\
\rowcolor[gray]{0.93}
\cellcolor[gray]{1} & & Conf & -3.70 & 23.09 & 23.10 & 23.37 & 91.9 & -4.43 & 22.41 & 23.04 & 22.83 & 92.7 \\[2pt]
      & 2000  & Semi & -0.73 & 8.91 & 8.29 & 8.94 & 92.7 & -0.82 & 8.41 & 8.08 & 8.45 & 93.8 \\
\rowcolor[gray]{0.93}
\cellcolor[gray]{1} & & Conf & -3.21 & 18.37 & 18.77 & 18.64 & 93.9 & -3.52 & 18.32 & 18.51 & 18.65 & 93.2 \\[2pt]
      & 4000  & Semi & -0.28 & 6.52 & 6.28 & 6.53 & 93.5 & -0.03 & 6.23 & 5.99 & 6.23 & 92.8 \\
\rowcolor[gray]{0.93}
\cellcolor[gray]{1} & & Conf & -1.27 & 14.64 & 15.43 & 14.69 & 94.0 & -1.81 & 14.93 & 14.99 & 15.04 & 93.3 \\[2pt]
      & 8000  & Semi & -0.07 & 4.54 & 4.78 & 4.53 & 96.2 & 0.08 & 4.36 & 4.55 & 4.35 & 94.7 \\
\rowcolor[gray]{0.93}
\cellcolor[gray]{1} & & Conf & -1.30 & 11.72 & 12.46 & 11.78 & 94.3 & -1.15 & 11.24 & 12.19 & 11.29 & 95.3 \\[2pt]
      & 16000 & Semi & -0.03 & 3.60 & 3.67 & 3.59 & 95.0 & -0.03 & 3.44 & 3.46 & 3.43 & 95.3 \\
\rowcolor[gray]{0.93}
\cellcolor[gray]{1}\multirow{-10}{*}[4pt]{Right}
      & & Conf & -0.62 & 10.00 & 10.19 & 10.01 & 95.1 & -0.76 & 9.48 & 9.85 & 9.50 & 94.9 \\[2pt]
\hline
\end{tabular}
\end{table}

Under both asymmetric error distributions, the semiparametric estimator has
smaller RMSE than the conformal estimator for $\zeta_\x(s_0)$ and
$\zeta_\w(s_0)$, at every $s_0$ and sample size. 
The conformal RMSE is
larger for both targets, and the gap remains visible at $n=16000$. The estimated
standard errors are generally close to the Monte Carlo standard deviations.
For both methods and both targets, the empirical coverage of the
confidence intervals is generally close to the nominal level of $95\%$.
This contrasts with the marginal case, where the conformal coverage declines as $n$ increases.
However, the conformal method achieves this coverage at the cost of
wider confidence intervals, as reflected in its larger SEs.
The smaller bandwidth $h_{\rm conf}$, relative to $h_{\rm semi}$,
reduces the effective sample size and contributes to the larger SEs.

Figure~\ref{fig:cond-zeta-rmse} plots $\sqrt{nh}$ times the RMSE of the estimators of $\zeta_\x(s_0)$, where $h=h_{\rm semi}$ for the semiparametric method and $h=h_{\rm conf}$ for the conformal method. The semiparametric curves lie below
the conformal curves except at a few small sample sizes. For both the semiparametric and conformal estimators, the curves decrease slowly or remain approximately flat as $n$
increases.
This behavior is consistent with
convergence rates of $(nh_{\rm semi})^{-1/2}$ and
$(nh_{\rm conf})^{-1/2}$ for the semiparametric and conformal
estimators, respectively.
Under the chosen bandwidths, these correspond to rates of $n^{-3/8}$
and $n^{-3/10}$, so the conformal estimator converges more slowly.
The height of the curves also varies across locations: it is lowest at
$s_0=-0.5$, intermediate at $s_0=0$, and highest at $s_0=0.5$. Figure~\ref{fig:cond-zeta-w-rmse} gives the same display for $\zeta_\w(s_0)$.

\begin{figure}[H]
  \centering
  \includegraphics[width=0.98\textwidth]{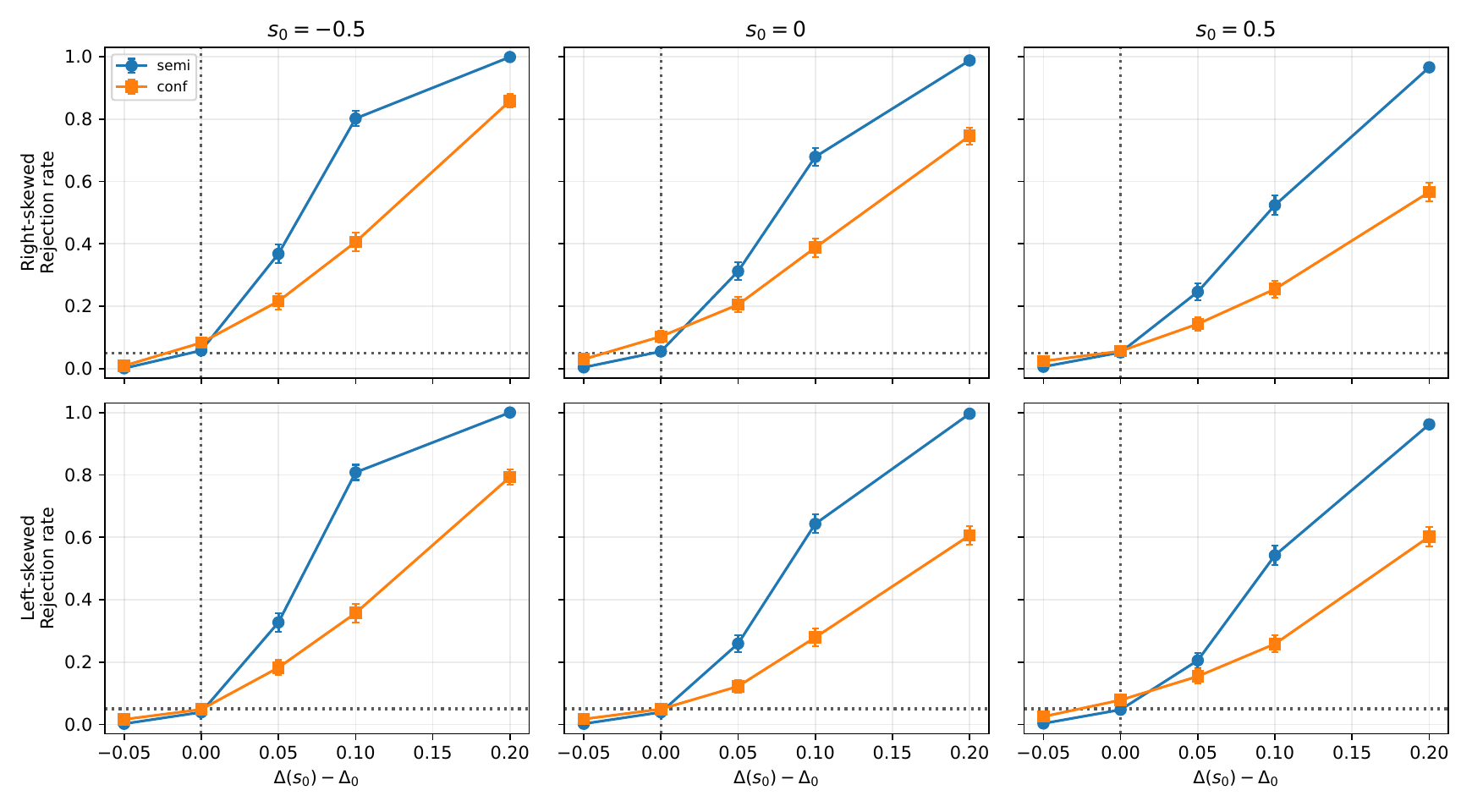}
  \caption{Rejection probabilities for testing
  $H_0:\Delta(s_0)\le\tau$ against $H_1:\Delta(s_0)>\tau$ with
  $\tau=0.20$, at $n=16000$. Rows correspond to right- and
  left-skewed errors, and columns to $s_0\in\{-0.5,0,0.5\}$.
  The horizontal dotted line denotes the nominal level $0.05$, the
  vertical dotted line marks the boundary gap $\Delta(s_0)-\tau=0$,
  and error bars are pointwise $95\%$ Monte Carlo confidence intervals.}
  \label{fig:cond-delta-power}
\end{figure}

\medskip\noindent\textbf{Inference based on $\Delta(\s_0)$.}
We test $H_0:\Delta(s_0)\le0.20$ against
$H_1:\Delta(s_0)>0.20$ at level $\alpha'=0.05$.
For each location and each gap
$\Delta(s_0;\beta)-0.20\in\{-0.05,0,0.05,0.10,0.20\}$,
$\beta$ is calibrated as follows.
\begin{center}
\begin{tabular}{lrrrrr}
\hline
gap& $-0.05$ & $0$ & $0.05$ & $0.10$ & $0.20$ \\
\hline
$s_0=-0.5$ & 0.53 & 0.59 & 0.65 & 0.70 & 0.80 \\
$s_0=0$    & 0.61 & 0.68 & 0.74 & 0.80 & 0.90 \\
$s_0=0.5$  & 0.68 & 0.76 & 0.83 & 0.89 & 1.00 \\
\hline
\end{tabular}
\end{center}
Tables~\ref{tab:cond-delta-m05}--\ref{tab:cond-delta-p05} report the results of the
semiparametric and conformal tests at the null boundary, where $\Delta(s_0)=0.20$.
Figure~\ref{fig:cond-delta-power} gives the rejection probabilities
across these gaps at $n=16000$. 

\begin{table}[H]
\centering
\caption{Finite-sample inference on $\Delta(s_0)$ under $H_0:
  \Delta(s_0)=\tau=0.20$ at $s_0=-0.5$ when the error distribution is left-skewed
  (Left) or right-skewed (Right).
Size is computed using the standard-normal critical value. The test level is $0.05$, and the confidence level is $0.95$.  For each sample size,
Semi and Conf denote the semiparametric and conformal estimators.
All entries are reported in units of
$0.01$ except for SD($T$).}
\label{tab:cond-delta-m05}
\setlength{\tabcolsep}{6pt}
\renewcommand{\arraystretch}{1.1}
\begin{tabular}{lrlrrrrr}
\hline
Error & $n$ & Method & Size & Bias & Cvrg & CI length & SD($T$) \\
\hline
      & 1000  & Semi & 4.9 & -0.1 & 94.9 & 45.0 & 0.990 \\
\rowcolor[gray]{0.93}
\cellcolor[gray]{1} & & Conf & 11.3 & 2.4 & 88.2 & 60.5 & 1.385 \\[2pt]
      & 2000  & Semi & 5.6 & 0.1 & 94.4 & 34.1 & 1.006 \\
\rowcolor[gray]{0.93}
\cellcolor[gray]{1} & & Conf & 10.9 & 2.6 & 90.7 & 52.3 & 1.200 \\[2pt]
      & 4000  & Semi & 4.8 & -0.1 & 94.4 & 25.8 & 1.002 \\
\rowcolor[gray]{0.93}
\cellcolor[gray]{1} & & Conf & 9.2 & 2.0 & 92.5 & 43.5 & 1.135 \\[2pt]
      & 8000  & Semi & 5.1 & -0.1 & 95.5 & 19.6 & 0.987 \\
\rowcolor[gray]{0.93}
\cellcolor[gray]{1} & & Conf & 5.5 & 1.2 & 94.7 & 35.8 & 1.021 \\[2pt]
      & 16000 & Semi & 3.9 & -0.3 & 94.3 & 14.9 & 1.012 \\
\rowcolor[gray]{0.93}
\cellcolor[gray]{1}\multirow{-10}{*}[4pt]{Left}
      & & Conf & 4.8 & 0.1 & 94.7 & 29.0 & 1.003 \\[2pt]
\hline
      & 1000  & Semi & 6.5 & -0.9 & 93.5 & 43.2 & 1.093 \\
\rowcolor[gray]{0.93}
\cellcolor[gray]{1} & & Conf & 11.2 & 3.3 & 89.2 & 59.4 & 1.461 \\[2pt]
      & 2000  & Semi & 6.8 & -0.1 & 93.6 & 32.0 & 1.054 \\
\rowcolor[gray]{0.93}
\cellcolor[gray]{1} & & Conf & 9.2 & 2.7 & 91.0 & 49.3 & 1.136 \\[2pt]
      & 4000  & Semi & 7.0 & 0.0 & 93.3 & 24.5 & 1.078 \\
\rowcolor[gray]{0.93}
\cellcolor[gray]{1} & & Conf & 8.5 & 2.3 & 92.7 & 41.8 & 1.060 \\[2pt]
      & 8000  & Semi & 5.4 & -0.3 & 93.8 & 18.8 & 1.066 \\
\rowcolor[gray]{0.93}
\cellcolor[gray]{1} & & Conf & 8.5 & 1.2 & 93.2 & 34.7 & 1.046 \\[2pt]
      & 16000 & Semi & 5.8 & 0.0 & 94.1 & 14.4 & 1.055 \\
\rowcolor[gray]{0.93}
\cellcolor[gray]{1}\multirow{-10}{*}[4pt]{Right}
      & & Conf & 8.3 & 1.3 & 94.8 & 28.4 & 1.007 \\[2pt]
\hline
\end{tabular}
\end{table}

\begin{table}[H]
      \centering
      \caption{Finite-sample inference on $\Delta(s_0)$ under $H_0:
        \Delta(s_0)=\tau=0.20$ at $s_0=0$ when the error distribution is left-skewed
        (Left) or right-skewed (Right).
      Size is computed using the standard-normal critical value. The test level is $0.05$, and the confidence level is $0.95$.  For each sample size,
      Semi and Conf denote the semiparametric and conformal estimators.
      All entries are reported in units of
      $0.01$ except for SD($T$).}
      \label{tab:cond-delta-0}
      \setlength{\tabcolsep}{6pt}
      \renewcommand{\arraystretch}{1.1}
      \begin{tabular}{lrlrrrrr}
      \hline
      Error & $n$ & Method & Size & Bias & Cvrg & CI length & SD($T$) \\
      \hline
            & 1000  & Semi & 5.3 & -0.2 & 96.7 & 54.6 & 0.932 \\
      \rowcolor[gray]{0.93}
      \cellcolor[gray]{1} & & Conf & 11.5 & 3.1 & 87.3 & 71.8 & 1.414 \\[2pt]
            & 2000  & Semi & 4.2 & -0.1 & 96.3 & 41.1 & 0.964 \\
      \rowcolor[gray]{0.93}
      \cellcolor[gray]{1} & & Conf & 10.2 & 2.7 & 89.9 & 62.8 & 1.198 \\[2pt]
            & 4000  & Semi & 5.5 & 0.2 & 94.5 & 31.2 & 1.001 \\
      \rowcolor[gray]{0.93}
      \cellcolor[gray]{1} & & Conf & 7.8 & 0.9 & 93.7 & 52.2 & 1.099 \\[2pt]
            & 8000  & Semi & 5.5 & 0.1 & 93.8 & 23.8 & 1.033 \\
      \rowcolor[gray]{0.93}
      \cellcolor[gray]{1} & & Conf & 6.1 & 0.6 & 94.1 & 43.4 & 1.045 \\[2pt]
            & 16000 & Semi & 3.9 & 0.0 & 95.9 & 18.1 & 0.988 \\
      \rowcolor[gray]{0.93}
      \cellcolor[gray]{1}\multirow{-10}{*}[4pt]{Left}
            & & Conf & 4.9 & 0.0 & 94.4 & 35.5 & 0.997 \\[2pt]
      \hline
            & 1000  & Semi & 6.7 & 0.3 & 95.0 & 52.4 & 1.033 \\
      \rowcolor[gray]{0.93}
      \cellcolor[gray]{1} & & Conf & 13.9 & 5.7 & 87.8 & 69.4 & 1.283 \\[2pt]
            & 2000  & Semi & 5.6 & -0.6 & 93.2 & 39.4 & 1.048 \\
      \rowcolor[gray]{0.93}
      \cellcolor[gray]{1} & & Conf & 12.9 & 4.2 & 89.3 & 59.7 & 1.210 \\[2pt]
            & 4000  & Semi & 6.3 & 0.4 & 94.8 & 29.6 & 1.011 \\
      \rowcolor[gray]{0.93}
      \cellcolor[gray]{1} & & Conf & 10.4 & 3.1 & 92.0 & 49.5 & 1.088 \\[2pt]
            & 8000  & Semi & 5.2 & -0.2 & 94.2 & 22.8 & 1.038 \\
      \rowcolor[gray]{0.93}
      \cellcolor[gray]{1} & & Conf & 9.4 & 2.8 & 93.7 & 41.5 & 1.031 \\[2pt]
            & 16000 & Semi & 5.5 & 0.2 & 94.5 & 17.5 & 1.023 \\
      \rowcolor[gray]{0.93}
      \cellcolor[gray]{1}\multirow{-10}{*}[4pt]{Right}
            & & Conf & 10.3 & 2.3 & 93.8 & 34.1 & 1.052 \\[2pt]
      \hline
      \end{tabular}
\end{table}

\begin{table}[H]
\centering
\caption{Finite-sample inference on $\Delta(s_0)$ under $H_0:
  \Delta(s_0)=\tau=0.20$ at $s_0=0.5$ when the error distribution is left-skewed
  (Left) or right-skewed (Right).
Size is computed using the standard-normal critical value. The test level is $0.05$, and the confidence level is $0.95$.  For each sample size,
Semi and Conf denote the semiparametric and conformal estimators.
All entries are reported in units of
$0.01$ except for SD($T$).}
\label{tab:cond-delta-p05}
\setlength{\tabcolsep}{6pt}
\renewcommand{\arraystretch}{1.1}
\begin{tabular}{lrlrrrrr}
\hline
Error & $n$ & Method & Size & Bias & Cvrg & CI length & SD($T$) \\
\hline
      & 1000  & Semi & 5.4 & -0.3 & 93.9 & 62.2 & 1.014 \\
\rowcolor[gray]{0.93}
\cellcolor[gray]{1} & & Conf & 11.9 & 3.5 & 86.1 & 80.9 & 1.597 \\[2pt]
      & 2000  & Semi & 5.3 & -0.1 & 95.5 & 46.6 & 1.001 \\
\rowcolor[gray]{0.93}
\cellcolor[gray]{1} & & Conf & 11.2 & 4.2 & 90.1 & 70.0 & 1.172 \\[2pt]
      & 4000  & Semi & 5.6 & -0.1 & 94.6 & 35.2 & 1.013 \\
\rowcolor[gray]{0.93}
\cellcolor[gray]{1} & & Conf & 8.1 & 2.3 & 92.6 & 59.3 & 1.083 \\[2pt]
      & 8000  & Semi & 4.8 & -0.1 & 94.8 & 27.0 & 1.000 \\
\rowcolor[gray]{0.93}
\cellcolor[gray]{1} & & Conf & 7.8 & 2.0 & 93.6 & 49.0 & 1.048 \\[2pt]
      & 16000 & Semi & 4.7 & -0.1 & 94.7 & 20.7 & 1.021 \\
\rowcolor[gray]{0.93}
\cellcolor[gray]{1}\multirow{-10}{*}[4pt]{Left}
      & & Conf & 7.8 & 2.2 & 94.5 & 40.4 & 1.014 \\[2pt]
\hline
      & 1000  & Semi & 5.7 & 0.4 & 93.5 & 62.4 & 1.016 \\
\rowcolor[gray]{0.93}
\cellcolor[gray]{1} & & Conf & 14.7 & 5.4 & 87.3 & 80.5 & 1.352 \\[2pt]
      & 2000  & Semi & 5.0 & -0.3 & 95.7 & 47.1 & 0.987 \\
\rowcolor[gray]{0.93}
\cellcolor[gray]{1} & & Conf & 9.6 & 2.3 & 90.5 & 70.5 & 1.173 \\[2pt]
      & 4000  & Semi & 4.6 & 0.0 & 96.4 & 35.8 & 0.951 \\
\rowcolor[gray]{0.93}
\cellcolor[gray]{1} & & Conf & 8.5 & 2.5 & 92.2 & 60.3 & 1.065 \\[2pt]
      & 8000  & Semi & 4.0 & 0.0 & 95.4 & 27.2 & 0.947 \\
\rowcolor[gray]{0.93}
\cellcolor[gray]{1} & & Conf & 6.6 & 1.1 & 94.0 & 49.1 & 1.039 \\[2pt]
      & 16000 & Semi & 5.2 & 0.2 & 95.4 & 20.9 & 0.977 \\
\rowcolor[gray]{0.93}
\cellcolor[gray]{1}\multirow{-10}{*}[4pt]{Right}
      & & Conf & 5.6 & 0.6 & 94.6 & 40.6 & 1.042 \\[2pt]
\hline
\end{tabular}
\end{table}
Tables~\ref{tab:cond-delta-m05}--\ref{tab:cond-delta-p05} show that
the semiparametric test is generally well calibrated under left- and right- skewed
errors. Confidence interval coverage is generally close
to the nominal $95\%$ level. The standard deviation of the standardized
statistic remains close to one.
The conformal test generally exhibits greater
size distortion and lower coverage despite wider confidence intervals.
Its calibration improves with sample size, although rejection
probabilities remain above the nominal $5\%$ level under right-skewed
errors at all three locations even at $n=16000$.
At some small sample sizes, the standardized conformal statistic also has
a standard deviation substantially above one, indicating that the distortion involves both bias and excess dispersion.

Figure~\ref{fig:cond-delta-power} shows that rejection probabilities
increase with the gap $\Delta(s_0)-\tau$ at all three locations.
At gaps of $0.10$ and $0.20$, the semiparametric test has higher
empirical power under both error distributions, with power close
to one at the larger gap. In contrast, the conformal test shows greater variation across
locations; under right-skewed errors, its power is substantially
lower at $s_0=0.5$ than at $s_0=-0.5$.
Together with the preceding results on test size, these findings
indicate that the semiparametric procedure combines better null
calibration with higher power at these alternatives.
Results for $n=1000$ and $n=4000$ are reported in
Appendix~\ref{sec:cond-delta-power}.

\begin{figure}[H]
  \centering
  \includegraphics[width=0.98\textwidth]{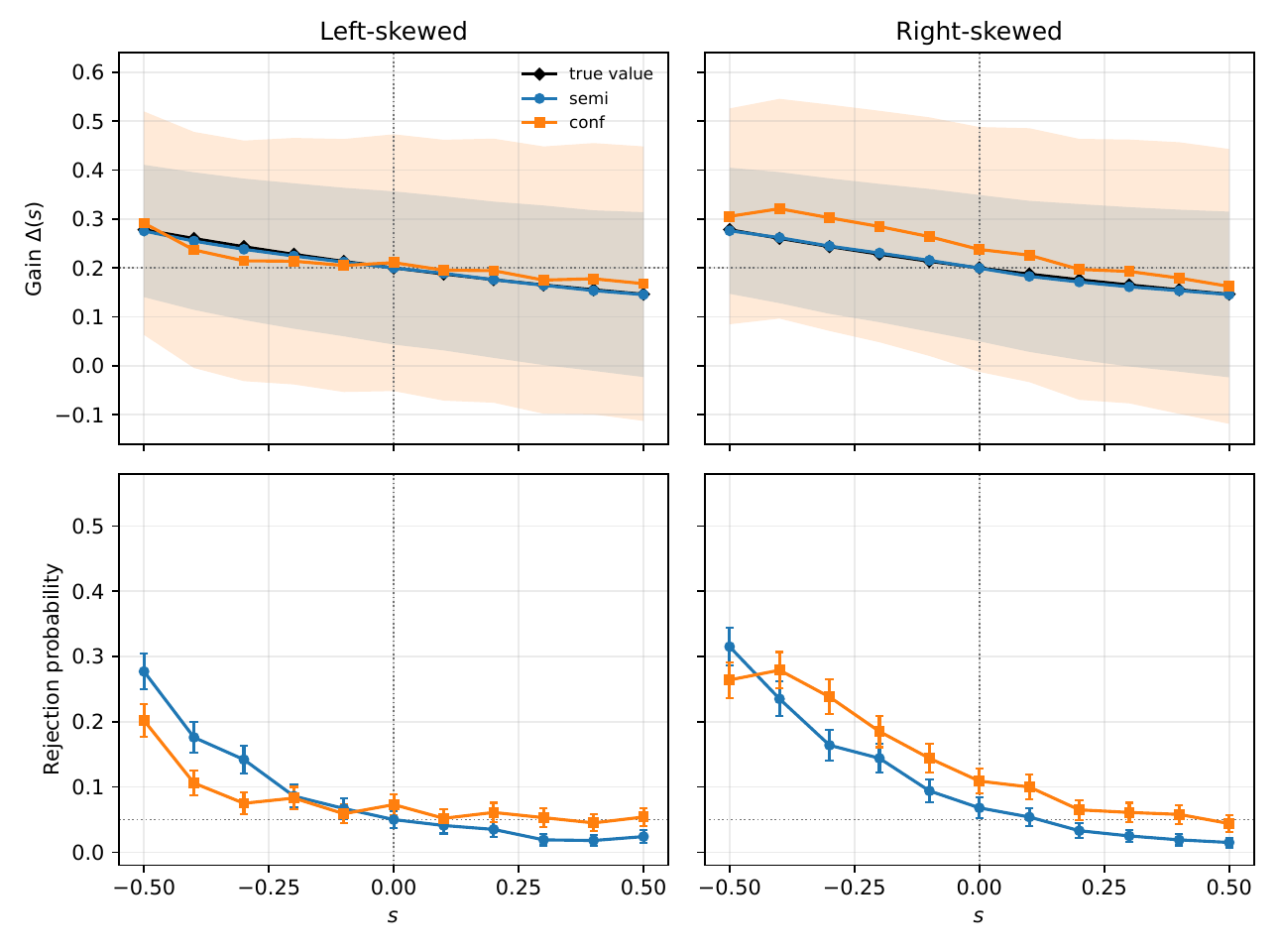}
  \caption{Columns correspond to
  left- and right-skewed errors. In the top row the black curve is the true value of $\Delta(s_0)$, the colored curves are Monte Carlo means, and
  the bands are averages of pointwise $95\%$ intervals. The horizontal
  dotted line is $0.20$. The bottom row is the rejection probability,
  with pointwise $95\%$ Monte Carlo error bars and 
  a horizontal line at $0.05$.}
  \label{fig:acquire-n4000}
\end{figure}

\medskip\noindent\textbf{Variation across the locations.}
To study how $\Delta(s_0)$ varies with $s_0$ in a single population, we hold $\beta=0.68$ fixed, so that $\Delta(0)=0.20$.
On the grid $s_0\in\{-0.5,-0.4,\ldots,0.5\}$ at $n=4000$, we estimate $\Delta(s_0)$ with
$R=1000$ replications under left- and right-skewed error distributions.
Figure~\ref{fig:acquire-n4000} reports the Monte Carlo means of the
semiparametric and conformal estimators, the average pointwise $95\%$ intervals, and the rejection
probabilities. Here, the average intervals are obtained by averaging the lower and upper endpoints separately over 1000 replications.

Figure~\ref{fig:acquire-n4000} shows that the semiparametric estimator
closely tracks the declining true gain $\Delta(s_0)$, with pointwise coverage
remaining near the nominal $95\%$ level.
At the boundary $s=0$, rejection probabilities are close to the
nominal level. They increase where the true gain exceeds the 
acquisition threshold and decrease where it falls below the threshold.
The conformal estimator tracks the true gain less closely.
Under right-skewed errors, it overestimates the gain with an inflated
rejection probability  throughout the grid.
Under left-skewed errors, rejection probabilities remain low
on the left
half of the grid and high on the right half of the grid, indicating limited power to identify locations
where acquisition is beneficial and frequent false positives 
at locations where acquisition yields no gain.

\section{Real data application}\label{sec:real}
In many research settings, a key question is whether the predictive
value of a covariate justifies the cost of collecting it. In this
section, we apply our method to several real datasets to address this
question. 

\subsection{NHANES}
We use data from the National Health
and Nutrition Examination Survey (NHANES;
\url{https://wwwn.cdc.gov/nchs/nhanes/}), a continuous 
cross-sectional survey of the civilian noninstitutionalized population of
the United States conducted by the National Center for Health Statistics. We use the 2011--2012, 2013--2014 and 2015--2016
cycles, with the sample containing
$n=5287$ participants. 
NHANES administered an
oral glucose tolerance test (OGTT): after the fasting blood draw, a
participant drinks a $75$\,g glucose solution and a second blood sample is
taken two hours later. It measures how
well the body clears a glucose load and is the reference quantity for
glucose tolerance: values of $140$\,mg/dL or above indicate impaired
glucose tolerance, and values of $200$\,mg/dL or above are in the diabetic
range. We take the two-hour glucose concentration as the outcome $Y$.
The relatively cheap covariate
vector $\X$ contains $13$ items from the questionnaire and the physical
examination: age, sex, race and Hispanic origin, the ratio of family
income to poverty level, education, BMI, waist circumference, height, systolic
and diastolic blood pressure, pulse, a
family history of diabetes, and an indicator of the survey cycle.

\begin{figure}[H]
  \centering
  \includegraphics[width=0.98\textwidth]{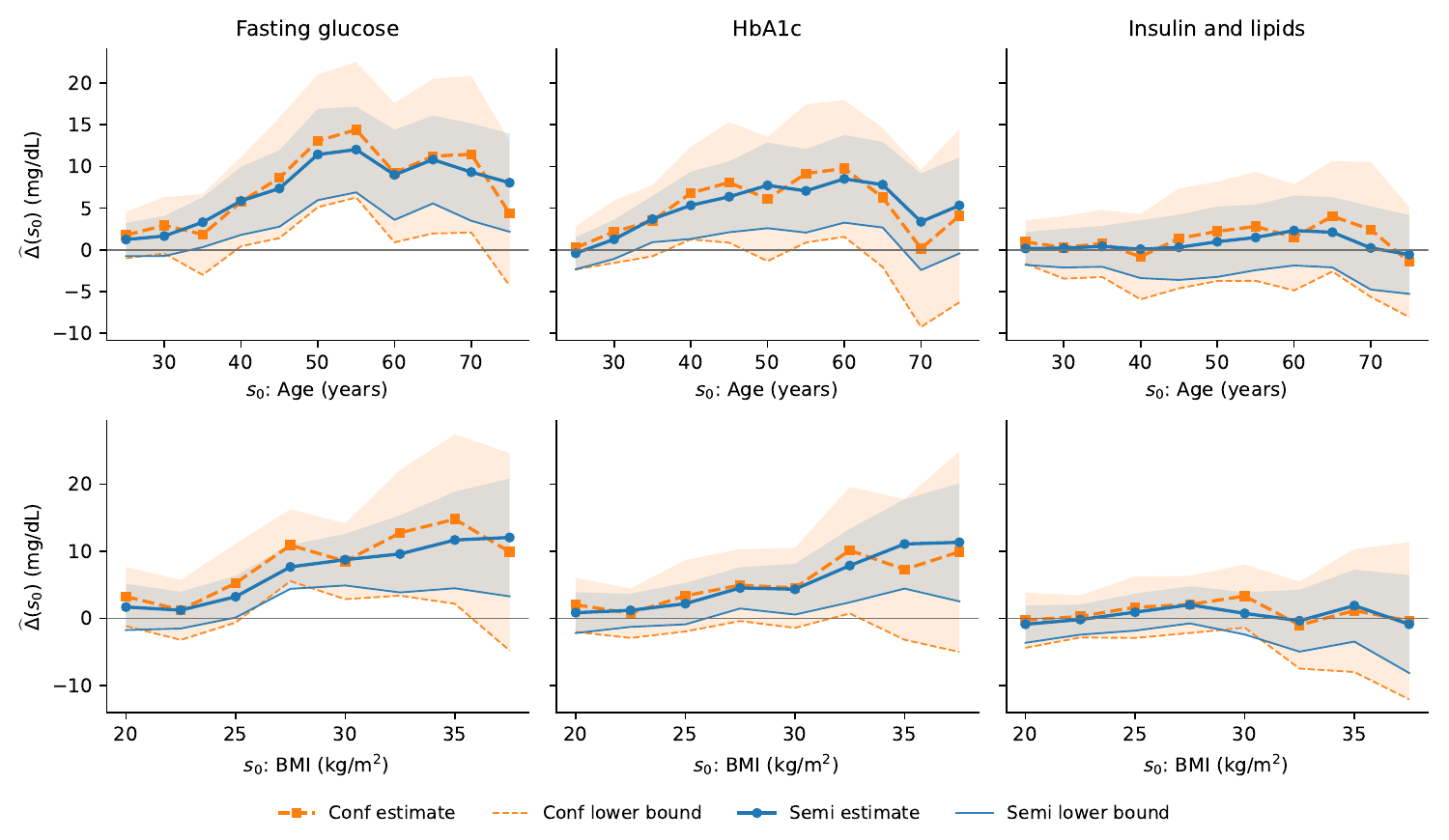}
  \caption{Estimated $\Delta(s_0)$ for the two-hour OGTT glucose. Bands are pointwise two-sided $90\%$ confidence intervals; their lower edge, drawn as a thin
  line without markers, is the one-sided $95\%$ lower bound. Columns correspond
  to the three choices of $\Z$; rows correspond to $S=$ age (top) and
  $S=$ BMI (bottom).}
  \label{fig:ogtt-delta}
\end{figure}

\begin{figure}[H]
  \centering
  \includegraphics[width=0.9\textwidth]{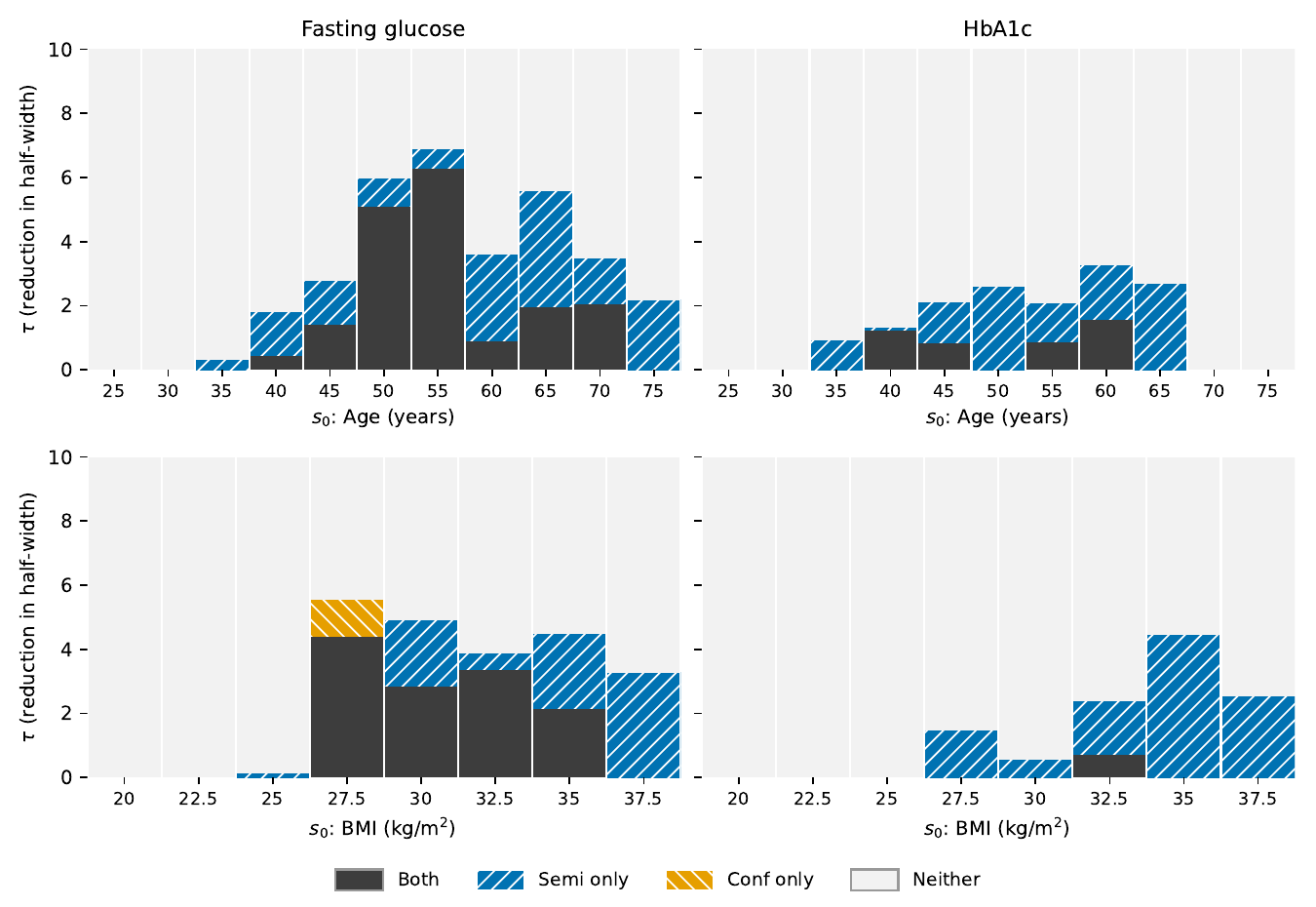}
  \caption{Decision map for the two-hour OGTT glucose, with $\Z$ fasting glucose (left) or HbA1c (right) and $S=$ age (top) or BMI (bottom). For each grid point $s_0$ and threshold $\tau$, the color shows which test rejects $H_0: \Delta(\s_0)\le\tau$ at level $0.05$.}
  \label{fig:ogtt-map}
\end{figure}

The practical question is how much laboratory information is needed to accuratey predict a person's two-hour glucose. 
We
consider three choices of the additional covariates $\Z$: (i) fasting
plasma glucose; (ii) glycated hemoglobin (HbA1c); (iii) fasting
insulin together with the lipid panel (triglycerides, LDL and HDL
cholesterol). 
Obtaining $\Z$ requires blood sampling and laboratory assays.
We take $S$ to be age or BMI and set prediction interval level at $1-\alpha=0.9$.
Both regression functions are estimated by gradient boosted trees. The
conditional density of the residual is estimated by the kernel
method. We still use a bandwidth of order $n^{-1/4}$ for the
semiparametric estimator and $n^{-2/5}$ for the conformal estimator.
Because a single random 
two-fold split adds noise, we repeat the cross-fitting over $20$ random
splits and aggregate by the median \citep{chernozhukov2018double}. 
We test $H_0:\Delta(s_0)\le 0$ with the one-sided
statistic at level $0.05$. 
Figure~\ref{fig:ogtt-delta} shows $\wh\Delta(s_0)$ and
$\wt\Delta(s_0)$ over the full grids with $\{25,30,\ldots,75\}$ for
age (years) and $\{20,22.5,\ldots,37.5\}$ for BMI (kg/m$^2$).  
Figure~\ref{fig:ogtt-map} compares the two tests through decision maps showing
which test rejects $H_0:\Delta(s_0)\le\tau$ as $s_0$ and the threshold
$\tau$ vary.

Figure~\ref{fig:ogtt-delta} shows that the estimated $\Delta(s_0)$
varies markedly with age and BMI.  
For fasting glucose, the estimated
reduction in prediction interval half width is small at younger ages
and lower BMI values, reaching on the order of $10$\,mg/dL at older
ages and higher BMI values. In terms of testing $H_0: \Delta(s_0)\le0$,
the semiparametric test rejects $H_0$ at ages
$35$--$75$ and BMI values of $25$--$37.5$, compared with
$40$--$70$ and $27.5$--$35$, respectively, for the conformal test.
HbA1c follows a similar pattern: the semiparametric test rejects at ages $35$--$65$
and BMI values of $27.5$--$37.5$, whereas the conformal test
rejects only at ages $40$--$45$ and $55$--$60$ and at a BMI of $32.5$.
For fasting glucose and HbA1c, the pointwise semiparametric test rejects
$H_0$ at $27$ of the $38$ grid points, compared with $16$
for the conformal test. At all $11$ points where the decisions differ,
only the semiparametric test rejects. Thus, with the same data, the semiparametric procedure
yields more precise estimates of the half-width reduction and is
more powerful in detecting 
 positive gains. 

 These age and BMI patterns suggest that questionnaire and examination
data can already predict two-hour glucose reasonably well for young adults
of normal weight, whereas fasting glucose or HbA1c measurements are
worth collecting for older adults and for 
overweight individuals. This is consistent with the emphasis on age
and adiposity in diabetes screening recommendations
\citep{uspstf2021screening,ada2026diagnosis}. 
For insulin and lipids, neither method detected a statistically
significant reduction in the half-width of the prediction interval. 

Figure~\ref{fig:ogtt-map} extends the comparison to positive thresholds
$\tau$, showing how large a reduction in prediction interval half width
each test supports at each grid point. The difference is most pronounced
for HbA1c: the semiparametric test still rejects at some grid points for
thresholds as large as $4.4$\,mg/dL, whereas the conformal test rejects at
none once $\tau$ reaches $1.6$\,mg/dL. For fasting glucose, the
semiparametric test generally supports larger thresholds, with an exception
at a BMI of $27.5$.

\subsection{Further applications}\label{sec:real-further}
The question of whether costly covariates sufficiently reduce prediction interval
width 
arises across many areas, such as civil engineering, environmental science, and medicine.
We illustrate the use of our method in these fields through applications
to concrete strength, soil organic carbon, and warfarin dosing.

\textbf{Concrete strength.}
Non-destructive tests allow concrete strength to be assessed without
destroying the specimen. The SonReb database
\citep{matthews2025sonreb} (\url{https://zenodo.org/records/15392443})
combines rebound number, ultrasonic pulse velocity and compressive strength
measurements from $53$ studies. Removing exact duplicates within each study
leaves $n=3{,}226$ records. The outcome $Y$ is compressive strength normalized
to a $150\times300$\,mm reference cylinder, with mean $28.8$\,MPa.
The cheap covariates $\X$ include the median rebound number, study
indicators, log specimen age, specimen types for the non-destructive and
compressive tests, hammer orientation, reinforcement, maximum aggregate
size, water--cement ratio and $28$-day design strength. 
The additional covariate $\Z$ is the average ultrasonic pulse velocity,
which requires ultrasonic equipment and additional operator time. Our
goal is to find out at which rebound numbers this measurement narrows the prediction interval
for compressive strength. Accordingly,
$S$ is the rebound number, with $s_0\in\{20,24,\ldots,44\}$.

\textbf{Soil organic carbon.}
The 2015 Land Use/Cover Area frame Survey (LUCAS) provides soil properties,
visible and near-infrared (vis--NIR) reflectance spectra, and ancillary
environmental data for samples from $28$ European countries
\citep{orgiazzi2018lucas,jones2020lucas}. These data are available from the
European Soil Data Centre (\url{https://esdac.jrc.ec.europa.eu/content/lucas2015-topsoil-data}).
We retain $n=20{,}383$ mineral-soil samples with organic carbon content at
most $120$\,g/kg and available spectra and climate covariates. The outcome
$Y$ is organic carbon content, with mean $26.4$\,g/kg.
The cheap covariates $\X$ comprise $29$ features available without a site
visit: projected coordinates, elevation, slope, the sine and cosine of
aspect, $19$ bioclimatic variables, climate zone, biogeographical region,
mapped soil type and country. The additional covariates $\Z$ are the first $30$ principal
components of the vis--NIR spectrum.
Obtaining $\Z$ requires soil sampling and a laboratory scan. Our question is in which
climates these measurements narrow the prediction interval for organic
carbon. We take $S$ to be mean annual
temperature, with $s_0\in\{2,4,\ldots,16\}$\,$^\circ$C. 

\textbf{Warfarin dose.}
The International Warfarin Pharmacogenetics Consortium collected clinical
and genetic data to study variation in warfarin dose requirements
\citep{iwpc2009}. We use its cleaned subset of $n=1780$ patients distributed
with the R package \texttt{simsl}
\citep{simsl} (\url{https://CRAN.R-project.org/package=simsl}).
The outcome $Y$ is the stable weekly dose, with mean $34.3$\,mg/week.
The cheap covariates $\X$ are eight routinely available characteristics:
weight, height, age, use of cytochrome P450 enzyme inducers, use of
amiodarone, sex, and indicators of Black or African American and of Asian
race. Weight, height and age are standardized.
We consider the VKORC1 and CYP2C9 genotypes separately as choices of $\Z$;
each requires an additional laboratory assay. Our question is for which patients
genotyping narrows the prediction interval for the stable dose. We
take $S$ to be standardized weight, with 
$s_0\in\{-1.5,-1,\ldots,1.5\}$.

\begin{figure}[H]
  \centering
  \includegraphics[width=0.98\textwidth]{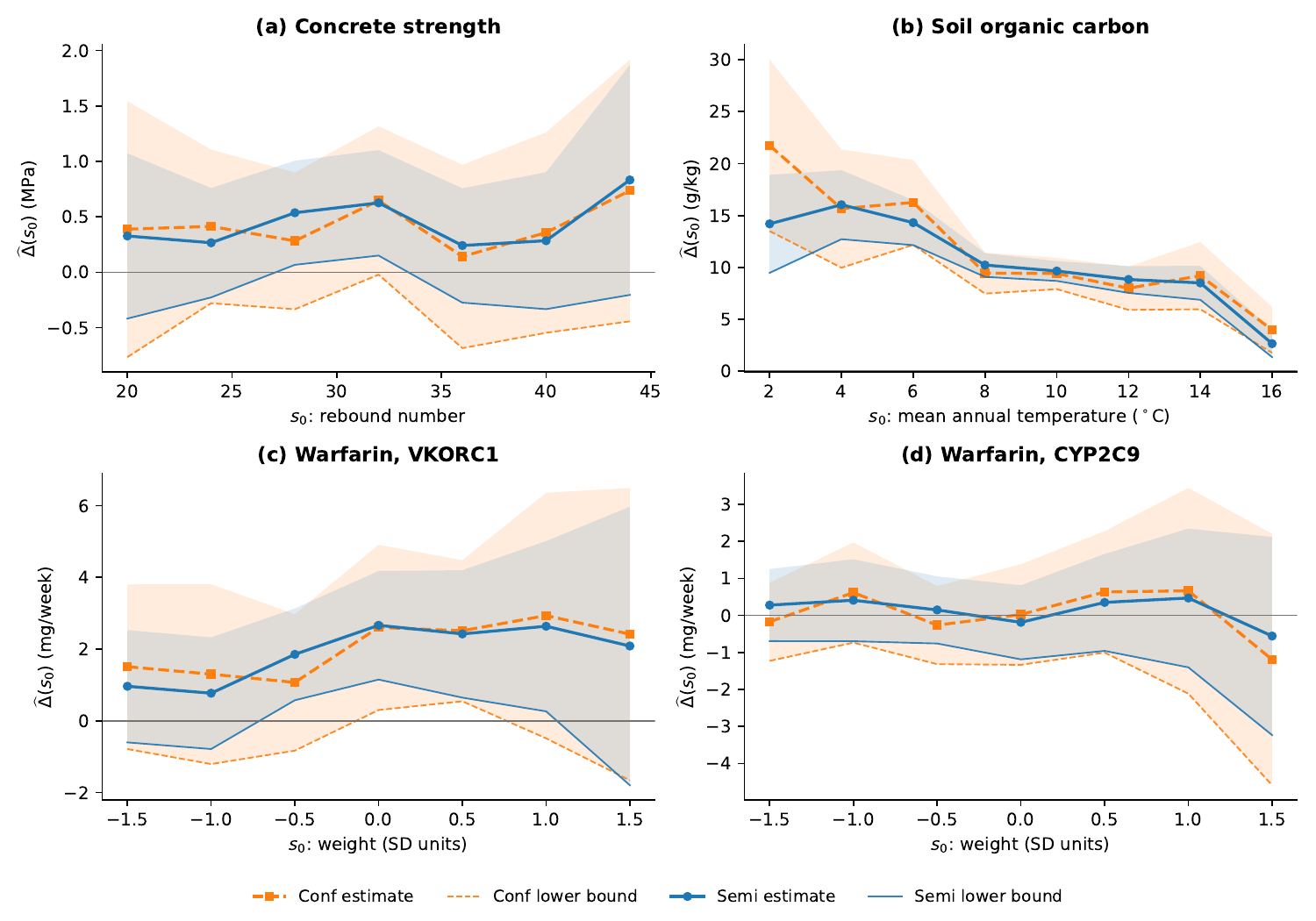}
  \caption{Estimated $\Delta(s_0)$. Bands are pointwise
  two-sided $90\%$ confidence intervals; their lower edge, drawn as a thin
  line without markers, is the one-sided $95\%$ lower bound.
  (a)~concrete compressive strength;
  (b)~soil organic carbon;
  (c)~warfarin dose, $\Z$ the VKORC1 genotype;
  (d)~warfarin dose, $\Z$ the CYP2C9 genotype.}
  \label{fig:further-delta}
\end{figure}

Figure~\ref{fig:further-delta} shows that, for concrete strength, the estimated half width
reduction from ultrasonic pulse velocity ranges from approximately $0.14$ to $0.83$\,MPa. The
semiparametric test rejects $H_0$ only at rebound numbers
$28$ and $32$, whereas the conformal test rejects nowhere on the grid.
\citet{shih2015improving} found that adding ultrasonic measurements improved
rebound-based prediction of concrete strength in their experimental data.
Our semiparametric
analysis further identifies where this incremental value is
statistically supported, whereas the conformal test detects
no significant gain at any point on the evaluated grid.

For soil organic carbon, the gain is largest in colder climates:
the semiparametric estimates are about $14$--$16$\,g/kg at
$2$--$6^\circ$C and decline to $2.7$\,g/kg at $16^\circ$C.
These results identify colder regions as the settings where soil sampling
and spectral scanning yield the largest estimated reduction in
interval half width.

For warfarin, adding VKORC1 genotype information to routine patient
characteristics yields estimated half-width reductions of about
$2$--$3$\,mg/week near the average weight. The semiparametric test rejects at standardized weights from
$-0.5$ to $1$, compared with $0$ to $0.5$ for the conformal test. The analysis therefore provides
evidence that VKORC1 testing can narrow dose prediction intervals for
patients in this weight range. Neither test detects a statistically
significant gain from CYP2C9 at any weight on the grid. 
This finding is consistent with \citet{rieder2005effect}, who found
that VKORC1 explained a larger proportion of variation in warfarin
maintenance dose than CYP2C9. Our analysis provides evidence that
adding VKORC1 genotype information narrows dose prediction intervals
for patients in this weight range. 

\begin{figure}[H]
  \centering
  \includegraphics[width=\textwidth]{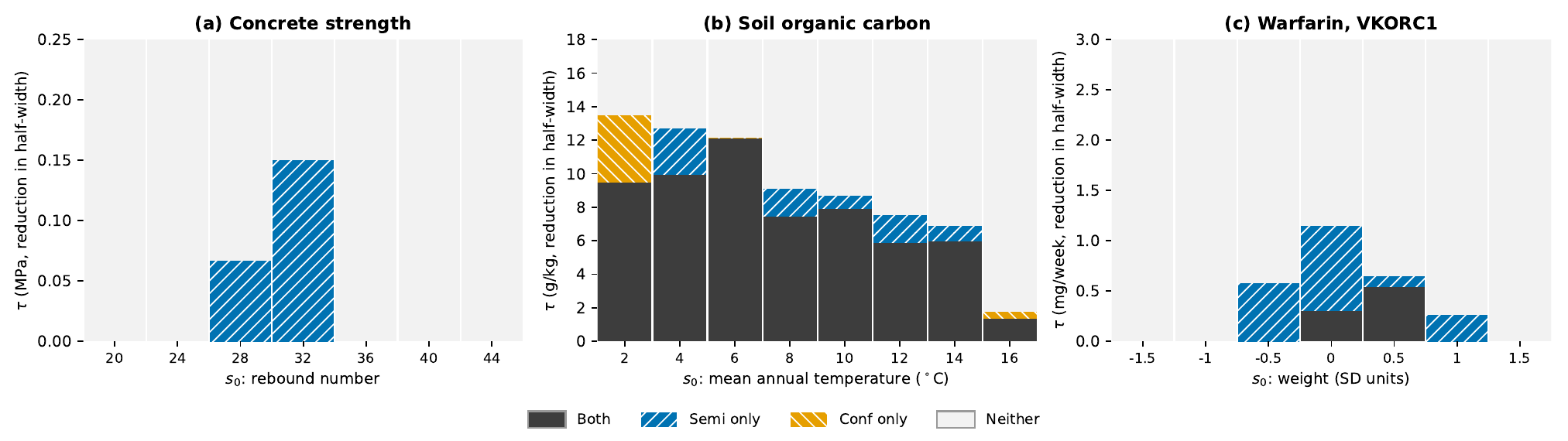}
  \caption{Decision maps for (a)~concrete compressive strength,
  (b)~soil organic carbon and (c)~the warfarin dose with $\Z$ the VKORC1
  genotype. For each grid point $s_0$ and threshold $\tau$,
  the color shows which test rejects
  $H_0:\Delta(s_0)\le\tau$ at level $0.05$.}
  \label{fig:further-map}
\end{figure}

Figure~\ref{fig:further-map} extends the comparison to positive thresholds
$\tau$. For concrete, only the semiparametric test rejects, at rebound
numbers $28$ and $32$, with lower bounds of about $0.07$ and
$0.15$\,MPa, respectively. These rejections therefore persist only for
small values of $\tau$.
For soil carbon, the semiparametric test supports larger thresholds at most
temperatures. The conformal test supports larger thresholds only at
$2^\circ$C and $16^\circ$C, the two endpoints of the grid.
For VKORC1, only the semiparametric test supports a reduction of
$1$\,mg/week, at the average weight ($s_0=0$). CYP2C9 is omitted because
neither test rejects even at $\tau=0$.

\clearpage
\pagenumbering{arabic}
{\centering
\section*{Appendix}%
\phantomsection
\makeatletter\def\@currentlabel{Appendix}\makeatother
\label{sec:appendix}}
\setcounter{equation}{0}\renewcommand{\theequation}{A\arabic{equation}}
\renewcommand{\theHequation}{A\arabic{equation}}
\setcounter{subsection}{0}\renewcommand{\thesubsection}{A.\arabic{subsection}}
\renewcommand{\theHsubsection}{A.\arabic{subsection}}
\setcounter{table}{0}\renewcommand{\thetable}{A.\arabic{table}}
\renewcommand{\theHtable}{A.\arabic{table}}
\setcounter{figure}{0}\renewcommand{\thefigure}{A.\arabic{figure}}
\renewcommand{\theHfigure}{A.\arabic{figure}}

\subsection{Simulation results under symmetric errors}
\label{sec:sim-sym}

As discussed in Remark~\ref{rem:sym}, when the error distribution is
symmetric, the first-order nuisance bias of the conformal estimator
vanishes, so the two methods are expected to behave similarly. We report
the corresponding Monte Carlo results here, using the same designs as in
Section~\ref{sec:sim} with $\epsilon_0=G$.

\begin{table}[H]
\centering
\caption{Bias, SD, SE, and RMSE for the estimators of $\zeta_\x$ and
$\zeta_\w$ under symmetric errors (symm).  For each sample size,
Semi and Conf denote the semiparametric and conformal estimators,
respectively. All entries measures are reported in units of
$0.01$.}
\label{tab:learned-cv-zeta-sym}
\setlength{\tabcolsep}{6pt}
\renewcommand{\arraystretch}{1.1}
\begin{tabular}{lrlrrrrrrrr}
\hline
& & & \multicolumn{4}{c}{$\zeta_\x$} & \multicolumn{4}{c}{$\zeta_\w$} \\
\cline{4-7}\cline{8-11}
Error & $n$ & Method & Bias & SD & SE & RMSE & Bias & SD & SE & RMSE \\
\hline
      & 1000  & Semi & 0.37 & 5.56 & 5.77 & 5.57 & 0.94 & 5.77 & 5.78 & 5.84 \\
\rowcolor[gray]{0.93}
\cellcolor[gray]{1} & & Conf & 0.65 & 5.36 & 5.71 & 5.39 & 1.61 & 5.59 & 5.71 & 5.81 \\[2pt]
      & 2000  & Semi & 0.33 & 4.16 & 4.00 & 4.18 & 0.48 & 4.15 & 3.98 & 4.18 \\
\rowcolor[gray]{0.93}
\cellcolor[gray]{1} & & Conf & 0.58 & 4.04 & 3.98 & 4.08 & 0.92 & 4.01 & 3.95 & 4.11 \\[2pt]
      & 4000  & Semi & 0.17 & 2.84 & 2.81 & 2.84 & 0.30 & 2.80 & 2.78 & 2.81 \\
\rowcolor[gray]{0.93}
\cellcolor[gray]{1} & & Conf & 0.30 & 2.79 & 2.80 & 2.80 & 0.60 & 2.74 & 2.77 & 2.80 \\[2pt]
      & 8000  & Semi & 0.06 & 1.97 & 1.97 & 1.97 & 0.17 & 1.91 & 1.95 & 1.92 \\
\rowcolor[gray]{0.93}
\cellcolor[gray]{1} & & Conf & 0.15 & 1.95 & 1.97 & 1.95 & 0.36 & 1.88 & 1.95 & 1.91 \\[2pt]
      & 16000 & Semi & 0.01 & 1.35 & 1.38 & 1.35 & 0.04 & 1.32 & 1.37 & 1.32 \\
\rowcolor[gray]{0.93}
\cellcolor[gray]{1}\multirow{-10}{*}[4pt]{Symm}
      & & Conf & 0.07 & 1.34 & 1.38 & 1.34 & 0.16 & 1.31 & 1.37 & 1.32 \\
\hline
\end{tabular}
\end{table}
Table~\ref{tab:learned-cv-zeta-sym} shows that the two estimators perform
similarly under symmetric errors. Although the semiparametric estimator
has a slight advantage in terms of Bias, this is consistent with
the explanation given in Remark~\ref{rem:sym}.

\begin{table}[H]
\centering
\caption{Finite-sample inference on $\Delta$ under $H_0: \Delta(\beta)=\tau=0.20$
and symmetric errors (symm). The test level is $0.05$, and the confidence level is $0.95$.  For each sample size,
Semi and Conf denote the semiparametric and conformal estimators,
respectively. All entries are reported in units of
$0.01$ except for SD($T$).}
\label{tab:delta_calib-sym}
\setlength{\tabcolsep}{6pt}
\renewcommand{\arraystretch}{1.1}
\begin{tabular}{lrlrrrrr}
\hline
Error & $n$ & Method & Size & Bias & Cvrg & CI length & SD($T$) \\
\hline
      & 1000  & Semi & 3.3 & -0.6 & 95.1 & 20.7 & 0.978 \\
\rowcolor[gray]{0.93}
\cellcolor[gray]{1} & & Conf & 2.3 & -1.0 & 95.6 & 20.4 & 0.933 \\[2pt]
      & 2000  & Semi & 4.8 & -0.3 & 93.3 & 14.3 & 1.021 \\
\rowcolor[gray]{0.93}
\cellcolor[gray]{1} & & Conf & 3.8 & -0.4 & 94.4 & 14.2 & 0.986 \\[2pt]
      & 4000  & Semi & 4.3 & -0.2 & 93.5 & 9.9 & 1.040 \\
\rowcolor[gray]{0.93}
\cellcolor[gray]{1} & & Conf & 3.5 & -0.3 & 94.8 & 9.9 & 1.011 \\[2pt]
      & 8000  & Semi & 4.2 & -0.1 & 95.2 & 7.0 & 1.044 \\
\rowcolor[gray]{0.93}
\cellcolor[gray]{1} & & Conf & 3.9 & -0.2 & 95.3 & 6.9 & 1.029 \\[2pt]
      & 16000 & Semi & 4.5 & -0.1 & 95.3 & 4.9 & 0.997 \\
\rowcolor[gray]{0.93}
\cellcolor[gray]{1}\multirow{-10}{*}[4pt]{Symm}
      & & Conf & 3.8 & -0.1 & 95.9 & 4.9 & 0.983 \\
\hline
\end{tabular}
\end{table}
Under symmetric errors, the semiparametric procedure is mildly
conservative, with rejection probabilities close to the nominal level,
whereas the conformal procedure is substantially more conservative. The
semiparametric confidence interval achieves coverage close to the
nominal level.

\subsection{Simulation results under heavy-tailed errors}
This subsection repeats the Monte Carlo design of the main numerical
experiments, with the single change that the base error is
$G\sim t(4)/\sqrt{2}$ rather than $G\sim N(0,1)$. The $t(4)$ law is scaled
to have variance one. The mixture probabilities, the skewness parameter
$a$, the mean function $m_0$, the sample sizes, and the number of
replications remain as in the main text. Estimation uses $\beta=0.2$.
Inference uses the same gaps
$\Delta(\beta)-\tau\in\{-0.05,0,0.05,0.10,0.20\}$ at
$\tau=0.20$. The corresponding
oracle-calibrated values of $\beta$ are $0.59$, $0.67$, $0.74$, $0.80$, and $0.91$, respectively.

\begin{figure}[H]
  \centering
  \includegraphics[width=0.98\textwidth]{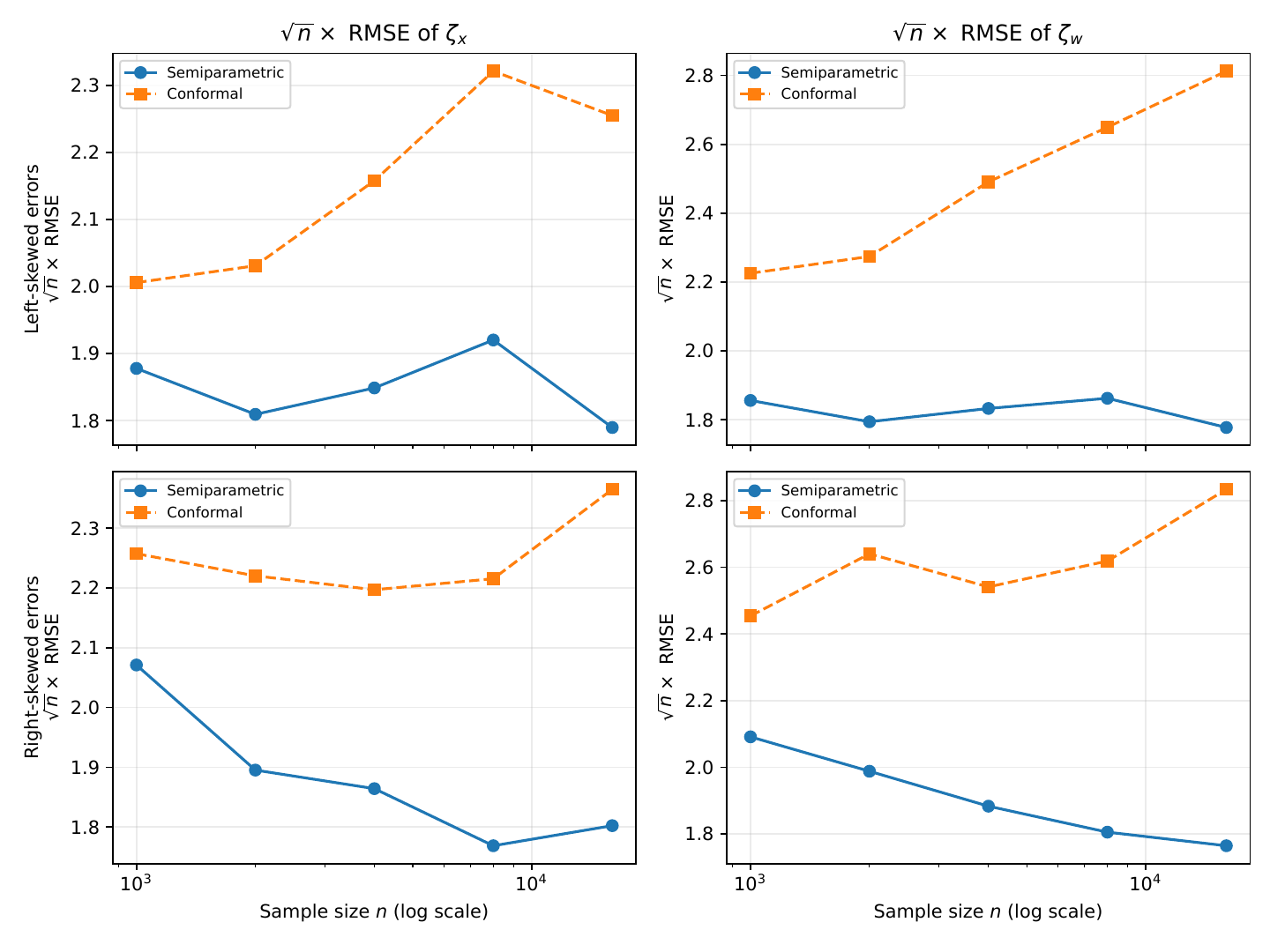}
  \caption{Monte Carlo RMSE multiplied by $\sqrt n$ for the semiparametric and conformal
  estimators of $\zeta_\x$ and $\zeta_\w$ under left-skewed (left)
  and right-skewed (right)
  heavy-tailed errors. The horizontal axis is logarithmic. Rows
  correspond to 
  the error direction and columns to the target.}
  \label{fig:t4-learned-cv-zeta-rmse}
\end{figure}

\medskip\noindent\textbf{Estimation of $\zeta_\v$.}
Table~\ref{tab:t4-learned-cv-zeta} reports bias, SD, the Monte Carlo mean of
the estimated standard error (SE), and RMSE. Figure~\ref{fig:t4-learned-cv-zeta-rmse}
reports $\sqrt n$ times RMSE against $n$.

\begin{table}[H]
\centering
\caption{Bias, SD, SE, and RMSE
for the estimators of $\zeta_\x$ and $\zeta_\w$ under left-skewed
(Left) and
right-skewed (Right) heavy-tailed errors. For each sample size,
Semi and Conf denote the semiparametric and conformal estimators,
respectively. All entries are reported in units of
$0.01$.}
\label{tab:t4-learned-cv-zeta}
\setlength{\tabcolsep}{6pt}
\renewcommand{\arraystretch}{1.1}
\begin{tabular}{lrlrrrrrrrr}
\hline
& & & \multicolumn{4}{c}{$\zeta_\x$} & \multicolumn{4}{c}{$\zeta_\w$} \\
\cline{4-7}\cline{8-11}
Error & $n$ & Method & Bias & SD & SE & RMSE & Bias & SD & SE & RMSE \\
\hline
      & 1000  & Semi & 0.50 & 5.92 & 5.92 & 5.94 & 0.79 & 5.82 & 5.68 & 5.87 \\
\rowcolor[gray]{0.93}
\cellcolor[gray]{1} & & Conf & 2.99 & 5.60 & 6.48 & 6.34 & 4.50 & 5.42 & 6.25 & 7.04 \\[2pt]
      & 2000  & Semi & 0.47 & 4.02 & 4.11 & 4.05 & 0.45 & 3.99 & 3.95 & 4.01 \\
\rowcolor[gray]{0.93}
\cellcolor[gray]{1} & & Conf & 2.44 & 3.83 & 4.60 & 4.54 & 3.50 & 3.69 & 4.45 & 5.09 \\[2pt]
      & 4000  & Semi & 0.29 & 2.91 & 2.90 & 2.92 & 0.33 & 2.88 & 2.78 & 2.90 \\
\rowcolor[gray]{0.93}
\cellcolor[gray]{1} & & Conf & 1.93 & 2.82 & 3.28 & 3.41 & 2.83 & 2.74 & 3.18 & 3.94 \\[2pt]
      & 8000  & Semi & 0.15 & 2.14 & 2.03 & 2.15 & 0.04 & 2.08 & 1.97 & 2.08 \\
\rowcolor[gray]{0.93}
\cellcolor[gray]{1} & & Conf & 1.49 & 2.12 & 2.32 & 2.60 & 2.16 & 2.03 & 2.27 & 2.96 \\[2pt]
      & 16000 & Semi & -0.01 & 1.42 & 1.43 & 1.41 & -0.02 & 1.41 & 1.39 & 1.41 \\
\rowcolor[gray]{0.93}
\cellcolor[gray]{1}\multirow{-10}{*}[4pt]{Left}
      & & Conf & 1.07 & 1.43 & 1.65 & 1.78 & 1.67 & 1.46 & 1.61 & 2.22 \\[2pt]
\hline
      & 1000  & Semi & -1.00 & 6.48 & 6.20 & 6.55 & -1.08 & 6.53 & 6.20 & 6.61 \\
\rowcolor[gray]{0.93}
\cellcolor[gray]{1} & & Conf & -3.49 & 6.23 & 6.85 & 7.14 & -4.54 & 6.30 & 6.92 & 7.76 \\[2pt]
      & 2000  & Semi & -0.60 & 4.20 & 4.28 & 4.24 & -0.81 & 4.37 & 4.27 & 4.44 \\
\rowcolor[gray]{0.93}
\cellcolor[gray]{1} & & Conf & -2.65 & 4.20 & 4.83 & 4.96 & -3.83 & 4.50 & 4.89 & 5.90 \\[2pt]
      & 4000  & Semi & -0.16 & 2.94 & 2.96 & 2.95 & -0.22 & 2.97 & 2.96 & 2.98 \\
\rowcolor[gray]{0.93}
\cellcolor[gray]{1} & & Conf & -1.83 & 2.95 & 3.40 & 3.47 & -2.60 & 3.07 & 3.44 & 4.02 \\[2pt]
      & 8000  & Semi & -0.05 & 1.98 & 2.08 & 1.98 & -0.08 & 2.02 & 2.07 & 2.02 \\
\rowcolor[gray]{0.93}
\cellcolor[gray]{1} & & Conf & -1.40 & 2.05 & 2.40 & 2.48 & -2.01 & 2.13 & 2.42 & 2.93 \\[2pt]
      & 16000 & Semi & -0.06 & 1.42 & 1.46 & 1.42 & -0.13 & 1.39 & 1.45 & 1.39 \\
\rowcolor[gray]{0.93}
\cellcolor[gray]{1}\multirow{-10}{*}[4pt]{Right}
      & & Conf & -1.13 & 1.49 & 1.70 & 1.87 & -1.67 & 1.49 & 1.71 & 2.24 \\
\hline
\end{tabular}
\end{table}

Table~\ref{tab:t4-learned-cv-zeta} shows the same pattern as in
the Gaussian design. The semiparametric
estimator has smaller RMSE for both $\zeta_\x$ and $\zeta_\w$ at all
displayed 
sample sizes and under both skewness directions, primarily owing to
smaller absolute 
bias. The conformal bias for $\zeta_\w$ remains visible even at
$n=16000$. 
Relative to Table~\ref{tab:learned-cv-zeta}, left-skewed RMSE is slightly
smaller under heavy tails, whereas right-skewed RMSE is slightly larger. 

Figure~\ref{fig:t4-learned-cv-zeta-rmse} rescales RMSE by $\sqrt n$.
The semiparametric curves for both targets stay in a band of about
$1.8$--$2.1$, which is consistent with a root-$n$ rate. The conformal
$\sqrt n\times$RMSE increases with $n$, most clearly for $\zeta_\w$. As in Figure~\ref{fig:learned-cv-zeta-rmse}, this is the
finite-sample effect of a remainder term that vanishes
more slowly than 
$n^{-1/2}$. 

\begin{figure}[H]
  \centering
  \includegraphics[width=0.98\textwidth]{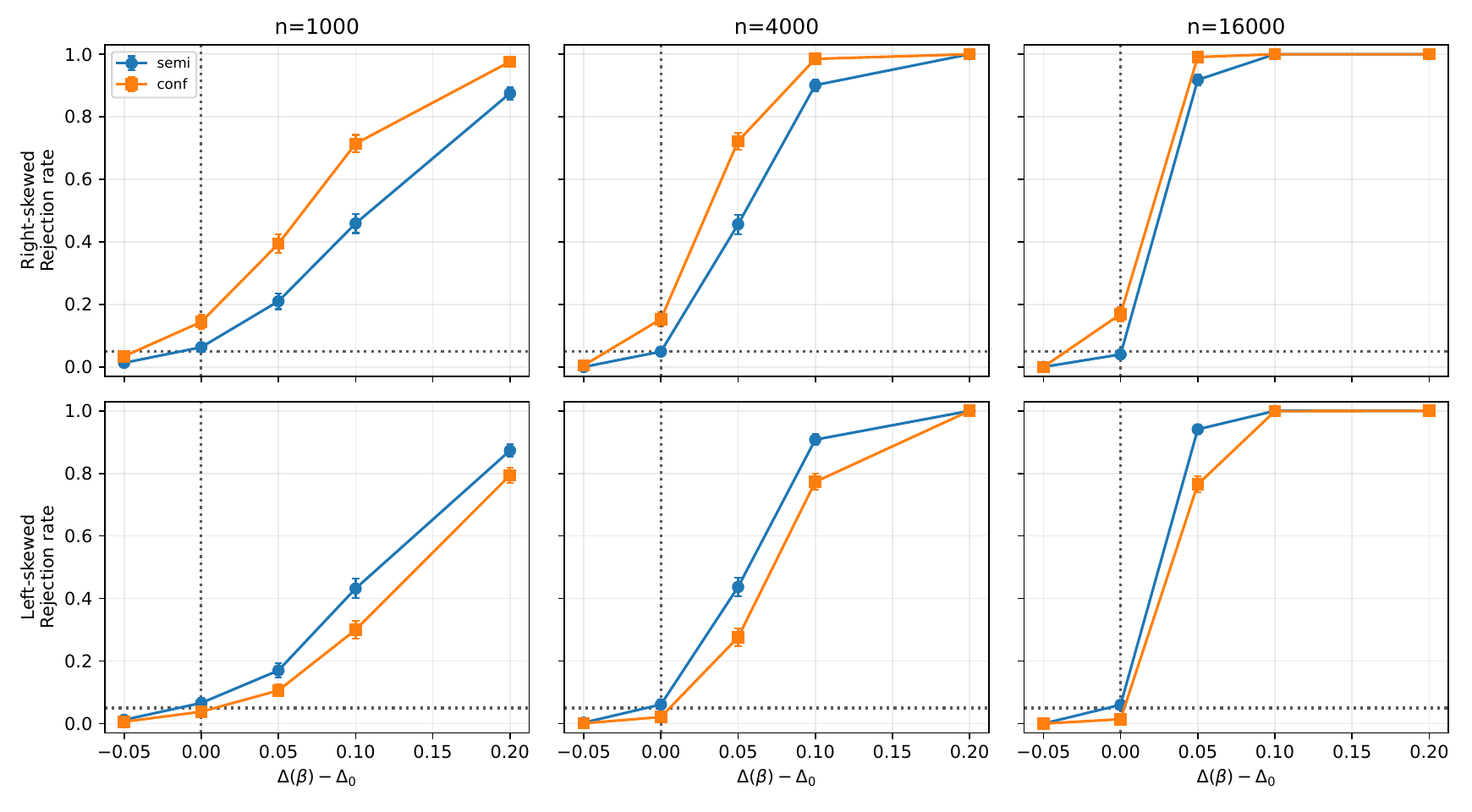}
  \caption{Rejection probabilities for testing
  $H_0:\Delta\leq\tau$ against $H_1:\Delta>\tau$ with
  $\tau=0.20$ under heavy-tailed errors, using the standard-normal
  critical value. Rows correspond to right- and left-skewed errors, columns
  to $n\in\{1000,4000,16000\}$, and the two curves in each panel represent the
  semiparametric and conformal methods. The horizontal dotted line denotes
  the nominal level $0.05$, the vertical dotted line marks the boundary gap
  $\Delta(\beta)-\tau=0$, and error bars are pointwise 95\% Monte Carlo
  confidence intervals.}
  \label{fig:t4-delta-power}
\end{figure}

\medskip\noindent\textbf{Inference based on $\Delta$.}
Table~\ref{tab:t4-delta-calib} assesses inferential calibration at the
boundary null 
$\Delta(\beta)=\tau=0.20$. Figure~\ref{fig:t4-delta-power} reports
rejection probabilities against the gap $\Delta(\beta)-\tau$.

\begin{table}[H]
\centering
\caption{Finite-sample calibration of inference on $\Delta$ at the
boundary null $\Delta(\beta)=\tau=0.20$ under heavy-tailed errors.
For each sample size,
Semi and Conf denote the semiparametric and conformal estimators,
respectively. Size is computed using the standard-normal critical value.
All entries are reported in units of
$0.01$ except for SD($T$). The nominal test and confidence levels are
$0.05$ and $0.95$, respectively.}
\label{tab:t4-delta-calib}
\setlength{\tabcolsep}{6pt}
\renewcommand{\arraystretch}{1.1}
\begin{tabular}{lrlrrrrr}
\hline
Error & $n$ & Method & Size & Bias & Cvrg & CI length & SD($T$) \\
\hline
      & 1000  & Semi & 6.6 & 0.4 & 94.8 & 25.8 & 1.026 \\
\rowcolor[gray]{0.93}
\cellcolor[gray]{1} & & Conf & 3.8 & -2.0 & 93.2 & 25.1 & 1.033 \\[2pt]
      & 2000  & Semi & 7.0 & 0.3 & 94.9 & 18.0 & 1.019 \\
\rowcolor[gray]{0.93}
\cellcolor[gray]{1} & & Conf & 2.3 & -1.5 & 93.0 & 17.5 & 1.021 \\[2pt]
      & 4000  & Semi & 6.1 & 0.3 & 94.5 & 12.5 & 1.027 \\
\rowcolor[gray]{0.93}
\cellcolor[gray]{1} & & Conf & 2.1 & -1.3 & 92.7 & 12.3 & 1.044 \\[2pt]
      & 8000  & Semi & 5.1 & 0.1 & 95.5 & 8.8 & 0.979 \\
\rowcolor[gray]{0.93}
\cellcolor[gray]{1} & & Conf & 0.7 & -1.3 & 91.2 & 8.7 & 1.020 \\[2pt]
      & 16000 & Semi & 6.0 & 0.0 & 94.6 & 6.2 & 1.011 \\
\rowcolor[gray]{0.93}
\cellcolor[gray]{1}\multirow{-10}{*}[4pt]{Left}
      & & Conf & 1.4 & -1.1 & 88.5 & 6.1 & 1.044 \\[2pt]
\hline
      & 1000  & Semi & 6.3 & 0.2 & 93.7 & 25.5 & 1.056 \\
\rowcolor[gray]{0.93}
\cellcolor[gray]{1} & & Conf & 14.4 & 4.0 & 90.5 & 25.1 & 1.007 \\[2pt]
      & 2000  & Semi & 5.7 & -0.2 & 94.2 & 17.6 & 1.045 \\
\rowcolor[gray]{0.93}
\cellcolor[gray]{1} & & Conf & 13.3 & 2.6 & 90.8 & 17.5 & 1.012 \\[2pt]
      & 4000  & Semi & 4.9 & -0.1 & 94.9 & 12.4 & 1.008 \\
\rowcolor[gray]{0.93}
\cellcolor[gray]{1} & & Conf & 15.3 & 2.0 & 90.8 & 12.4 & 0.985 \\[2pt]
      & 8000  & Semi & 4.8 & 0.0 & 94.5 & 8.7 & 1.011 \\
\rowcolor[gray]{0.93}
\cellcolor[gray]{1} & & Conf & 16.9 & 1.6 & 89.8 & 8.7 & 1.013 \\[2pt]
      & 16000 & Semi & 4.0 & -0.1 & 95.2 & 6.2 & 0.979 \\
\rowcolor[gray]{0.93}
\cellcolor[gray]{1}\multirow{-10}{*}[4pt]{Right}
      & & Conf & 16.9 & 1.1 & 89.1 & 6.2 & 1.016 \\
\hline
\end{tabular}
\end{table}
Table~\ref{tab:t4-delta-calib} confirms the main findings of
Table~\ref{tab:delta_calib}: the semiparametric procedure is generally
better 
calibrated, whereas the conformal procedure is conservative under left-skewed
errors and liberal under right-skewed errors, with undercoverage in both cases.
Relative to the design with a Gaussian base error,
the heavy-tailed design shows mild semiparametric size inflation under
left-skewed errors and less severe conformal size distortion under
right-skewed errors at larger sample sizes.

Figure~\ref{fig:t4-delta-power} shows the same monotone increase in rejection
probability as the gap becomes positive and as $n$ increases. At the
boundary, Table~\ref{tab:t4-delta-calib} already records the size: the
conformal procedure is oversized under right-skewed errors and conservative
under left-skewed errors, while the semiparametric size stays near the
nominal level. Power is already substantial at a gap of $0.05$ for
$n=4000$ and $n=16000$, and is close to one at a gap of $0.20$. As in the
Gaussian design, the two error distributions are reflections of one another.

\medskip\noindent\textbf{Symmetric errors.}
Tables~\ref{tab:sym-t4-learned-cv-zeta} and
\ref{tab:sym-t4-delta-calib} report results for the symmetric-error
design with $\epsilon_0=G\sim t(4)/\sqrt{2}$.

\begin{table}[H]
\centering
\caption{Bias, SD, SE, and RMSE
for the estimators of $\zeta_\x$ and $\zeta_\w$ under symmetric
heavy-tailed errors. For each sample size,
Semi and Conf denote the semiparametric and conformal estimators,
respectively. All entries are reported in units of
$0.01$.}
\label{tab:sym-t4-learned-cv-zeta}
\setlength{\tabcolsep}{6pt}
\renewcommand{\arraystretch}{1.1}
\begin{tabular}{lrlrrrrrrrr}
\hline
& & & \multicolumn{4}{c}{$\zeta_\x$} & \multicolumn{4}{c}{$\zeta_\w$} \\
\cline{4-7}\cline{8-11}
Error & $n$ & Method & Bias & SD & SE & RMSE & Bias & SD & SE & RMSE \\
\hline
      & 1000  & Semi & -0.13 & 6.53 & 7.00 & 6.52 & 0.39 & 6.67 & 7.04 & 6.68 \\
\rowcolor[gray]{0.93}
\cellcolor[gray]{1} & & Conf & 0.47 & 6.32 & 6.93 & 6.33 & 1.36 & 6.49 & 6.95 & 6.62 \\[2pt]
      & 2000  & Semi & 0.11 & 4.66 & 4.85 & 4.66 & 0.28 & 4.74 & 4.88 & 4.75 \\
\rowcolor[gray]{0.93}
\cellcolor[gray]{1} & & Conf & 0.54 & 4.57 & 4.82 & 4.60 & 0.93 & 4.60 & 4.84 & 4.69 \\[2pt]
      & 4000  & Semi & 0.31 & 3.26 & 3.36 & 3.27 & 0.43 & 3.35 & 3.38 & 3.37 \\
\rowcolor[gray]{0.93}
\cellcolor[gray]{1} & & Conf & 0.54 & 3.21 & 3.35 & 3.26 & 0.79 & 3.31 & 3.36 & 3.40 \\[2pt]
      & 8000  & Semi & 0.06 & 2.32 & 2.35 & 2.32 & 0.17 & 2.35 & 2.36 & 2.35 \\
\rowcolor[gray]{0.93}
\cellcolor[gray]{1} & & Conf & 0.19 & 2.30 & 2.35 & 2.31 & 0.41 & 2.33 & 2.36 & 2.36 \\[2pt]
      & 16000 & Semi & 0.16 & 1.62 & 1.65 & 1.63 & 0.21 & 1.60 & 1.66 & 1.62 \\
\rowcolor[gray]{0.93}
\cellcolor[gray]{1}\multirow{-10}{*}[4pt]{Symm}
      & & Conf & 0.26 & 1.60 & 1.65 & 1.62 & 0.37 & 1.58 & 1.66 & 1.63 \\
\hline
\end{tabular}
\end{table}

Relative to Table~\ref{tab:learned-cv-zeta-sym}, RMSE is larger under
heavy-tailed errors, but the two methods continue to perform similarly
for both targets. The absolute biases of the conformal estimators of
$\zeta_\x$ and $\zeta_\w$ are much smaller than those under skewed errors in
Table~\ref{tab:t4-learned-cv-zeta}.

\begin{table}[H]
\centering
\caption{{Finite-sample calibration of inference on $\Delta$ at the
boundary null $\Delta(\beta)=\tau=0.20$ under symmetric heavy-tailed
errors. For each sample size,
Semi and Conf denote the semiparametric and conformal estimators,
respectively. Size is computed using the
standard-normal critical value.
All entries are reported in units of
$0.01$ except for SD($T$). 
The nominal test and confidence levels are $0.05$ and $0.95$,
respectively.}}
\label{tab:sym-t4-delta-calib}
\setlength{\tabcolsep}{6pt}
\renewcommand{\arraystretch}{1.1}
\begin{tabular}{lrlrrrrr}
\hline
Error & $n$ & Method & Size & Bias & Cvrg & CI length & SD($T$) \\
\hline
      & 1000  & Semi & 3.4 & -0.5 & 95.7 & 22.6 & 0.965 \\
\rowcolor[gray]{0.93}
\cellcolor[gray]{1} & & Conf & 2.4 & -0.8 & 96.4 & 22.2 & 0.929 \\[2pt]
      & 2000  & Semi & 4.5 & -0.2 & 95.3 & 15.5 & 0.989 \\
\rowcolor[gray]{0.93}
\cellcolor[gray]{1} & & Conf & 3.6 & -0.3 & 95.4 & 15.3 & 0.964 \\[2pt]
      & 4000  & Semi & 3.2 & -0.2 & 95.8 & 10.7 & 0.986 \\
\rowcolor[gray]{0.93}
\cellcolor[gray]{1} & & Conf & 2.5 & -0.3 & 96.0 & 10.6 & 0.953 \\[2pt]
      & 8000  & Semi & 4.2 & -0.1 & 95.6 & 7.5 & 0.987 \\
\rowcolor[gray]{0.93}
\cellcolor[gray]{1} & & Conf & 3.8 & -0.1 & 96.3 & 7.4 & 0.959 \\[2pt]
      & 16000 & Semi & 4.2 & -0.0 & 95.1 & 5.2 & 0.993 \\
\rowcolor[gray]{0.93}
\cellcolor[gray]{1}\multirow{-10}{*}[4pt]{Symmetric}
      & & Conf & 4.1 & -0.1 & 95.3 & 5.2 & 0.980 \\
\hline
\end{tabular}
\end{table}
The results in Table~\ref{tab:sym-t4-delta-calib} are consistent with
those under symmetric Gaussian errors in Table~\ref{tab:delta_calib-sym}.
Together, these results suggest that the size and coverage distortions
of the conformal procedure observed in the main text and in the
heavy-tailed experiments are associated with skewness rather than
heavy tails alone.

\subsection{Conditional estimation of $\zeta_\w(s_0)$}
\label{sec:cond-zeta-w}

\begin{figure}[H]
  \centering
  \includegraphics[width=0.98\textwidth]{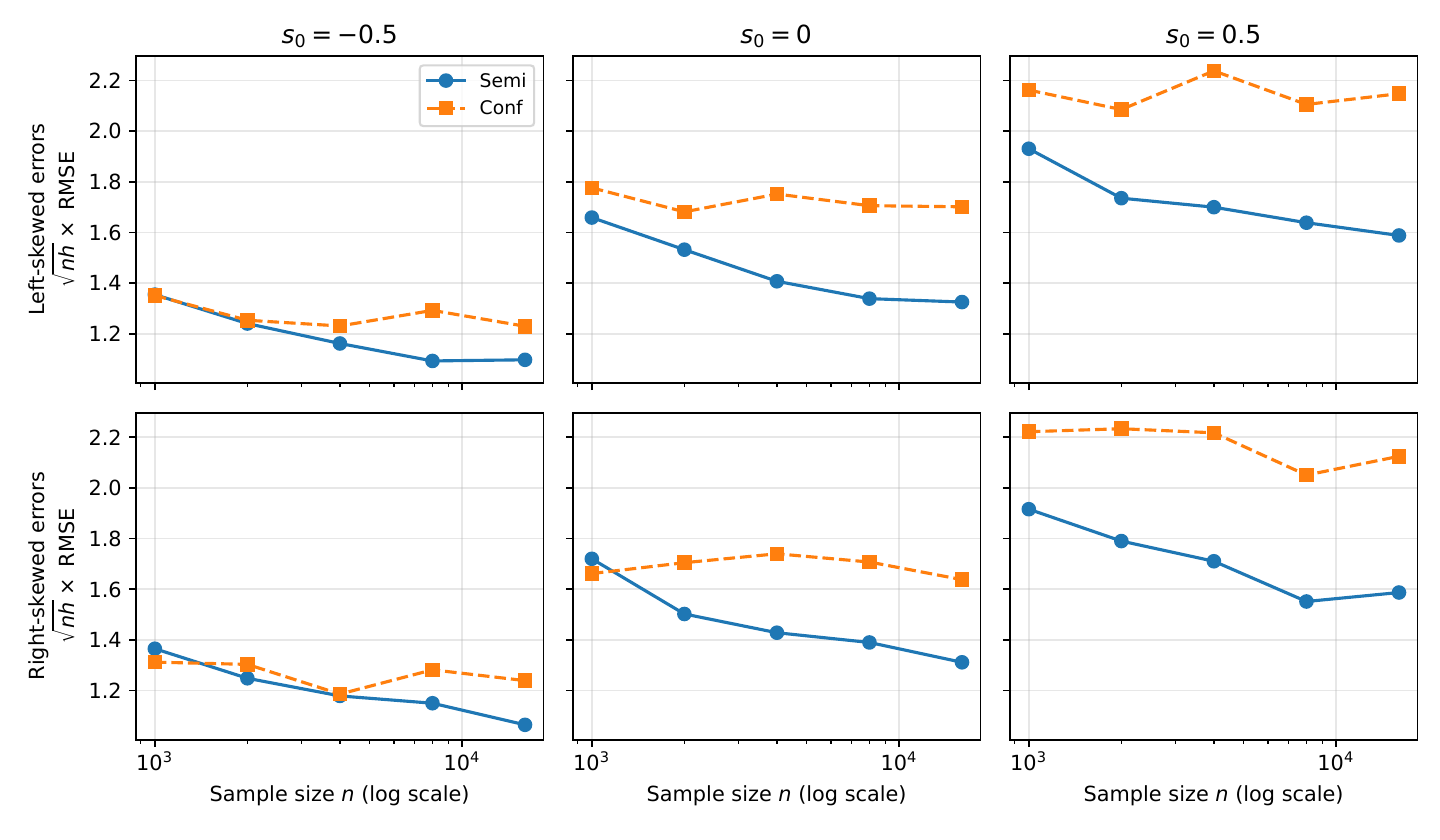}
  \caption{Monte Carlo $\sqrt{nh}$ times RMSE of the semiparametric and
  conformal estimators of $\zeta_\w(s_0)$. The horizontal
  axis is logarithmic.  Rows correspond
  to right- and left-skewed errors and columns to
  $s_0\in\{-0.5,0,0.5\}$.}
  \label{fig:cond-zeta-w-rmse}
\end{figure}

\subsection{Rejection probabilities across gaps}
\label{sec:cond-delta-power}

\begin{figure}[H]
  \centering
  \includegraphics[width=0.98\textwidth]{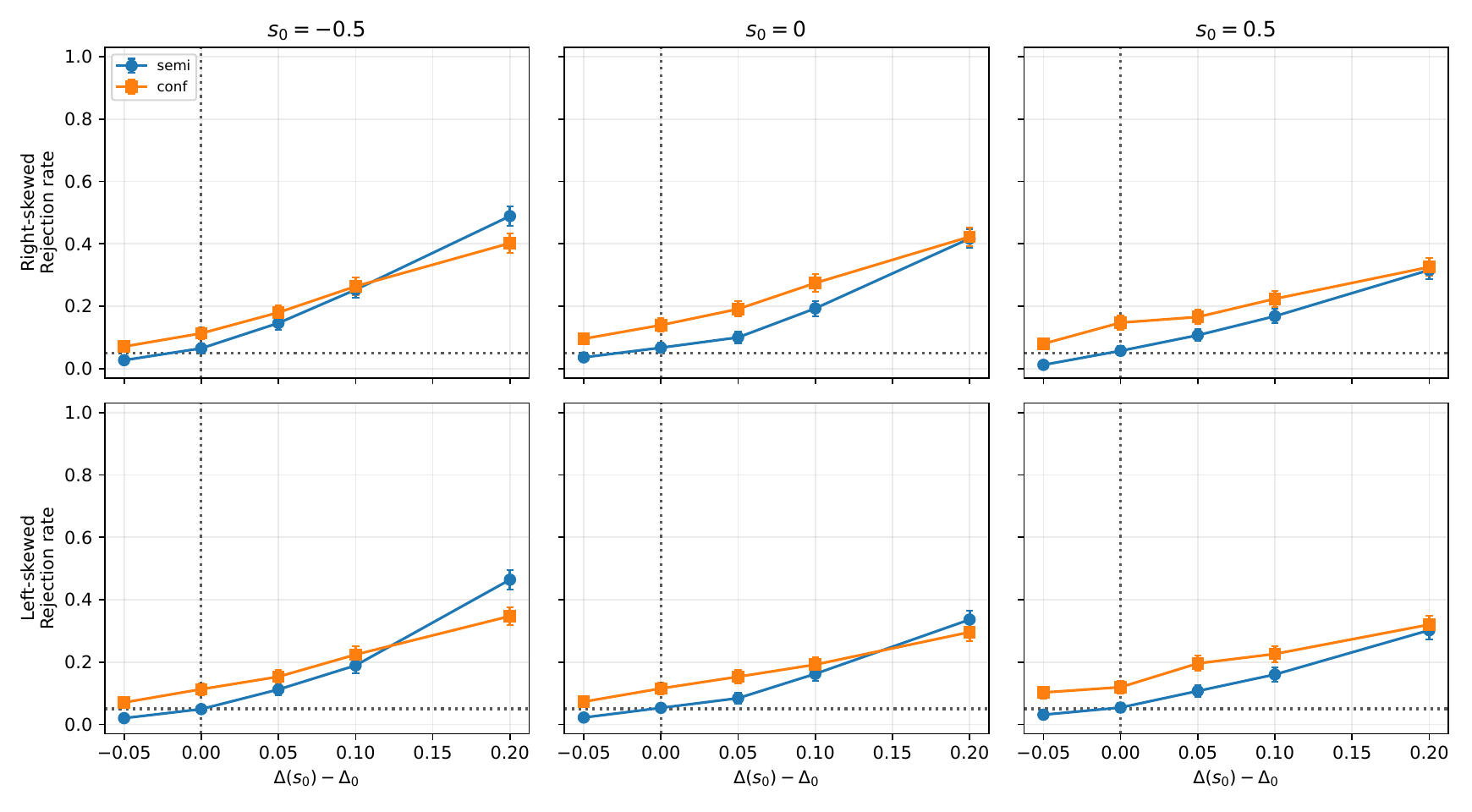}
  \caption{Rejection probabilities for testing
  $H_0:\Delta(s_0)\le\tau$ against $H_1:\Delta(s_0)>\tau$ with
  $\tau=0.20$, at $n=1000$. The layout matches
  Figure~\ref{fig:cond-delta-power}.}
  \label{fig:cond-delta-power-n1000}
\end{figure}

\begin{figure}[H]
  \centering
  \includegraphics[width=0.98\textwidth]{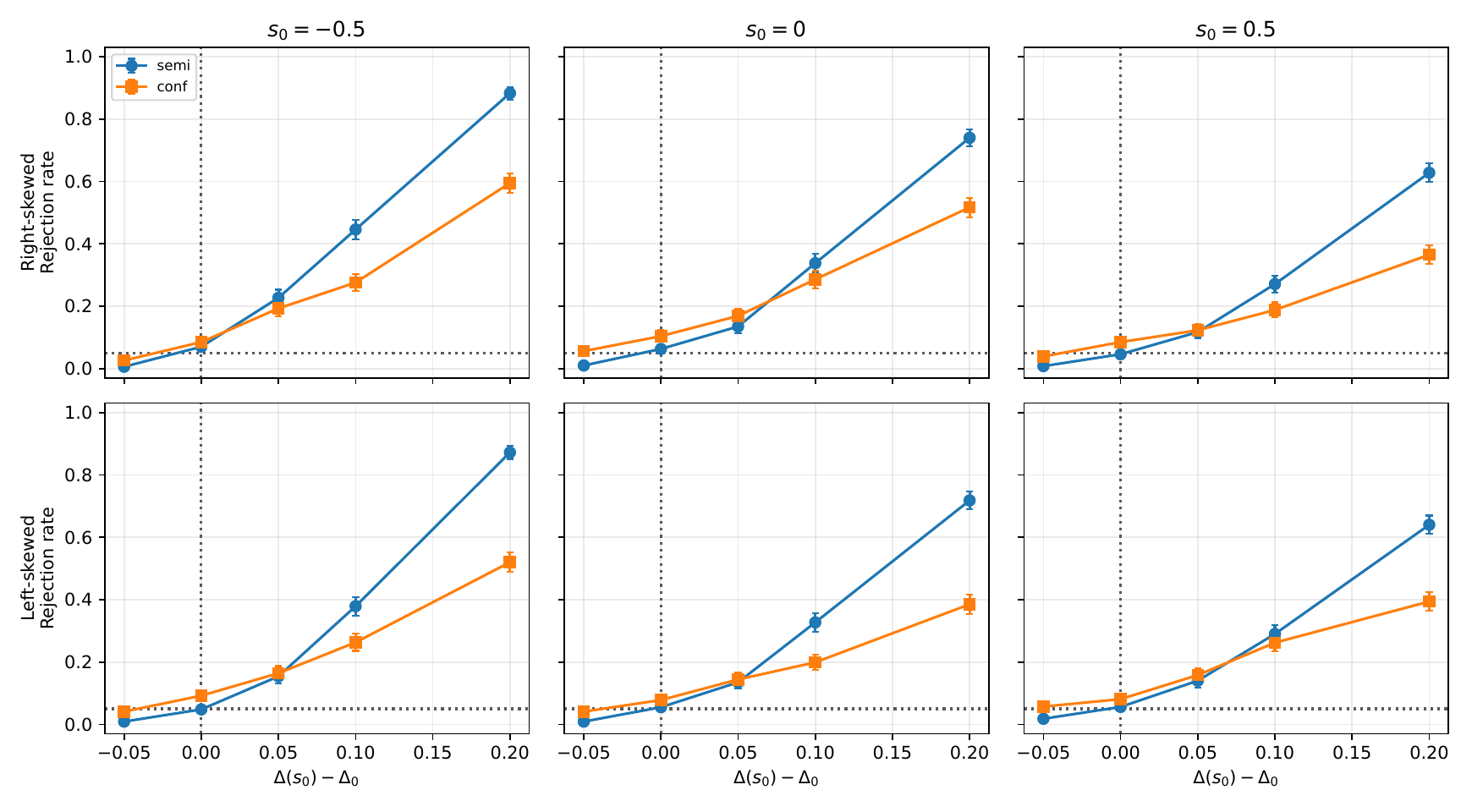}
  \caption{Rejection probabilities for testing
  $H_0:\Delta(s_0)\le\tau$ against $H_1:\Delta(s_0)>\tau$ with
  $\tau=0.20$, at $n=4000$. The layout matches
  Figure~\ref{fig:cond-delta-power}.}
  \label{fig:cond-delta-power-n4000}
\end{figure}

Figures~\ref{fig:cond-delta-power-n1000} and~\ref{fig:cond-delta-power-n4000}
repeat Figure~\ref{fig:cond-delta-power} at $n=1000$ and $n=4000$.

\subsection{Derivation of the efficient influence function for $\zeta_\x$
when $\s_0$ does not appear}\label{sec:phix}

In this case, the model is $f_\X(\x)f_{\epsilon_\x\mid
  \X}\{y-m_\x(\x),\x\}$, where $\int \epsilon_\x f_{\epsilon_\x\mid
  \X}(\epsilon_\x,\x)d\epsilon_\x=0$.
Let $f_{\epsilon_\x\mid \X}(\epsilon_\x,\x,\bg_1)$,
$f_\X(\x,\bg_2)$ and $m_\x(\x,\bg_3)$  be sub-models of 
$\eta_1(\epsilon_\x,\x)\equiv f_{\epsilon_\x\mid
  \X}(\epsilon_\x,\x)$, $\eta_2(\x)\equiv f_\X(\x)$ and
$m_\x(\x)$. 
Now the scores are
\bse
\S_{\bg_3}(\o,\bg_3,\eta_1,\eta_2)&=&-d\log f_{\epsilon_\x\mid
  \X}(\epsilon_\x,\x)/d\epsilon_\x \frac{\partial
m_\x(\x,\bg_3)}{\partial\bg_3},\\
\S_{\bg_1}(\o,m_\x,\bg_1,\eta_2)&=&\frac{\partial f_{\epsilon_\x\mid
  \X}\{y-m_\x(\x),\x,\bg_1\}/\partial\bg_1 }{f_{\epsilon_\x\mid
  \X}\{y-m_\x(\x),\x,\bg_1\}},\\
\S_{\bg_2}(\x,\bg_2)&=& \frac{\partial f_\X(\x,\bg_2)/\partial\bg_2}{f_\X(\x)}.
\ese
We can easily see that
the tangent space is
$\calT=\Lambda_m+(\Lambda_1\oplus\Lambda_2)
=\Lambda_1\oplus\Lambda_2\oplus\Lambda_3$,
where
\bse
\Lambda_m&=&\{a(\x) d\log f_{\epsilon_\x\mid
\X}(\epsilon_\x,\x)/d\epsilon_\x \},\\
\Lambda_1&=&\left[a(\epsilon_\x,\x): E\{a(\epsilon_\x,\x)\mid \x\}=0, 
E\{\epsilon_\x a(\epsilon_\x,\x)\mid \x\}=0
\right]\\
\Lambda_2&=&[a(\x): E\{a(\X)\}=0],\\
\Lambda_3&=&\{a(\x)\epsilon_\x\}.
\ese

Next, consider 
\bse
1-\alpha&=&\pr\{r(\O.m_\x)<\zeta(m_\x,\eta_1,\eta_2)\}\\
&=&
\int_{\o:r(\o,m_\x)<\zeta(m_\x,\eta_1,\eta_2)}\eta_2(\x) \eta_1\{y-m_\x(\x),\x\}d\mu(\o),
\ese
where we write $\zeta$ as $\zeta(m_\x,\eta_1,\eta_2)$ to emphasize that
$\zeta$ is a functional of $m_\x, \eta_1,\eta_2$.
Taking derivative with respect to $\bg_3$ leads to
\bse
\0&=&\int I\{r(\o,\bg_3)<\zeta(\bg_3,\eta_1,\eta_2)\}
\S_{\bg_3}(y,\x,\bg_3,\eta_1,\eta_2)f_\O(\o,\bg_3,\eta_1,\eta_2)d\o\\
&&+\int\delta\{\zeta(\bg_3,\eta_1,\eta_2)-r(\o,\bg_3)\}
\left\{\frac{\partial\zeta(\bg_3,\eta_1,\eta_2)}{\partial\bg_3}
-\frac{\partial r(\o,\bg_3)}{\partial\bg_3}\right\}
f_\O(\o,\bg_3,\eta_1,\eta_2)d\o\\
&=&E\left[I\{ r(\O,\bg_3)<\zeta(\bg_3,\eta_1,\eta_2)\}\S_{\bg_3}(\O,\bg_3,\eta_1,\eta_2)\right]\\
&&+\frac{\partial\zeta(\bg_3,\eta_1,\eta_2)}{\partial\bg_3}
\int\delta\{\zeta(\bg_3,\eta_1,\eta_2)-r(\o,\bg_3)\}
f_\O(\o,\bg_3,\eta_1,\eta_2)d\o\\
&&-\int\delta\{\zeta(\bg_3,\eta_1,\eta_2)-r(\o,\bg_3)\}
\frac{\partial r(\o,\bg_3)}{\partial\bg_3}
f_\O(\o,\bg_3,\eta_1,\eta_2)d\o,
\ese
hence
\bse
\frac{\partial\zeta(\bg_3,\eta_1,\eta_2)}{\partial\bg_3}
&=&\frac{\int\delta\{\zeta-r(\o,\bg_3)\}
\frac{\partial r(\o,\bg_3)}{\partial\bg_3}
f_\O(\o,\bg_3,\eta_1,\eta_2)d\o
}{\int\delta\{\zeta(\bg_3,\eta_1,\eta_2)-r(\o,\bg_3)\}
f_\O(\o,\bg_3,\eta_1,\eta_2)d\o}\\
&&-\frac{ E\left[I\{
    r(\O,\bg_3)<\zeta\}\S_{\bg_3}(\O,\bg_3,\eta_1,\eta_2)\right] 
}{\int\delta\{\zeta(\bg_3,\eta_1,\eta_2)-r(\o,\bg_3)\}
f_\O(\o,\bg_3,\eta_1,\eta_2)d\o}\\
&=&\frac{E[\delta\{\zeta-r(\O,\bg_3)\}
\frac{\partial r(\O,\bg_3)}{\partial\bg_3}]
-E[I\{
    r(\O,\bg_3)<\zeta\}\S_{\bg_3}(\O,\bg_3,\eta_1,\eta_2)]
}{E[\delta\{\zeta-r(\O,\bg_3)\}]}.
\ese
Similarly, replacing $\eta_1$ with its parametric submodel,
and taking derivative with respect to $\bg_1$, we have
\bse
\0&=&\int I\{r(\o,m_\x)<\zeta(m_\x,\bg_1,\eta_2)\}
\S_{\bg_1}(\o,m_\x,\bg_1,\eta_2)f_\O(\o,m_\x,\bg_1,\eta_2)d\o\\
&&+\int\delta\{\zeta(m_\x,\bg_1,\eta_2)-r(\o,m_\x)\}
\frac{\partial\zeta(m_\x,\bg_1,\eta_2)}{\partial\bg_1}
f_\O(\o,m_\x,\bg_1,\eta_2)d\o\\
&=&E\left[I\{ r(\O,m_\x)<\zeta\}\S_{\bg_1}(\O,m_\x,\bg_1,\eta_2)\right]
+\frac{\partial\zeta(m_\x,\bg_1,\eta_2)}{\partial\bg_1}
\int\delta\{\zeta-r(\o,m_\x)\}
f_\O(\o,m_\x,\bg_1,\eta_2)d\o,
\ese
hence
\bse
\frac{\partial\zeta(m_\x,\bg_1,\eta_2)}{\partial\bg_1}
=-
\frac{
E[I\{ r(\O,m_\x)<\zeta\}\S_{\bg_1}(\O, m_\x,\bg_1,\eta_2)]
}{E[\delta\{\zeta-r(\O,m_\x)\}]}.
\ese
Finally, replacing $\eta_2$ with its parametric submodel, and 
taking derivative with respect to $\bg_2$, we get
\bse
\0&=&\int I\{r(\o,m_\x)<\zeta(m_\x,\eta_1,\bg_2)\}
\S_{\bg_2}(\x,\bg_2)f_\O(\o,m_\x,\eta_1,\bg_2)d\o\\
&&+\int\delta\{\zeta(m_\x,\eta_1,\bg_2)-r(\o,m_\x)\}
\frac{\partial\zeta(m_\x,\eta_1,\bg_2)}{\partial\bg_2}
f_\O(\o,m_\x,\eta_1,\bg_2)d\o\\
&=&E\left[I\{ r(\O,m_\x)<\zeta\}\S_{\bg_2}(\X,\bg_2)\right]
+\frac{\partial\zeta(m_\x,\eta_1,\bg_2)}{\partial\bg_2}
\int\delta\{\zeta-r(\o,m_\x)\}
f_\O(\o,m_\x,\eta_1,\bg_2)d\o,
\ese
hence
\bse
\frac{\partial\zeta(m_\x,\eta_1,\bg_2)}{\partial\bg_2}
=\frac{-E[I\{ r(\O,m_\x)<\zeta\}\S_{\bg_2}(\X,\bg_2)]
}{E[\delta\{\zeta-r(\O,m_\x)\}]}.
\ese

Let $\phi_{\x,\rm{eff}}(\o,m_\x,\zeta)$ be the efficient influence function for estimating
$\zeta$. Then it has the form
$\phi_{\x,\rm{eff}}(\o,m_\x,\zeta)=a_1(\o)+a_2(\x)+\epsilon_\x a_3(\x)$, where 
$E\{\epsilon_\x a_1(\O)\mid \x\}=E\{a_1(\O)\mid \x\}=0$ and $E\{a_2(\X)\}=0$.
In addition, at the true
parameter values, 
$\phi_\x(\o,m_\x,\zeta)$ satisfies
\bse
\frac{\partial\zeta(\bg_3,\eta_1,\eta_2)}{\partial\bg_3}&=&E\{\phi_{\x,\rm{eff}}(\O,m_\x,\zeta)\S_{\bg_3}(\O)\},\\
\frac{\partial\zeta(m_\x,\bg_1,\eta_2)}{\partial\bg_1}&=&E\{\phi_{\x,\rm{eff}}(\O,m_\x,\zeta)\S_{\bg_1}(\O)\},\\
\frac{\partial\zeta(m_\x,\eta_1,\bg_2)}{\partial\bg_2}&=&E\{\phi_{\x,\rm{eff}}(\O,m_\x,\zeta)\S_{\bg_2}(\X)\}.
\ese

This implies that $a_1(\o,\bg_3,\zeta),a_2(\x), a_3(\x)$ must satisfy 
\bse
&&\frac{E[\delta\{\zeta-r(\O,\bg_3)\}
{\partial r(\O,\bg_3)}/{\partial\bg_3}]
-E[I\{
    r(\O,\bg_3)<\zeta\}\S_{\bg_3}(\O,\bg_3,\eta_1,\eta_2)]
}{E[\delta\{\zeta-r(\O,\bg_3)\}]}\\
&=&E\left\{\{a_1(\O,\bg_3,\zeta)+a_2(\X)+\epsilon_\x a_3(\X)\}
  [
-d\log f_{\epsilon_\x\mid
  \X}(\epsilon_\x,\x)/d\epsilon_\x \frac{\partial
m_\x(\x,\bg_3)}{\partial\bg_3}]\right\}\\
&=&E\left\{\{a_1(\O,\bg_3,\zeta)+\epsilon_\x a_3(\X)\}
  [
-d\log f_{\epsilon_\x\mid
  \X}(\epsilon_\x,\x)/d\epsilon_\x \frac{\partial
m_\x(\x,\bg_3)}{\partial\bg_3}]\right\}\\
&=&-E\left\{a_1(\O,\bg_3,\zeta)
  d\log f_{\epsilon_\x\mid
  \X}(\epsilon_\x,\x)/d\epsilon_\x \frac{\partial
m_\x(\x,\bg_3)}{\partial\bg_3}\right\}+E\left\{ a_3(\X) \frac{\partial
m_\x(\x,\bg_3)}{\partial\bg_3}
\right\},\\
&&
\frac{
-E[I\{ r(\O,m_\x)<\zeta\}\S_{\bg_1}(\O, m_\x,\bg_1,\eta_2)]
}{E[\delta\{\zeta-r(\O,m_\x)\}]}\\
&=&E\left[\{a_1(\O,m_\x,\zeta)+a_2(\X)+\epsilon_\x a_3(\X)\}
\S_{\bg_1}(\O)\right]\\
&=&E\{a_1(\O,m_\x,\zeta)\S_{\bg_1}(\O)\},\\
&&\frac{
-E[I\{ r(\O,m_\x)<\zeta\}\S_{\bg_2}(\X,\bg_2)]
}{E[\delta\{\zeta-r(\O,m_\x)\}]}\\
&=&E\left[\{a_1(\O,m_\x,\zeta)+a_2(\X)+\epsilon_\x a_3(\X)\}
\S_{\bg_2}(\X)\right]\\
&=&E\left\{a_2(\X)
\S_{\bg_2}(\X)\right\}.
\ese
Because $\S_{\bg_2}(\X)$ can  be an arbitrary mean zero
  function of $\X$, $\S_{\bg_1}(\O)$ can be any function that
  satisfies $E\{\S_{\bg_1}(\O)\mid \X\}=
  E\{\epsilon_\x\S_{\bg_1}(\O)\mid \X\}=\0$ and $\partial m_\x(\X,\bg_3)/\partial\bg_3$ can
  be any function of $\X$, 
these requirements directly lead to
\bse
a_2(\X)&=&\frac{(1-\alpha)
-E[I\{ r(\O,m_\x)<\zeta\}\mid \X]
}{E[\delta\{\zeta-r(\O)\}]},\n\\
a_1(\O)&=&\frac{
E[I\{
  r(\O,m_\x)<\zeta\}\mid \X]-
I\{
  r(\O,m_\x)<\zeta\}}{E[\delta\{\zeta-r(\O,m_\x)\}]}
+\frac{\epsilon_\x E[\epsilon_\x I\{
  r(\O,m_\x)<\zeta\}\mid \X ]}{ E[\delta\{\zeta-r(\O,m_\x)\}]
  E(\epsilon_\x^2\mid \X)},\n\\
a_3(\X)
&=&-\frac{E[\epsilon_\x I\{
  r(\O,m_\x)<\zeta\}\mid \X ]}{ E[\delta\{\zeta-r(\O,m_\x)\}]
  E(\epsilon_\x^2\mid \X)}-\frac{ E[\mathrm{sign}(\epsilon_\x)\delta\{\zeta-r(\O,m_\x)\}\mid
\X]}{E[\delta\{\zeta-r(\O,m_\x)\}]}.
\ese
Here $r(\O,m_\x)=|\epsilon_\x|$ and $\partial|\epsilon_\x|/\partial\epsilon_\x=\mathrm{sign}(\epsilon_\x)$.
In summary, our efficient influence function for estimating $\zeta$ is
\bse
\phi_{\x,\rm{eff}}(\O,\zeta, m_\x,f_{\epsilon_\x\mid\x})=\frac{(1-\alpha)-I\{
  r(\O,m_\x)<\zeta\}-\epsilon_\x E[\mathrm{sign}(\epsilon_\x)\delta\{\zeta-r(\O,m_\x)\}\mid
\X]
}{E[\delta\{\zeta-r(\O,m_\x)\}]},
\ese
 where the dependence on $f_{\epsilon_\x\mid\x}$ is via $E[\sign(\epsilon_\x)\delta\{\zeta-r(\O,m_\x)\}\mid
\X]$, which is equals
$f_{\epsilon_\x\mid\x}(\zeta,\x)-f_{\epsilon_\x\mid\x}(-\zeta,\x)$.

\subsection{Supporting Lemmas}
  \begin{Lem}
    \label{lemm:conditional}
    (Conditional convergence implies unconditional convergence)
    Let $\{\X_m\}$ and $\{\Y_m\}$ be sequences of random vectors.
      If, for $\epsilon_m\to0$,
      $\pr(\|\X_m\|>\epsilon_m\mid \Y_m)\xrightarrow{p}0$,
      then $\pr(\|\X_m\|>\epsilon_m)\to0$.
      In particular, $\pr(\|\X_m\|>\epsilon_m)\to0$
        if
      $E(\|\X_m\|^q/\epsilon_m^q\mid \Y_m)\xrightarrow{p}0$ for some
      $q\ge1$.
  \end{Lem}
  Proof: 
  For any $\epsilon>0$,
  \[
  \pr(\|\X_m\|>\epsilon_m)=
  E\{\pr(\|\X_m\|>\epsilon_m\mid \Y_m)\}\to0,
  \]
  as the sequence $\{\pr(\|\X_m\|>\epsilon_m\mid \Y_m)\}$ is uniformly integrable.
  For the second claim, note that
  \[
  \pr(\|\X_m\|>\epsilon_m\mid \Y_m)
  \le E(\|\X_m\|^q/\epsilon_m^q\mid \Y_m)\wedge1
  \xrightarrow{p}0
  \]
  by Markov's inequality.
  \qed
\begin{Lem}\label{lemm:semi0-onefold}
Let
 $D_1\equiv\{\O_i:1\le i\le n_1\}$ and
 $D_2\equiv\{\O_i:n_1+1\le i\le n\}$. For $\v\in\{\x,\w\}$, under
 Assumptions~\ref{ass:eps0}--\ref{ass:lip0}, assume
$\wh m_\v(\cdot), \wh f_{\epsilon_\v\mid\v}$ are estimated using
  $D_2$ and
 $\sup_\v|\wh m_\v(\v)-m_\v(\v)|=o_p(n^{-1/4})$
       and  
      $\sup_{\zeta}\sup_\v \left|\wh f_{\epsilon_\v\mid\v}(\zeta,\v)-f_{\epsilon_\v\mid\v}(\zeta,\v)\right|=o_p(n^{-1/4})$. 
  Then $\wh\zeta_\v$ satisfies
  \be
  \label{equ:onefold-linear0}
  n_1^{1/2}(\wh\zeta_\v-\zeta_\v)=-n_1^{-1/2}g_{\v}^{-1}\sumia \phi_\v(\O_i,\zeta_\v,m_\v,f_{\epsilon_\v\mid\v})+o_p(1).
  \ee
  Further it
  satisfies 
  \bse
  n_1^{1/2}(\wh\zeta_\v-\zeta_\v)
  \to N\big(0, \sigma_\v\big)
  \ese
  in distribution
  as $n\to\infty$.
  Here
  $\sigma_\v=
  g_{\v}^{-2}
  E\{\phi_\v^2(\O_i, \zeta_\v, m_\v,f_{\epsilon_\v\mid\v})\}$, and
  $g_{\v}=f_{\epsilon_\v}(\zeta_\v)+f_{\epsilon_\v}(-\zeta_\v)$.
  \end{Lem}

\begin{Lem}\label{lemm:conformal0-onefold}
Let $D_1\equiv\{\O_i:1\le i\le n_1\}$ and
 $D_2\equiv\{\O_i:n_1+1\le i\le n\}$.  For
 $\v\in\{\x,\w\}$, under Assumption \ref{ass:lip20}, let
 $a_n\equiv\sup_\v|\wh m_\v-m_\v|=o_p(1)$, where $\wh m_\v$ is
 estimated using $D_2$.
 Then $\wt \zeta_\v$
 satisfies
 \be
 \label{equ:onefold-theo2-linear0}
 n_1^{1/2}(\wt\zeta_\x-\zeta_\x)
 =
 -g_\x^{-1}n^{1/2}B(a_n)+
 g_\x^{-1}n_1^{-1/2}\sumia
 \psi_{1-\alpha}(|\epsilon_{\x,i}|-\zeta_\x)+O_p(a_n^2 n^{1/2})+o_p(1).
 \ee
 where $g_{\v}=f_{\epsilon_\v}(\zeta_\v)+f_{\epsilon_\v}(-\zeta_\v)$, $B(a_n)= E
 [\{f_{\epsilon_\x\mid\x}(\zeta_\x,\X)-f_{\epsilon_\x\mid\x}(-\zeta_\x,\X)\}\{\wh
 m_\x(\X)-m_\x(\X)\}]=O_p(a_n)$, and $\psi_{1-\alpha}(t)=1-\alpha-I(t<0)$.
\end{Lem}

\begin{Lem}\label{lemm:semi-onefold}
 Let $D_1\equiv\{\O_i:1\le i\le n_1\}$ and
$D_2\equiv\{\O_i:n_1+1\le i\le n\}$.  For
$\v\in\{\x,\w\}$, under
Assumptions~\ref{ass:eps}--\ref{ass:kernel}, assume 
$nh^{d_s}\to\infty$, $nh^{4t+d_s}\to0$, $\sup_\v|\wh
m_\v(\v)-m_\v(\v)|=o_p\{(nh^{d_s})^{-1/4}\}$ and
$\sup_{\zeta}\sup_\v \left|\wh
  f_{\epsilon_\v\mid\v}(\zeta,\v)-f_{\epsilon_\v\mid\v}(\zeta,\v)\right|=o_p\{(nh^{d_s})^{-1/4}\}$,
where $\wh m_\v$ and $\wh f_{\epsilon_\v\mid\v}$ are estimated
using $D_2$.
Then $\wh\zeta_\v(\s_0)$ satisfies
\be
\label{equ:onefold-theo1-linear}
&&(n_1h^{d_s})^{1/2}\{\wh\zeta_\v(\s_0)-\zeta_\v(\s_0)-b_\v^{\rm semi}\}\notag\\
&=&-\frac{(n_1h^{-d_s})^{-1/2}\sumia\bigl\{\phi_\v(\O_i,\zeta_\v(\s_0),m_\v,f_{\epsilon_\v\mid\v})K_h(\S_i-\s_0)-E(\phi_\v K_h)\bigr\}}{g_{\v,0}f_\S(\s_0)}+o_p(1),
\ee
where $E(\phi_\v K_h)=E\{\phi_\v(\O_i,\zeta_\v(\s_0),m_\v,f_{\epsilon_\v\mid\v})K_h(\S_i-\s_0)\}$ and
\bse
b_{\v}^{\rm semi}\equiv h^{t}
\left\{g_{\v,0}f_\S(\s_0)\right\}^{-1}
\frac{\partial^t E[\phi_\v\{\O_i,\zeta_\v(\s),f_{\epsilon_\v\mid\v}\}\mid\s_0]f_{\S}(\s_0)}{\partial\s_0^t}\int\u^tK(\u)d\u+o(h^t).
\ese
Further it
satisfies
\bse
(n_1h^{d_s})^{1/2}\{\wh\zeta_\v(\s_0)-\zeta_\v(\s_0)-b_{\v}^{\rm semi}\}
\to N\big(0,\frac{\int K^2(\u)d\u}{f_\S(\s_0)}\sigma_\v\big)
\ese
in distribution
as $n\to\infty$,
where
\bse
\sigma_\v=g_{\v,0}^{-2}
E[\phi_\v^2\{\O_i,\zeta_\v(\s_0),f_{\epsilon_\v\mid\v}\}\mid\s_0],
\ese
where $\phi_\v=(1-\alpha)-I\{
r(\O,m_\v)<\zeta\}-\epsilon_\v E[\sign(\epsilon_\v)\delta\{\zeta-r(\O,m_\v)\}\mid\V]$ represents the numerator of $\phi_\v$ and
$g_{\v,0}=f_{\epsilon_\v\mid\s}\{\zeta_\v(\s_0),\s_0\}+f_{\epsilon_\v\mid\s}\{-\zeta_\v(\s_0),\s_0\}$.
\end{Lem}

\begin{Lem}\label{lemm:conformal-onefold}
Let $D_1\equiv\{\O_i:1\le i\le n_1\}$ and
 $D_2\equiv\{\O_i:n_1+1\le i\le n\}$.  For
 $\v\in\{\x,\w\}$, under Assumptions~\ref{ass:kernel}
 and~\ref{ass:lip2}, assume $nh^{d_s}\to\infty$ and $\sup_\v|\wh
 m_\v-m_\v|=o_p\{(nh^{d_s})^{-1/2}\}$, where $\wh m_\v$ is estimated
 using $D_2$. 
 Then $\wt\zeta_\v(\s_0)$
 satisfies
 \be
 \label{equ:onefold-theo2-linear}
&&(n_1h^{d_s})^{1/2}\{\wt\zeta_\v(\s_0)-\zeta_\v(\s_0)-b_\v^{\rm conf}\}\notag\\
&=&\frac{(n_1h^{-d_s})^{-1/2}\sumia\bigl[\psi_{1-\alpha}\{|\epsilon_{\v,i}|-\zeta_\v(\s_0)\}K_h(\S_i-\s_0)-E(\psi_{\v,1-\alpha}K_h)\bigr]}{g_{\v,0}f_\S(\s_0)}+o_p(1),
\ee
where $E(\psi_{\v,1-\alpha}K_h)=E\bigl[\psi_{1-\alpha}\{|\epsilon_{\v,i}|-\zeta_\v(\s_0)\}K_h(\S_i-\s_0)\bigr]$, $\epsilon_{\v,i}=Y_i-m_\v(\V_i)$ and
\bse
b_{\v}^{\rm conf}=\{g_{\v,0}f_\S(\s_0)\}^{-1}h^t\frac{\partial^t\bigl[E\{\psi_{1-\alpha}(|\epsilon_\v|-\zeta_\v(\s_0))\mid\s\}f_\S(\s)\bigr]}{\partial\s^t}\bigg|_{\s=\s_0}\int\u^tK(\u)d\u+o(h^t).
\ese
Further, it satisfies, as $n\to\infty$,
\bse
(n_1h^{d_s})^{1/2}\{\wt\zeta_\v(\s_0)-\zeta_\v(\s_0)-b_\v^{\rm conf}\}
\xrightarrow{d}N\left(0,\frac{\int K^2(\u)d\u}{f_\S(\s_0)}g_{\v,0}^{-2}\alpha(1-\alpha)\right),
\ese
where $\psi_{1-\alpha}(t)=1-\alpha-I(t<0)$ and
$g_{\v,0}=f_{\epsilon_\v\mid\s}\{\zeta_\v(\s_0),\s_0\}+f_{\epsilon_\v\mid\s}\{-\zeta_\v(\s_0),\s_0\}$.
\end{Lem}

  \begin{Lem} (Bound the terms  $R_1(\s_0), R_2(\s_0)$) \label{lemm:1} 
  Under Assumptions~\ref{ass:eps}--\ref{ass:kernel}, assume $\wh
  \zeta_\x(\s_0)-\zeta_\x(\s_0)=o_p(1)$. Let
  $\phi_\x(\O, \zeta, m_\x,f_{\epsilon_\x\mid\x})\equiv(1-\alpha)-I\{
    r(\O,m_\x)<\zeta\}-\epsilon_\x E[\delta\{\zeta-r(\O,m_\x)\}\mid
  \X]$, then
    $R_1(\s_0)=o_p(1), R_2(\s_0)=o_p(1)$, where
   $R_1(\s_0), R_2(\s_0)$ are defined in \eqref{eq:R1}, \eqref{eq:R2} respectively.
  \end{Lem}
  Proof:
  Let
  $D_1\equiv\{\O_i:1\le i\le n_1\}$ and
  $D_2\equiv\{\O_i:n_1+1\le i\le n\}$. By construction,
  $\wh m_\x$ and $\wh f_{\epsilon_\x\mid\x}$ are estimated from $D_2$ and they are non-stochastic conditional on $D_2$. Note that
  \bse
  &&(n_1h^{d_s})^{-1/2}E\{R_1(\s_0)\mid D_2\}\\&=& E\left\{\left(\phi_\x\{\O_i, \wh \zeta_\x(\s_0),\wh m_\x,\wh
  f_{\epsilon_\x\mid\x}\}-E[\phi_\x\{\O_i, \wh \zeta_\x(\s_0),\wh m_\x,\wh
  f_{\epsilon_\x\mid\x}\}\mid\S_i]\right)K_h(\S_i-\s_0)\mid D_2\right\}\\
  &&-E\left\{\left(\phi_\x\{\O_i, \zeta_\x(\s_0), \wh m_\x,\wh
  f_{\epsilon_\x\mid\x}\}-E[\phi_\x\{\O_i, \zeta_\x(\s_0), \wh m_\x,\wh
  f_{\epsilon_\x\mid\x}\}\mid\S_i]\right)K_h(\S_i-\s_0)\mid D_2\right\}.
  \ese
  By 
  \bse
  &&E\left\{\left(\phi_\x\{\O_i, \wh \zeta_\x(\s_0),\wh m_\x,\wh
  f_{\epsilon_\x\mid\x}\}-E[\phi_\x\{\O_i, \wh \zeta_\x(\s_0),\wh m_\x,\wh
  f_{\epsilon_\x\mid\x}\}\mid\S_i]\right)\mid \S_i\right\}=0\\
  &&E\left\{\left(\phi_\x\{\O_i, \zeta_\x(\s_0),\wh m_\x,\wh
  f_{\epsilon_\x\mid\x}\}-E[\phi_\x\{\O_i, \zeta_\x(\s_0),\wh m_\x,\wh
  f_{\epsilon_\x\mid\x}\}\mid\S_i]\right)\mid \S_i\right\}=0,
  \ese
  then $E\{R_1(\s_0)\mid D_2\}=0$. 
  Similarly, $E\{R_2(\s_0)\mid D_2\}=0$. It follows that $E\{R_1(\s_0)\}=E\{R_2(\s_0)\}=0$.
  Thus, to prove the results, 
  we only need to analyze the second moments of the summands in
  $R_1(\s_0), R_2(\s_0)$. 
  
  Note that
  \be
  \label{eq:lemm2-1}
  &&E\left\{\left([\phi_\x\{\O_i, \wh \zeta_\x(\s_0),\wh m_\x,\wh
  f_{\epsilon_\x\mid\x}\}- \phi_\x\{\O_i, \zeta_\x(\s_0), \wh m_\x,\wh
  f_{\epsilon_\x\mid\x}\}]K_h(\S_i-\s_0)\right)^2\mid D_2\right\}\notag\\
  &=& E[\{(\wh\epsilon_\x \wh E[\sign(\epsilon_\x)\delta\{\zeta_\x(\s_0)-r(\O_i,\wh m_\x)\}\mid\X_i]-\wh\epsilon_\x \wh E[\sign(\epsilon_\x)\delta\{\wh\zeta_\x(\s_0)-r(\O_i,\wh m_\x)\}\mid\X_i]\notag\\
  &&+I\{r(\O_i,\wh m_\x)<\zeta_\x(\s_0)\}-I\{r(\O_i, \wh m_\x)<\wh\zeta_\x(\s_0)\}) K_h(\S_i-\s_0)\}^2\mid D_2]\notag\\
  &\le& 2E[\{(\wh\epsilon_\x \wh E[\sign(\epsilon_\x)\delta\{\wh\zeta_\x(\s_0)-r(\O_i,\wh m_\x)\}\mid\X_i]\\
  &&-\wh\epsilon_\x \wh E[\sign(\epsilon_\x)\delta\{\zeta_\x(\s_0)-r(\O_i,\wh m_\x)\}\mid\X_i])K_h(\S_i-\s_0)\}^2\mid D_2]\notag\\
  && +2E\{\left(\left[I\{r(\O_i,\wh m_\x)<\wh
      \zeta_\x(\s_0)\}-I\{r(\O_i, \wh
      m_\x)<\zeta_\x(\s_0)\}\right]K_h(\S_i-\s_0)\right)^2\mid D_2\}\notag\\
  &=&
  O_p\{|\wh\zeta_\x(\s_0)-\zeta_\x(\s_0)|h^{-d_s}\}+O_p\{|\wh\zeta_\x(\s_0)-\zeta_\x(\s_0)|^2
  h^{-d_s}\}+o_p(n^{-1/2}h^{-d_s})\notag\\
  &=&O_p\{|\wh\zeta_\x(\s_0)-\zeta_\x(\s_0)|h^{-d_s}\}+o_p(n^{-1/2}h^{-d_s}).
  \ee
  We explain the last equality in detail. For $a<b$,
  $\left\{I(t<a)-I(t<b)\right\}^2
  = I\left(a\le t < b\right).$
  Then 
  \bse
  &&E\left(\left[I\{r(\O_i,\wh m_\x)<\wh\zeta_\x(\s_0)\}-I\{r(\O_i, \wh m_\x)<\zeta_\x(\s_0)\}\right]^2\mid \X_i,D_2\right)\\
  &=& P\left[\{\wh\zeta_\x(\s_0)\wedge\zeta_\x(\s_0)\} \le r(\O_i,\wh m_\x) < \{\wh\zeta_\x(\s_0)\vee\zeta_\x(\s_0)\}\mid \X_i, D_2\right]\\
  &=& f_{\epsilon_\x\mid\x}(\xi)|\wh\zeta_\x(\s_0)-\zeta_\x(\s_0)|
  \ese
  for some $\xi$ between $\wh\zeta_\x(\s_0)$ and $\zeta_\x(\s_0)$.
  Since $|\wh\zeta_\x(\s_0)-\zeta_\x(\s_0)|=o_p(1)$, for sufficiently large $n$,
  $\xi$ lies in the $c_2$-neighborhood of $\zeta_\x(\s_0)$ with probability
  approaching 1. By the Lipschitz continuity in Assumption~\ref{ass:lip},
  $\sup_\x f_{\epsilon_\x\mid\x}(\xi)\le C_2+Lc_2<\infty$, so
  \bse
  &&E\left\{\left(\left[I\{r(\O_i,\wh m_\x)<\wh\zeta_\x(\s_0)\}-I\{r(\O_i, \wh m_\x)<\zeta_\x(\s_0)\}\right]K_h(\S_i-\s_0)\right)^2\mid D_2\right\}\\
  &\le& (C_2+Lc_2)|\wh\zeta_\x(\s_0)-\zeta_\x(\s_0)| E\left\{K_h(\S_i-\s_0)^2\mid D_2\right\}
  = O_p\{|\wh\zeta_\x(\s_0)-\zeta_\x(\s_0)|h^{-d_s}\}.
  \ese
  For the second term, by $\sup_\x|\wh
  m_\x-m_\x|=o_p\{(nh^{d_s})^{-1/4}\}$, $\sup_\x|\wh
  f_{\epsilon_\x\mid\x}-f_{\epsilon_\x\mid\x}|=o_p\{(nh^{d_s})^{-1/4}\}$
  and the Lipschitz continuity of $f_{\epsilon_\x\mid\x}$ in
  Assumption~\ref{ass:lip}, we have
  \bse
  &&\left|\wh E[\sign(\epsilon_\x)\delta\{\wh\zeta_\x(\s_0)-r(\O_i,\wh m_\x)\}\mid\X_i, D_2]-\wh
    E[\sign(\epsilon_\x)\delta\{\zeta_\x(\s_0)-r(\O_i,\wh m_\x)\}\mid\X_i,  D_2]\right|\\
  &\le&
  \left|\wh E[\sign(\epsilon_\x)\delta\{\wh\zeta_\x(\s_0)-r(\O_i,\wh m_\x)\}\mid\X_i, D_2]-E[\sign(\epsilon_\x)\delta\{\wh\zeta_\x(\s_0)-r(\O_i, m_\x)\}\mid\X_i, D_2]\right|\\
  &+&\left| E[\sign(\epsilon_\x)\delta\{\wh\zeta_\x(\s_0)-r(\O_i, m_\x)\}\mid\X_i, D_2]-
  E[\sign(\epsilon_\x)\delta\{\zeta_\x(\s_0)-r(\O_i, m_\x)\}\mid\X_i,  D_2]\right|
  \\
  &+&\left|E[\sign(\epsilon_\x)\delta\{\zeta_\x(\s_0)-r(\O_i, m_\x)\}\mid\X_i,  D_2]-
    \wh E[\sign(\epsilon_\x)\delta\{\zeta_\x(\s_0)-r(\O_i,\wh m_\x)\}\mid\X_i,  D_2]\right|\\
  &\le& L|\wh\zeta_\x(\s_0)-\zeta_\x(\s_0)|+o_p\{(nh^{d_s})^{-1/4}\}.
  \ese
  Therefore,
  \bse
  &&E[\{(\wh\epsilon_\x [\wh E[\sign(\epsilon_\x)\delta\{\wh\zeta_\x(\s_0)-r(\O,\wh m_\x)\}\mid\X]\\
  &&-\wh E[\sign(\epsilon_\x)\delta\{\zeta_\x(\s_0)-r(\O,\wh m_\x)\}\mid\X]])K_h(\S_i-\s_0)\}^2\mid D_2]\\
  &\le& \{L|\wh\zeta_\x(\s_0)-\zeta_\x(\s_0)|+o_p\{(nh^{d_s})^{-1/4}\}\}^2
   E\left\{\wh\epsilon_\x^2 K_h(\S_i-\s_0)^2\mid D_2\right\}\\
   &=& O_p\{|\wh\zeta_\x(\s_0)-\zeta_\x(\s_0)|^2 h^{-d_s}\}+o_p\{(nh^{d_s})^{-1/2}h^{-d_s}\}.
  \ese
  Combining with Lemma~\ref{lemm:conditional}, we have
  \begin{align}
  \label{equ: remain-1}
  h^{-d_s}E\{R_1^2(\s_0)\}&\le E\left([\phi_\x\{\o_i, \wh \zeta_\x(\s_0),\wh m_\x,\wh
  f_{\epsilon_\x\mid\x}\}- \phi_\x\{\o_i, \zeta_\x(\s_0), \wh m_\x,\wh
      f_{\epsilon_\x\mid\x}\}]K_h(\S_i-\s_0)\right)^2\nonumber\\
    &=O_p\{|\wh\zeta_\x(\s_0)-\zeta_\x(\s_0)|h^{-d_s}\}+o_p\{(nh^{d_s})^{-1/2}h^{-d_s}\}.
  \end{align}
  Thus, $R_1(\s_0)=O_p(|\wh\zeta_\x(\s_0)-\zeta_\x(\s_0)|^{1/2})+o_p\{(nh^{d_s})^{-1/4}\}=o_p(1)$.
  
  Similarly, to handle $E\{R_2(\s_0)^2\}$, we have
  \be\label{eq:remain-2}
  && E\{h^{-d_s}R_2(\s_0)^2\mid D_2\}\n\\
  &\le&E\{\left([\phi_\x\{\O_i, \zeta_\x(\s_0),\wh m_\x,\wh f_{\epsilon_\x\mid\x}\}-\phi_\x\{\O_i, \zeta_\x(\s_0),  m_\x, f_{\epsilon_\x\mid\x}\}]K_h(\S_i-\s_0)\right)^2\mid D_2\}\n\\
  &=&E[\{ (I\{r(\O_i,\wh m_\x)<\zeta_\x(\s_0)\}-I\{r(\O_i,
  m_\x)<\zeta_\x(\s_0)\}\n\\
  &+&\wh\epsilon_\x \wh E[\sign(\epsilon_\x)\delta\{\zeta_\x(\s_0)-r(\O, \wh m_\x)\}\mid\X]-\epsilon_\x E[\sign(\epsilon_\x)\delta\{\zeta_\x(\s_0)-r(\O, m_\x)\}\mid\X])K_h(\S_i-\s_0)\}^2 \mid D_2]\n\\
  &\leq&2E\{\left([I\{r(\O_i,\wh m_\x)<\zeta_\x(\s_0)\}-I\{r(\O_i, m_\x)<\zeta_\x(\s_0)\}]K_h(\s_i-\s_0)\right)^2\mid D_2\}\n
  \\
  &&+2E[\{(\wh \epsilon_\x \wh
  E[\sign(\epsilon_\x)\delta\{\zeta_\x(\s_0)-r(\O,\wh
  m_\x)\}\mid\X]-\epsilon_\x
  E[\sign(\epsilon_\x)\delta\{\zeta_\x(\s_0)-r(\O, m_\x)\}\mid\X])\n
  \\
  &&K_h(\S_i-\s_0)\}^2\mid D_2]\n\\
  &=&o_p\{(nh^{d_s})^{-1/4} h^{-d_s}\}+o_p\{(nh^{d_s})^{-1/2} h^{-d_s}\}\n\\
  &=&o_p\{(nh^{d_s})^{-1/4} h^{-d_s}\},
  \ee
  where the second last equality can be obtained directly from Assumptions~\ref{ass:eps}--\ref{ass:kernel}.
  Combining with Lemma~\ref{lemm:conditional}, we have
    $E\{R_2(\s_0)^2\}=o_p\{(nh^{d_s})^{-1/4}\}$.
   Thus, $R_2(\s_0)= o_p\{(nh^{d_s})^{-1/8}\} = o_p(1)$.
   \qed

   \begin{Lem} (Bound the terms $R_{30},R_{40}$)
    \label{lemm:R30-R40}
    Under Assumption~\ref{ass:lip20}, suppose
    $a_n\equiv\sup_\x|\wh m_\x(\x)-m_\x(\x)|=o_p(1)$,
    $\wt\zeta_\x-\zeta_\x=o_p(1)$.
    Then the terms $R_{30}$ and $R_{40}$ defined in
    \eqref{eq:R30} and \eqref{eq:R40}, respectively, satisfy  $R_{30}=o_p(n^{-1/2})$, $R_{40}=o_p(n^{-1/2})$.
    \end{Lem}
    Proof:
    Let
    $D_1\equiv\{\O_i:1\le i\le n_1\}$ and
    $D_2\equiv\{\O_i:n_1+1\le i\le n\}$. By construction,
    $\wh m_\x$ is estimated from $D_2$ and it is non-stochastic conditional on $D_2$. Note that $E(R_{30}\mid D_2)=0$ and $E(R_{40}\mid D_2)=0$.
  In addition,
    \begin{align*}E(R_{30}^2\mid D_2)&\le n_1^{-1}E\left([\psi_{1-\alpha}\{|Y_i-\wh m_\x(\X_i)|-\wt\zeta_\x\}-\psi_{1-\alpha}\{|Y_i- \wh m_\x(\X_i)|-\zeta_\x\}]^2\mid D_2\right)\\
    &=n_1^{-1}E\left(\left[I\{r(\O_i,\wh m_\x)<\wt
    \zeta_\x\}-I\{r(\O_i, \wh
    m_\x)<\zeta_\x\}\right]^2\mid D_2\right)\\
      &=O_p(n_1^{-1}|\wh\zeta_\x-\zeta_\x|).
    \end{align*}
    Combining with Lemma~\ref{lemm:conditional}, we have
    $R_{30}(\s_0)=O_p(n_1^{-1/2}|\wt\zeta_\x-\zeta_\x|^{1/2})=o_p(n^{-1/2})$.
    We explain the last equality in detail. For $a<b$,
    $\left\{I(t<a)-I(t<b)\right\}^2
    = I\left(a\le t < b\right).$
    Then 
    \bse
    &&E\left(\left[I\{r(\O_i,\wh m_\x)<\wt\zeta_\x\}-I\{r(\O_i, \wh m_\x)<\zeta_\x\}\right]^2\mid D_2\right)\\
    &=& P\left\{(\wt\zeta_\x\wedge\zeta_\x) \le r(\O_i,\wh m_\x) < (\wt\zeta_\x\vee\zeta_\x)\mid D_2\right\}= f_{\epsilon_\x}(\xi)|\wt\zeta_\x-\zeta_\x|
    \ese
    for some $\xi$ between $\wt\zeta_\x$ and $\zeta_\x$.
    Since $|\wt\zeta_\x-\zeta_\x|=o_p(1)$, for sufficiently large $n$,
    $\xi$ lies in the $c_3$-neighborhood of $\zeta_\x$ with probability
    approaching 1. By Assumption~\ref{ass:lip20}, $\sup_\x f_{\epsilon_\x}(\xi)<\infty$, so
    \bse
    &&E\left\{\left(\left[I\{r(\O_i,\wh m_\x)<\wh\zeta_\x\}-I\{r(\O_i, \wh m_\x)<\zeta_\x\}\right]\right)^2\mid D_2\right\}= O_p(|\wt\zeta_\x-\zeta_\x|).
    \ese
  
    Similarly, to handle $E(R_{40}^2)$, we have
    \begin{align*}E(R_{40}^2\mid
      D_2)&\le n_1^{-1}E\left([\psi_{1-\alpha}\{|Y_i-\wh m_\x(\X_i)|-\zeta_\x\}-\psi_{1-\alpha}\{|Y_i- m_\x(\X_i)|-\zeta_\x\}]^2\mid D_2\right)\\
      &=n_1^{-1}E\left(\left[I\{r(\O_i,\wh m_\x)<
      \zeta_\x\}-I\{r(\O_i,
      m_\x)<\zeta_\x\}\right]^2\mid D_2\right)\\
        &=O_p(n_1^{-1}|\wh m_\x-m_\x|).
      \end{align*}
    Combining with Lemma~\ref{lemm:conditional}, we have $R_{40}= o_p(n^{-1/2})$.
    \qed   


   \begin{Lem}(Consistency of $\wh \zeta_\x(\s_0)$)
   \label{lemm:consistency}
   Under Assumptions~\ref{ass:eps}--\ref{ass:kernel} and
   $nh^{d_s}\to\infty$, $h\to0$, $\wh\zeta_\x(\s_0)\xrightarrow{p}\zeta_\x(\s_0)$.
   \end{Lem}
   Proof:
   Define $H_0\equiv[\zeta_\x(\s_0)-c,
   \zeta_\x(\s_0)+c]$ with $0<c<(c_1\wedge c_2)$. We prove that, uniformly for $\zeta\in H_0$,
   \begin{align}
     \label{equ:cons-0}
   \frac{1}{n_1}\sumia \phi_\x(\o_i,\zeta,\wh
   m_\x,\wh f_{\epsilon_\x\mid\x})K_h(\s_i-\s_0)\xrightarrow{p}E\{\phi_\x(\O_i,\zeta,m_\x,f_{\epsilon_\x\mid\x})\mid \s_0\}f_{\S}(\s_0).
   \end{align}
   We write
   \be
   &&\frac{1}{n_1}\sumia\phi_\x(\o_i,\zeta,\wh
   m_\x,\wh f_{\epsilon_\x\mid\x})K_h(\s_i-\s_0)-E\{\phi_\x(\O_i,\zeta,m_\x,f_{\epsilon_\x\mid\x})\mid \s_0\}f_{\S}(\s_0)\n\\
   &=&\frac{1}{n_1}\sumia \{\phi_\x(\o_i,\zeta,\wh
   m_\x,\wh f_{\epsilon_\x\mid\x})-\phi_\x(\o_i,\zeta,m_\x,f_{\epsilon_\x\mid\x})\}K_h(\s_i-\s_0)\label{eq:p1}\\
   &+&\frac{1}{n_1}\sumia\phi_\x(\o_i,\zeta,m_\x,f_{\epsilon_\x\mid\x})K_h(\s_i-\s_0)-E\{\phi_\x(\O_i,\zeta,m_\x,f_{\epsilon_\x\mid\x})K_h(\S_i-\s_0)\}\label{eq:p2}\\
   &+&E\{ \phi_\x(\O_i,\zeta,m_\x,f_{\epsilon_\x\mid\x})K_h(\S_i-\s_0)\}-E\{\phi_\x(\O_i,\zeta,m_\x,f_{\epsilon_\x\mid\x})\mid \s_0\}f_{\S}(\s_0). \label{eq:p3}
   \ee
   To handle \eqref{eq:p1}, let
   $D_1\equiv\{\O_i:1\le i\le n_1\}$ and
   $D_2\equiv\{\O_i:n_1+1\le i\le n\}$. By construction,
   $\wh m_\x$ and $\wh f_{\epsilon_\x\mid\x}$ are estimated from $D_2$
   and they are non-stochastic conditional on $D_2$. From a similar
   analysis as for \eqref{eq:remain-2} and convergence rate of $\wh m_\x$ and $\wh f_{\epsilon_\x\mid\x}$, we
   have 
   \bse
   &&E\left(\left[\frac{1}{n_1}\sumia \{ \phi_\x(\O_i,\zeta,\wh
   m_\x,\wh f_{\epsilon_\x\mid\x})-\phi_\x(\O_i,\zeta,m_\x,f_{\epsilon_\x\mid\x})\}K_h(\S_i-\s_0)\right]^2\mid D_2\right)\\&=&o_p\{(nh^{d_s})^{-5/4}\}=o_p(1). 
   \ese
   Then by Lemma~\ref{lemm:conditional}, 
   \begin{align}
     \label{equ:cons-1}
     \frac{1}{n_1}\sumia \{\phi_\x(\o_i,\zeta,\wh
   m_\x,\wh f_{\epsilon_\x\mid\x})-\phi_\x(\o_i,\zeta,m_\x,f_{\epsilon_\x\mid\x})\}K_h(\s_i-\s_0)\xrightarrow{p} 0.
   \end{align}

   \eqref{eq:p2} is an empirical process term.
   Define a class of functions
   \bse
   \calF_n\equiv\left\{f(\o,\s,\zeta)=\phi_\x(\o,\zeta,m_\x,f_{\epsilon_\x\mid\x})K\{(\s-\s_0)/h\},\zeta\in H_0\right\}.
   \ese
   By the continuity of $f_{\epsilon_\x\mid\x}$(Assumption~\ref{ass:lip}), there exists $B_1>0$ such that
   $\sup_{\zeta\in H_0}E[\sign(\epsilon_\x)\delta\{\zeta-r(\O,m_\x)\}\mid
   \X]\leq B_1$.
   Then we define an envelope function
   $$F_n(\o,\s)\equiv\Psi(\o)\left|K\{(\s-\s_0)/h\}\right|,$$
   where $\Psi(\o)\equiv\{1+B_1\left|\epsilon_\x\right|\}$.
   It follows that $F_n(\o,\s)\geq\sup_{f\in\calF_n}\left|f(\o,\s,\zeta)\right|$. Moreover,
   \bse
    E\{F_n(\O,\S)^2\}
   &=&E[\Psi^2(\O)K^2\{(\S-\s_0)/h\}]\\
   &=&\int E\{\Psi^2(\O)\mid\s\}K^2\{(\s-\s_0)/h\}f_{\S}(\s)d\s\\
   &=&h^{d_s}\int E\{\Psi^2(\O)\mid\s_0+h\s\}K^2(\s)f_{\S}(\s_0+h\s)d\s.
   \ese
   By $E(\epsilon_\x^2\mid \s_0+h\s) <\infty$ (Assumption~\ref{ass:eps}
   and $h\to 0$), we have $E\{\Psi^2(\O)\mid\s_0+h\s\}<\infty$. It
   follows that $\norm{F_n}_{P,2}\asymp h^{d_s/2}$.
   
   Next we bound the bracketing number.
   Note that
   \bse
   &&\norm{\{f(\o,\s,\zeta_1)- f(\o,\s,\zeta_2)\}}_{P,2}\\
   &=&\norm{\{\phi_\x(\o,\zeta_1,m_\x,f_{\epsilon_\x\mid\x})-\phi_\x(\o,\zeta_2,m_\x,f_{\epsilon_\x\mid\x})\}K\{(\s-\s_0)/h\}}_{P,2}\\
   &\leq&\norm{[I\{r(\o,m_\x)<\zeta_2\}-I\{r(\o,m_\x)<\zeta_1\}]K\{(\s-\s_0)/h\}}_{P,2}\\&&+\norm{\epsilon_\x\left(E[\sign(\epsilon_\x)\delta\{\zeta_2-r(\O,m_\x)\}\mid\X]-E[\sign(\epsilon_\x)\delta\{\zeta_1-r(\O,m_\x)\}\mid\X]\right)K\{(\s-\s_0)/h\}}_{P,2}\\
   &\lesssim&h^{d_s/2}\left|\zeta_2-\zeta_1\right|^{1/2}+h^{d_s/2}\left|\zeta_2-\zeta_1\right|\\
   &\asymp&
   h^{d_s/2}\left|\zeta_1-\zeta_2\right|^{1/2}.
   \ese
   Let $\{\zeta_j\}_{j=1}^N$ be an $\eta$-network for $H_0$. For each $f(\o,\s,\zeta)\in\calF_n$, 
   there exists $j$ such that $\zeta\in[\zeta_{j-1},\zeta_{j}]$. Let
   \bse
   \Delta_j=\sup_{\zeta\in[\zeta_{j-1},\zeta_j]}\left|f(\o,\s,\zeta)-f(\o,\s,\zeta_j)\right|\lesssim h^{d_s/2}\left|\zeta_{j-1}-\zeta_j\right|^{1/2}\lesssim \eta^{1/2}\norm{F_n}_{P,2}.
   \ese
   For each $\zeta\in[\zeta_{j-1},\zeta_j]$, $f(\o,\s,\zeta_{j})-\Delta_j\le f(\o,\s,\zeta)\le f(\o,\s,\zeta_{j})+\Delta_j.$
   Let $\eta= \epsilon^2$, then the bracketing number satisfies
   $$N_{[]}\left(\epsilon\norm{F_n}_{P,2}, \calF_n, L_2(P)\right)\leq N\left(\eta,H_0,\left|\cdot\right|\right)\lesssim 2c/\eta+1\lesssim\epsilon^{-2}.$$
   It follows that the bracketing integral
   \bse
   J_{[]}\left(a,\calF_n\mid F_n,L_2(P)\right)=\int_0^{a}\left\{1+\log N_{[]}\left(\epsilon\norm{F_n}_{P,2}, \calF_n, L_2(P)\right)\right\}^{1/2}d\epsilon\lesssim a [1+\{\log(1/a)\}^{1/2}].
   \ese
   By the maximal inequality, (Theorem~2.14.16 of \cite{wellner2013weak})
   \bse
   \norm{\sup_{f\in\calF_n}|G_{n_1}(f)|}_{P,1}\lesssim J_{[]}(1,\calF_n\mid F,L_2(P))\norm{F_n}_{P,2}\lesssim h^{d_s/2}.
   \ese
   where $G_{n_1}(f)=n_1^{-1/2}\sum_{i=1}^{n_1}f(\o_i,\s_i,\zeta)-E\{f(\O,\S,\zeta)\}$. 
   By Markov's inequality,
   $\sup_{f\in\calF_n}|G_{n_1}(f)|=O_p(h^{d_s/2})$.
   Thus, 
   \begin{align}
     \label{equ:cons-2}
   \frac{1}{n_1}\sumia\wt
     \phi_\x(\o_i,\zeta,m_\x,f_{\epsilon_\x\mid\x})K_h(\s_i-\s_0)-E\{\wt
     \phi_\x(\O_i,\zeta,m_\x,f_{\epsilon_\x\mid\x})K_h(\S_i-\s_0)\}
   =O(h^{-d_s/2}n_1^{-1/2})
     \xrightarrow{p}0
   \end{align}
   since $n_1\asymp n$ and $nh^{d_s}\to\infty$. 
   Lastly, to handle \eqref{eq:p3},
   \bse 
   &&\lim_{h\to 0}E\{\phi_\x(\O_i,\zeta,m_\x,f_{\epsilon_\x\mid\x})K_h(\S_i-\s_0)\}\\&=&\lim_{h\to 0}E\left[E\{\phi_\x(\O_i,\zeta,m_\x,f_{\epsilon_\x\mid\x})\mid \S\}K_h(\S-\s_0)\right]\\
   &=&\lim_{h\to 0}\int E\{\phi_\x(\O_i,\zeta,m_\x,f_{\epsilon_\x\mid\x})\mid \s\}K_h(\s-\s_0)f_{\S}(\s)d\s\\
   &=&\lim_{h\to 0}\int E\{\phi_\x(\O_i,\zeta,m_\x,f_{\epsilon_\x\mid\x})\mid \s_0+h\s\}K(\s)f_{\S}(\s_0+h\s)d\s\\
   &=&E\{\phi_\x(\O_i,\zeta,m_\x,f_{\epsilon_\x\mid\x})\mid \s_0\}f_{\S}(\s_0).
   \ese
   Thus, \eqref{eq:p3} approaches 0 when $h\to0$.
   Combining results on \eqref{eq:p1}--\eqref{eq:p3},  we have proven
   \eqref{equ:cons-0}.
   
   Next, we discuss the identification of the
   root of the
     estimating equation.
   Note that $$E\{\phi_\x(\O_i,\zeta_\x(\s_0),m_\x,f_{\epsilon_\x\mid\x})\mid \s_0\}f_{\S}(\s_0)=0.$$
   By $ E(\epsilon_\x \mid \X) = 0$,
   $$E\{\phi_\x(\O_i,\zeta,m_\x,f_{\epsilon_\x\mid\x})\mid \s_0\}f_{\S}(\s_0)=\{1-\alpha-\pr(\left|\epsilon_\x\right|<\zeta\mid \s_0)\}f_{\S}(\s_0),$$
   which is strictly decreasing in $\zeta$, so that $\zeta_\x(\s_0)$ is the unique zero. 
   By \eqref{equ:cons-0}, the convergence holds uniformly over $\zeta\in H_0$. Hence, in particular, at the two endpoints $\zeta_\x(\s_0)\pm c$, with probability approaching~$1$,
   \bse
   \frac{1}{n_1}\sumia\phi_\x(\o_i,\zeta_\x(\s_0)-c,\wh m_\x,\wh f_{\epsilon_\x\mid\x})K_h(\s_i-\s_0)>0,\\
   \frac{1}{n_1}\sumia\phi_\x(\o_i,\zeta_\x(\s_0)+c,\wh m_\x,\wh f_{\epsilon_\x\mid\x})K_h(\s_i-\s_0)<0.
   \ese
   Since $I\{r(\o_i,\wh m_\x)<\zeta\}$ is non-decreasing in $\zeta$ for each $i$, the empirical sum is non-increasing in $\zeta$, 
   Therefore, its zero $\wh\zeta_\x(\s_0)$ lies in $\left(\zeta_\x(\s_0)-c,\,\zeta_\x(\s_0)+c\right)$ with probability approaching $1$. Since $c\in(0,c_1\wedge c_2)$ is arbitrary, we conclude $\wh\zeta_\x(\s_0)\xrightarrow{p}\zeta_\x(\s_0)$. 
   \qed

  \begin{Lem}{(Basic Corollary of \cite{hjort2011asymptotics})}

    Suppose $A_n(s)$ is convex and can be represented as $\frac{1}{2} s^{\prime} V s+U_n^{\prime} s+C_n+ r_n(s)$, where $V$ is symmetric and positive definite, $U_n$ is stochastically bounded, $C_n$ is arbitrary, and $r_n(s)$ goes to zero in probability for each $s$. Then $\alpha_n$, the argmin of $A_n$, is only $o_p(1)$ away from $\beta_n=-V^{-1} U_n$, the argmin of $\frac{1}{2} s^{\prime} V s+U_n^{\prime} s+C_n$. If also $U_n \xrightarrow{d} U$ then $\alpha_n \xrightarrow{d}-V^{-1} U$.
  \end{Lem}  

\subsection{Proof of Lemma \ref{lemm:semi0-onefold}}
 We prove the result for $\v=\x$; the proof for $\v=\w$ is identical.
Recall that
\bse
\phi_\x(\O, \zeta,m_\x, f_{\epsilon_\x\mid\x})=(1-\alpha)-I\{
  r(\O,m_\x)<\zeta\}-\epsilon_\x E[\sign(\epsilon_\x)\delta\{\zeta-r(\O,m_\x)\}\mid
\X].
\ese
Note that $n_1/n\to1$. The estimator $\wh\zeta_\x$
satisfies
\bse
\sumia \phi_\x(\o_i, \wh\zeta_\x,\wh m_\x,\wh
f_{\epsilon_\x\mid\x})=0.
\ese
To address the discontinuity of $\phi_\x$, we use the decomposition
\bse
&&n_1^{-1/2}\sumia \phi_\x(\o_i, \wh\zeta_\x,\wh m_\x,\wh
f_{\epsilon_\x\mid\x})\\
&=&n_1^{-1/2}\sumia \phi_\x(\o_i, \zeta_\x,\wh m_\x,\wh
f_{\epsilon_\x\mid\x})+n_1^{-1/2}\sumia E\{\phi_\x(\O_i,\wh\zeta_\x,\wh m_\x,\wh
f_{\epsilon_\x\mid\x})\}\\
&-&n_1^{-1/2}\sumia E\{\phi_\x(\O_i,\zeta_\x,\wh m_\x,\wh
f_{\epsilon_\x\mid\x})\}+R_{10},
\ese
where
\bse
R_{10}&\equiv&n_1^{-1/2}\sumia\bigl[\phi_\x(\o_i,\wh\zeta_\x,\wh m_\x,\wh
f_{\epsilon_\x\mid\x})-\phi_\x(\o_i,\zeta_\x,\wh m_\x,\wh
f_{\epsilon_\x\mid\x})\\
&&-E\{\phi_\x(\O_i,\wh\zeta_\x,\wh m_\x,\wh
f_{\epsilon_\x\mid\x})-\phi_\x(\O_i,\zeta_\x,\wh m_\x,\wh
f_{\epsilon_\x\mid\x})\}\bigr].
\ese
  Following the same argument as in  Lemma~\ref{lemm:consistency} with
  the kernel weight replaced by 1, we obtain
  $\wh\zeta_\x-\zeta_\x=o_p(1)$.
Following the same argument as Lemma~\ref{lemm:1}, with the kernel weight
replaced by 1, we obtan $R_{10}=o_p(1)$.
Therefore, a Taylor expansion gives
\bse
0&=&n_1^{-1/2}\sumia \phi_\x(\o_i,\zeta_\x,\wh m_\x,\wh
f_{\epsilon_\x\mid\x})+\left[\frac{\partial E\{\phi_\x(\O_i,\zeta,\wh m_\x,\wh
f_{\epsilon_\x\mid\x})\}}{\partial \zeta}\bigg|_{\zeta=\zeta_\x^\ast}
+o_p(1)\right]n_1^{1/2}(\wh\zeta_\x-\zeta_\x)+R_{10}\\
&=&n_1^{-1/2}\sumia \phi_\x(\o_i,\zeta_\x,\wh m_\x,\wh
f_{\epsilon_\x\mid\x})+\left[\frac{\partial E\{\phi_\x(\O_i,\zeta,m_\x,
f_{\epsilon_\x\mid\x})\}}{\partial \zeta}\bigg|_{\zeta=\zeta_\x^\ast}
+o_p(1)\right]n_1^{1/2}(\wh\zeta_\x-\zeta_\x)+o_p(1),
\ese
where $\zeta_\x^\ast$ lies between $\zeta_\x$ and $\wh\zeta_\x$.

Next, we replace the nuisance estimators by their population limits. We
write
\bse
&&n_1^{-1/2}\sumia \phi_\x(\o_i,\zeta_\x,\wh m_\x,\wh
f_{\epsilon_\x\mid\x})\\
&=&n_1^{-1/2}\sumia \phi_\x(\o_i,\zeta_\x,m_\x,
f_{\epsilon_\x\mid\x})+n_1^{-1/2}\sumia E\{\phi_\x(\O_i,\zeta_\x,\wh m_\x,\wh
f_{\epsilon_\x\mid\x})\}\\
&-&n_1^{-1/2}\sumia E\{\phi_\x(\O_i,\zeta_\x,m_\x,
f_{\epsilon_\x\mid\x})\}+R_{20},
\ese
where
\bse
R_{20}&\equiv&n_1^{-1/2}\sumia [\phi_\x(\o_i,
\zeta_\x,\wh m_\x,\wh f_{\epsilon_\x\mid\x})-\phi_\x(\o_i,
\zeta_\x,  m_\x, f_{\epsilon_\x\mid\x})\n\\
&&-E\{\phi_\x(\O_i, \zeta_\x,\wh m_\x,\wh f_{\epsilon_\x\mid\x})-\phi_\x(\O_i, \zeta_\x, m_\x, f_{\epsilon_\x\mid\x})\mid\s_i\}].
\ese
 Following the same argument as Lemma~\ref{lemm:1}, with the kernel weight
replaced by 1, we obtan $R_{20}=o_p(1)$.
 Let $\eta=(m_\x,f_{\epsilon_\x\mid\x})$. Using the
Gateaux derivative  expansion in $\eta$,
\bse
&&n_1^{-1/2}\sumia \phi_\x(\o_i,\zeta_\x,\wh m_\x,\wh
f_{\epsilon_\x\mid\x})\\
&=&n_1^{-1/2}\sumia \phi_\x(\o_i,\zeta_\x,m_\x,
f_{\epsilon_\x\mid\x})
+n_1^{1/2}\partial_\eta E\{\phi_\x(\O_i,\zeta_\x,m_\x,
f_{\epsilon_\x\mid\x})\}[\wh\eta-\eta]\\
&&+\frac{1}{2}n_1^{1/2}\partial_\eta^2
E\{\phi_\x(\O_i,\zeta_\x,m_\x^\ast,
f_{\epsilon_\x\mid\x}^\ast)\}[\wh\eta-\eta,\wh\eta-\eta]+R_{20}\\
&=&n_1^{-1/2}\sumia \phi_\x(\o_i,\zeta_\x,m_\x,
f_{\epsilon_\x\mid\x})+o_p(1),
\ese
where $m_\x^\ast$ and $f_{\epsilon_\x\mid\x}^\ast$ lie on the line
segments between the estimated and true nuisance functions. The last
equality follows from the 
the convergence rate of $\wh m_\x$, $\wh f_{\epsilon_\x\mid\x}$ and 
Neyman-orthogonality
\bse
\partial_\eta E[\phi_\x\{\O_i,\zeta_\x,m_\x,
f_{\epsilon_\x\mid\x}\}][\wh\eta-\eta]=0.
\ese
Thus,
\bse
n_1^{-1/2}\sumia \phi_\x(\o_i,\zeta_\x,m_\x,
f_{\epsilon_\x\mid\x})+o_p(1)
=-\left[
\frac{\partial E\{\phi_\x(\O_i,\zeta,m_\x,
f_{\epsilon_\x\mid\x})\}}{\partial \zeta}\bigg|_{\zeta=\zeta_\x^\ast}
+o_p(1)\right]n_1^{1/2}(\wh\zeta_\x-\zeta_\x).
\ese
By consistency, $\zeta_\x^\ast\xrightarrow{p}\zeta_\x$. Moreover,
\bse
\frac{\partial E\{\phi_\x(\O_i,\zeta,m_\x,
f_{\epsilon_\x\mid\x})\}}{\partial \zeta}\bigg|_{\zeta=\zeta_\x}
=-\{f_{\epsilon_\x}(\zeta_\x)+f_{\epsilon_\x}(-\zeta_\x)\}= -g_\x,
\ese
because $E(\epsilon_\x\mid\X)=0$ leads to the expectation of the
orthogonal correction term to be zero. Hence
\be
\label{equ:asylinear-semi0}
n_1^{1/2}(\wh\zeta_\x-\zeta_\x)
=-g_\x^{-1}n_1^{-1/2}\sumia
\phi_\x(\o_i,\zeta_\x,m_\x,f_{\epsilon_\x\mid\x})+o_p(1).
\ee
Finally, since
$E\{\phi_\x(\O_i,\zeta_\x,m_\x,f_{\epsilon_\x\mid\x})\}=0$ and
$E\{\phi_\x^2(\O_i,\zeta_\x,m_\x,f_{\epsilon_\x\mid\x})\}<\infty$,
the central limit theorem yields
\bse
n_1^{1/2}(\wh\zeta_\x-\zeta_\x)
\to N\left(0,g_\x^{-2}E\{\phi_\x^2(\O_i,\zeta_\x,m_\x,
f_{\epsilon_\x\mid\x})\}\right).
\ese
This proves the lemma.
\qed

\subsection{Proof of Theorem \ref{th:semi0}}
Let $n_j=|\mathcal I_j|$, $j=1,2$.  By construction,
$n_j/n\to1/2$.  Apply Lemma~\ref{lemm:semi0-onefold} to each
evaluation fold, using the nuisance estimators trained on the other
fold.  For $j=1,2$, this gives
\begin{align}
n_j^{1/2}(\wh\zeta_\v^{(j)}-\zeta_\v)
={}&-g_\v^{-1}n_j^{-1/2}
\sum_{i\in\mathcal I_j}
\phi_\v(\O_i,\zeta_\v,m_\v,
f_{\epsilon_\v\mid\v})+o_p(1).
\label{equ:foldwise-linear0}
\end{align}
Equivalently,
\[
\wh\zeta_\v^{(j)}-\zeta_\v
=-g_\v^{-1}n_j^{-1}
\sum_{i\in\mathcal I_j}
\phi_\v(\O_i,\zeta_\v,m_\v,
f_{\epsilon_\v\mid\v})+o_p(n_j^{-1/2}).
\]
Since
$\wh\zeta_\v=(\wh\zeta_\v^{(1)}+
\wh\zeta_\v^{(2)})/2$, averaging the two expansions yields
\begin{align*}
n^{1/2}(\wh\zeta_\v-\zeta_\v)
={}&-\frac{g_\v^{-1}n^{1/2}}{2}
\left(
\frac{1}{n_1}\sum_{i\in\mathcal I_1}\phi_{\v,i}
+\frac{1}{n_2}\sum_{i\in\mathcal I_2}\phi_{\v,i}
\right)+o_p(1)\\
={}&-g_\v^{-1}n^{-1/2}\sumi\phi_{\v,i}+o_p(1),
\end{align*}
where
$\phi_{\v,i}=\phi_\v(\O_i,\zeta_\v,m_\v,
f_{\epsilon_\v\mid\v})$.  The last equality follows from
$n_1=\lfloor n/2\rfloor$ and $n_2=n-n_1$; the discrepancy caused by
an odd $n$ is $o_p(1)$.  This proves
\eqref{equ:theo1-linear0}.  Finally, the central limit theorem gives
\[
n^{1/2}(\wh\zeta_\v-\zeta_\v)
\xrightarrow{d}N(0,\sigma_\v).
\]
\qed

\subsection{Proof of Lemma \ref{lemm:conformal0-onefold}}

We prove the result for $\v=\x$; the proof for $\v=\w$ is identical.
The conformal prediction based estimator $\wt\zeta_\x$ satisfies
\bse
n_1^{-1}\sumia\psi_{1-\alpha}\{|y_i-\wh m_\x(\x_i)|-\wt\zeta_\x\}=0.
\ese
We first verify that $\wt\zeta_\x$ is consistent. Let
\bse
\wt\zeta_\x^{\,o}
\equiv \inf\left\{\zeta:\ n_1^{-1}\sumia
I(|\epsilon_{\x,i}|\le \zeta)\ge 1-\alpha\right\}
\ese
be the oracle empirical $(1-\alpha)$ quantile based on
$|\epsilon_{\x,i}|$. Since
\bse
\left||y_i-\wh m_\x(\x_i)|-|\epsilon_{\x,i}|\right|
\le |\wh m_\x(\x_i)-m_\x(\x_i)|\le a_n,
\ese
the empirical distribution functions of
$|y_i-\wh m_\x(\x_i)|$ and $|\epsilon_{\x,i}|$ are uniformly within an
$a_n$ horizontal shift. Hence the corresponding empirical quantiles satisfy
\bse
|\wt\zeta_\x-\wt\zeta_\x^{\,o}|\le a_n.
\ese
By the Glivenko--Cantelli theorem and the continuity of the distribution of
$|\epsilon_\x|$ at $\zeta_\x$ with positive density
$g_\x=f_{\epsilon_\x}(\zeta_\x)+f_{\epsilon_\x}(-\zeta_\x)>0$, we have
$\wt\zeta_\x^{\,o}\xrightarrow{p}\zeta_\x$. Together with $a_n=o_p(1)$, this
gives
\bse
\wt\zeta_\x-\zeta_\x=o_p(1).
\ese
Next, we consider the linear expansion, 
\bse
0&=&n_1^{-1}\sumia\psi_{1-\alpha}\{|y_i-\wh m_\x(\x_i)|-\wt\zeta_\x\}\\&=&n_1^{-1}\sumia\psi_{1-\alpha}\{|y_i- \wh m_\x(\x_i)|-\zeta_\x\}-\pr\{|y_i-\wh m_\x(\x_i)|<\wt\zeta_\x\}+\pr\{|y_i- \wh m_\x(\x_i)|<\zeta_\x\}+R_{30}\\
 &=&n_1^{-1}\sumia\psi_{1-\alpha}\{|y_i- \wh m_\x(\x_i)|-\zeta_\x\}-\frac{\partial \pr\{|y_i- \wh m_\x(\x_i)|<\zeta_\x\}}{\partial \zeta_\x}(\wt\zeta_\x-\zeta_\x)+R_{30}\\
 &=&n_1^{-1}\sumia\psi_{1-\alpha}\{|y_i- \wh m_\x(\x_i)|-\zeta_\x\}-\frac{\partial \pr\{|y_i- m_\x(\x_i)|<\zeta_\x\}}{\partial \zeta_\x}(\wt\zeta_\x-\zeta_\x)+R_{30}\\
 &=&n_1^{-1}\sumia\psi_{1-\alpha}\{|y_i- \wh m_\x(\x_i)|-\zeta_\x\}-g_{\x}(\wt\zeta_\x-\zeta_\x)+R_{30},
\ese
where
 \be\label{eq:R30}
 R_{30}&\equiv& n_1^{-1}\sumia\psi_{1-\alpha}\{|y_i-\wh
 m_\x(\x_i)|-\wt\zeta_\x\}-n_1^{-1}\sumia\psi_{1-\alpha}\{|y_i- \wh
 m_\x(\x_i)|-\zeta_\x\}\n\\
 &&+\pr\{|y_i-\wh m_\x(\x_i)|<\wt\zeta_\x\}-\pr\{|y_i- \wh m_\x(\x_i)|<\zeta_\x\}.
 \ee
In addition, 
\bse
&&n_1^{-1}\sumia\psi_{1-\alpha}\{|y_i- \wh m_\x(\x_i)|-\zeta_\x\}\\&=&n_1^{-1}\sumia\psi_{1-\alpha}\{|y_i- m_\x(\x_i)|-\zeta_\x\}-\pr\{|y_i-\wh m_\x(\x_i)|<\zeta_\x\}+\pr\{|y_i- m_\x(\x_i)|<\zeta_\x\}+R_{40}\\
&=&n_1^{-1}\sumia\psi_{1-\alpha}\{|y_i- m_\x(\x_i)|-\zeta_\x\}\\&&-E [\{f_{\epsilon_\x\mid\x}(\zeta_\x,\X)-f_{\epsilon_\x\mid\x}(-\zeta_\x,\X)\}\{\wh m_\x(\X)-m_\x(\X)\}]+O_p(a_n^2)+R_{40},
\ese
where
 \be\label{eq:R40}
 R_{40}&\equiv& n_1^{-1}\sumia\psi_{1-\alpha}\{|y_i- \wh
 m_\x(\x_i)|-\zeta_\x\}-n_1^{-1}\sumia\psi_{1-\alpha}\{|y_i-
 m_\x(\x_i)|-\zeta_\x\}\n\\
 &&+\pr\{|y_i-\wh m_\x(\x_i)|<\zeta_\x\}-\pr\{|y_i- m_\x(\x_i)|<\zeta_\x\}.
 \ee
By Lemma~\ref{lemm:R30-R40},
$R_{30}=o_p(n^{-1/2})$ and $R_{40}=o_p(n^{-1/2})$.
Finally, combining the above two results,
\bse
n_1^{1/2}(\wt\zeta_\x-\zeta_\x)
&=&g_\x^{-1}n_1^{-1/2}\sumia
\psi_{1-\alpha}(|\epsilon_{\x,i}|-\zeta_\x)
\\&&-g_\x^{-1}E
[\{f_{\epsilon_\x\mid\x}(\zeta_\x,\X)-f_{\epsilon_\x\mid\x}(-\zeta_\x,\X)\}\{\wh
m_\x(\X)-m_\x(\X)\}]+O_p(a_n^2 n^{1/2})+o_p(1).
\ese
Note that, if the distribution of $\epsilon_\x\mid\X$ is symmetrical,
the first order term of $\wh m_\x$ will disappear.
\qed

\subsection{Proof of Theorem \ref{th:conformal0}}
Let $n_j=|\mathcal I_j|$, $j=1,2$.  By construction,
$n_j/n\to1/2$.  Apply Lemma~\ref{lemm:conformal0-onefold} to each
evaluation fold, using the nuisance estimator trained on the other
fold.  For $j=1,2$, this gives
\begin{align*}
n_j^{1/2}(\wt\zeta_\v^{(j)}-\zeta_\v)
={}&-g_\v^{-1}n_j^{1/2}B_\v^{(-j)}
+g_\v^{-1}n_j^{-1/2}
\sum_{i\in\mathcal I_j}\psi_{\v,i}+O_p\bigl\{(a_n^{(-j)})^2n_j^{1/2}\bigr\}+o_p(1),
\end{align*}
where
$\psi_{\v,i}=\psi_{1-\alpha}(|\epsilon_{\v,i}|-\zeta_\v)$.
Equivalently,
\begin{align*}
\wt\zeta_\v^{(j)}-\zeta_\v
={}&-g_\v^{-1}B_\v^{(-j)}
+g_\v^{-1}n_j^{-1}
\sum_{i\in\mathcal I_j}\psi_{\v,i}
+O_p\bigl\{(a_n^{(-j)})^2\bigr\}+o_p(n_j^{-1/2}).
\end{align*}
Since
$\wt\zeta_\v=(\wt\zeta_\v^{(1)}+
\wt\zeta_\v^{(2)})/2$, averaging the two expansions yields
\begin{align*}
n^{1/2}(\wt\zeta_\v-\zeta_\v)
={}&-g_\v^{-1}n^{1/2}\overline B_\v
+g_\v^{-1}n^{-1/2}\sumi\psi_{\v,i}
+O_p(\overline a_n^2n^{1/2})+o_p(1).
\end{align*}
As in the proof of Theorem~\ref{th:semi0}, the discrepancy caused by
an odd $n$ is $o_p(1)$.  This proves
\eqref{equ:theo2-linear0}.
\qed

\subsection{Proof of Lemma \ref{lemm:semi-onefold}}
We prove the result for $\v=\x$; the proof for $\v=\w$ is identical.
\bse
\phi_\x(\O, \zeta,m_\x, f_{\epsilon_\x\mid\x})=(1-\alpha)-I\{
  r(\O,m_\x)<\zeta\}-\epsilon_\x E[\sign(\epsilon_\x)\delta\{\zeta-r(\O,m_\x)\}\mid
\X].
\ese
First, we point out that $n_1/n\to1$.  
The estimator $\wh\zeta_\x(\s_0)$ satisfies
\bse
\sumia \phi_\x\{\o_i, \wh\zeta_\x(\s_0),\wh m_\x,\wh f_{\epsilon_\x\mid\x}\}K_h(\s_i-\s_0)=0.
\ese

To address the discontinuity of $\phi_\x$, we perform the following decomposition:
\bse
&&(n_1h^{-d_s})^{-1/2}\sumia \phi_\x\{\o_i, \wh\zeta_\x(\s_0), \wh m_\x,\wh f_{\epsilon_\x\mid\x}\}K_h(\s_i-\s_0)\\
&=&(n_1h^{-d_s})^{-1/2}\sumia \phi_\x\{\o_i, \zeta_\x(\s_0), \wh m_\x,\wh
f_{\epsilon_\x\mid\x}\}K_h(\s_i-\s_0)\\
&+&(n_1h^{-d_s})^{-1/2}\sumia E[\phi_\x\{\o_i, \wh \zeta_\x(\s_0),\wh m_\x,\wh
f_{\epsilon_\x\mid\x}\}\mid\s_i]K_h(\s_i-\s_0)\\
&-&(n_1h^{-d_s})^{-1/2}\sumia E[\phi_\x\{\o_i, \zeta_\x(\s_0),\wh m_\x,\wh
f_{\epsilon_\x\mid\x}\}\mid\s_i]K_h(\s_i-\s_0)\\
&+&R_1(\s_0),
\ese
where 
\be\label{eq:R1}
R_1(\s_0)&\equiv&(n_1h^{-d_s})^{-1/2}\sumia (\phi_\x\{\o_i, \wh \zeta_\x(\s_0),\wh m_\x,\wh
f_{\epsilon_\x\mid\x}\}-\phi_\x\{\o_i, \zeta_\x(\s_0), \wh m_\x,\wh
f_{\epsilon_\x\mid\x}\}\n\\
&&-E[\phi_\x\{\o_i, \wh \zeta_\x(\s_0),\wh m_\x,\wh
f_{\epsilon_\x\mid\x}\}-\phi_\x\{\o_i, \zeta_\x(\s_0), \wh m_\x,\wh
f_{\epsilon_\x\mid\x}\}\mid\s_i])K_h(\s_i-\s_0).
\ee
We perform Taylor expansion to obtain
\bse
0&=&(n_1h^{-d_s})^{-1/2}\sumia \phi_\x\{\o_i, \wh\zeta_\x(\s_0), \wh m_\x,\wh f_{\epsilon_\x\mid\x}\}K_h(\s_i-\s_0)\\
&=&(n_1h^{-d_s})^{-1/2}\sumia \phi_\x\{\o_i, \zeta_\x(\s_0), \wh m_\x,\wh
f_{\epsilon_\x\mid\x}\}K_h(\s_i-\s_0)\\
&& +\left[\frac{\partial  E\{\phi_\x(\O_i, \zeta, \wh m_\x,\wh
  f_{\epsilon_\x\mid\x})\mid \s_0\}
f_\S(\s_0)
}{\partial \zeta}\big|_{\zeta=\zeta_\x^*(\s_0)}+
  O_p\{h^t+(nh^{d_s})^{-1/2}\}\right]
(n_1h^{d_s})^{1/2}\{\wh\zeta_\x(\s_0)-\zeta_\x(\s_0)\}\\
&&+ R_1(\s_0)\\
&=& (n_1h^{-d_s})^{-1/2}\sumia \phi_\x\{\o_i, \zeta_\x(\s_0), \wh m_\x,\wh
f_{\epsilon_\x\mid\x}\}K_h(\s_i-\s_0)\\
&&+\left[\frac{\partial  E\{\phi_\x(\O_i, \zeta,  m_\x,
  f_{\epsilon_\x\mid\x})\mid \s_0\}
f_\S(\s_0)
}{\partial \zeta}\big|_{\zeta=\zeta_\x^*(\s_0)}+o_p(1)\right]
(n_1h^{d_s})^{1/2}\{\wh\zeta_\x(\s_0)-\zeta_\x(\s_0)\}+ R_1(\s_0),
\ese
where $\zeta_\x^*(\s_0)$ is a point between $\zeta_\x(\s_0)$ and $\wh\zeta_\x(\s_0)$,
and $R_1(\s_0)=o_p(1)$ by Lemmas~\ref{lemm:1} and \ref{lemm:consistency}.
Then we perform a simliar decomposition:
\bse
&&(n_1h^{-d_s})^{-1/2}\sumia \phi_\x\{\o_i, \zeta_\x(\s_0), \wh m_\x,\wh f_{\epsilon_\x\mid\x}\}K_h(\s_i-\s_0)\\
&=&(n_1h^{-d_s})^{-1/2}\sumia \phi_\x\{\o_i, \zeta_\x(\s_0), m_\x,
f_{\epsilon_\x\mid\x}\}K_h(\s_i-\s_0)\\
&+&(n_1h^{-d_s})^{-1/2}\sumia E[\phi_\x\{\o_i, \zeta_\x(\s_0),\wh m_\x,\wh
f_{\epsilon_\x\mid\x}\}\mid\s_i]K_h(\s_i-\s_0)\\
&-&(n_1h^{-d_s})^{-1/2}\sumia E[\phi_\x\{\o_i, \zeta_\x(\s_0), m_\x,
f_{\epsilon_\x\mid\x}\}\mid\s_i]K_h(\s_i-\s_0)\\
&+&R_2(\s_0),
\ese
where 
\be\label{eq:R2}
R_2(\s_0)&\equiv&(n_1h^{-d_s})^{-1/2}\sumia (\phi_\x\{\o_i,
\zeta_\x(\s_0),\wh m_\x,\wh f_{\epsilon_\x\mid\x}\}-\phi_\x\{\o_i,
\zeta_\x(\s_0),  m_\x, f_{\epsilon_\x\mid\x}\}\n\\
&&-E[\phi_\x\{\o_i, \zeta_\x(\s_0),\wh m_\x,\wh f_{\epsilon_\x\mid\x}\}-\phi_\x\{\o_i, \zeta_\x(\s_0), m_\x, f_{\epsilon_\x\mid\x}\}\mid\s_i])K_h(\s_i-\s_0).
\ee
Let $\eta=(m_\x, f_{\epsilon_\x\mid\x})$ denote the nuisance parameters and $\sup_\x\norm{\wh \eta-\eta}=\sup_\x\norm{(\wh m_\x, \wh f_{\epsilon_\x\mid\x})-(m_\x, f_{\epsilon_\x\mid\x})}=o_p\{(nh^{d_s})^{-1/4}\}$. 
Now using the Gateaux derivative for $\eta$, we obtain
\bse
&&(n_1h^{-d_s})^{-1/2}\sumia  \phi_\x\{\o_i, \zeta_\x(\s_0),\wh m_\x,\wh
f_{\epsilon_\x\mid\x}\}K_h(\s_i-\s_0)\\
&=&(n_1h^{-d_s})^{-1/2}\sumia  \phi_\x\{\o_i, \zeta_\x(\s_0),  m_\x, f_{\epsilon_\x\mid\x}\}K_h(\s_i-\s_0)\\
&&+(n_1h^{-d_s})^{-1/2}\sumia \partial_{\eta} E[
\phi_\x\{\O_i,\zeta_\x(\s_0), m_\x,
f_{\epsilon_\x\mid\x}\}\mid\s_i][\wh\eta-\eta]K_h(\s_i-\s_0)\\
&&+
  \frac{1}{2}(n_1h^{-d_s})^{-1/2}\sumia \partial_{\eta}^2 E[
\phi_\x\{\O_i,\zeta_\x(\s_0),
m_\x^\ast,f_{\epsilon_\x\mid\x}^\ast\}\mid\s_i][\wh\eta-\eta,\wh\eta-\eta]K_h(\s_i-\s_0)+R_2(\s_0)\\
&=&(n_1h^{-d_s})^{-1/2}\sumia \phi_\x\{\o_i, \zeta_\x(\s_0),  m_\x, f_{\epsilon_\x\mid\x}\}K_h(\s_i-\s_0)\\
&&+(n_1h^{d_s})^{1/2}\left(\partial_{\eta} E[\phi_\x\{\O_i,\zeta_\x(\s_0), m_\x,f_{\epsilon_\x\mid\x}\}\mid\s_0][\wh\eta-\eta]f_{\S}(\s_0)+O_p\{h^t+(nh^{d_s})^{-1/2}\}o_p(nh^{d_s})^{-1/4}\}\right)
\\
&&+o_p(n_1^{1/2}(nh^{d_s})^{-1/2}h^{d_s/2})+o_p(1)\\
&=&(n_1h^{-d_s})^{-1/2}\sumia \phi_\x\{\o_i, \zeta_\x(\s_0),  m_\x, f_{\epsilon_\x\mid\x}\}K_h(\s_i-\s_0)+o_p(1).
\ese
where $m_\x^\ast$ and $f_{\epsilon_\x\mid\x}^\ast$ lie on the line
segments between $\wh m_\x$ and $m_\x$, and between $\wh
f_{\epsilon_\x\mid\x}$ and $f_{\epsilon_\x\mid\x}$,
respectively. 
From Lemma~\ref{lemm:1}, $R_2(\s_0)=o_p(1)$.
By the Neyman-orthogonality of the influence function, we have $\partial_{\eta} E[\phi_\x\{\O_i,\zeta_\x(\s_0), m_\x,f_{\epsilon_\x\mid\x}\}\mid\s_0][\wh\eta-\eta]=0$.
We thus get
\bse
&&(n_1h^{-d_s})^{-1/2}\sumia \phi_\x\{\o_i, \zeta_\x(\s_0), m_\x, 
f_{\epsilon_\x\mid\x}\}K_h(\s_i-\s_0) +o_p(1)\\
&=&-\left\{\frac{\partial  E\{\phi_\x(\O_i, \zeta,  m_\x,
  f_{\epsilon_\x\mid\x})\mid \s_0\}
}{\partial \zeta}\big|_{\zeta=\zeta_\x^*(\s_0)}f_\S(\s_0) +o_p(1)\right\} (n_1h^{d_s})^{1/2}\{\wh\zeta_\x(\s_0)-\zeta_\x(\s_0)\}.
\ese
From the consistency of $\wh\zeta_\x(\s_0)$, $\wh\zeta_\x(\s_0)-\zeta_\x(\s_0)=o_p(1)$. Then
\be\label{equ:asylinear-semi}
&&(n_1h^{-d_s})^{-1/2}\sumia \phi_\x\{\o_i, \zeta_\x(\s_0), m_\x, 
f_{\epsilon_\x\mid\x}\}K_h(\s_i-\s_0) +o_p(1)\\
&=&-\left[\frac{\partial  E\{\phi_\x(\O_i, \zeta_\x(\s_0), m_\x,
  f_{\epsilon_\x\mid\x})\mid \s_0\}
}{\partial \zeta_\x(\s_0)}f_\S(\s_0) +o_p(1)\right]
(n_1h^{d_s})^{1/2}\{\wh\zeta_\x(\s_0)-\zeta_\x(\s_0)\}.\n
\ee
Note that 
  \be\label{eq:bias-semi}
&&E[n_1^{-1}\sumia  \phi_\x\{\O_i, \zeta_\x(\s_0), m_\x,
f_{\epsilon_\x\mid\x}\}K_h(\S_i-\s_0)] \\
&=&E[\phi_\x\{\O_i,\zeta_\x(\s_0), m_\x, f_{\epsilon_\x\mid\x}\}\mid\s_0]f_\S(\s_0)\notag\\
&+&h^{t}\frac{\partial^t \{E[\phi_\x\{\O_i,
  \zeta_\x(\s), m_\x,f_{\epsilon_\x\mid\x}\}\mid\s]f_{\S}(\s)\}}{\partial \s^t}\mid_{\s=\s_0}\int \u^tK(\u)d\u+o(h^t)\notag\\
  &=&h^{t}\frac{\partial^t E[\phi_\x\{\O_i,\zeta_\x(\s), m_\x, f_{\epsilon_\x\mid\x}\}\mid\s]f_{\S}(\s)}{\partial \s^t}\mid_{\s=\s_0}\int \u^tK(\u)d\u+o(h^t),
\ee
and 
\bse
&&\var[n_1^{-1/2}\sumia\phi_\x\{\O_i, \zeta_\x(\s_0), m_\x, 
f_{\epsilon_\x\mid\x}\}K_h(\S_i-\s_0)] \\
&=&\var[\phi_\x\{\O_i, \zeta_\x(\s_0), m_\x, 
f_{\epsilon_\x\mid\x}\}K_h(\S_i-\s_0)] \\
&=&E[\phi_\x^2\{\O_i, \zeta_\x(\s_0), m_\x, 
f_{\epsilon_\x\mid\x}\}K_h^2(\s_i-\s_0)]+O(h^{2t})\\
&=& h^{-d_s}E[\phi_\x^2\{\O_i, \zeta_\x(\s_0), m_\x,
f_{\epsilon_\x\mid\x}\}\mid\s_0]f_{\S}(\s_0)\int
K^2(\u)d\u+O(h^{2t}+h^{-d_s+2}).
\ese
Thus, $\wh\zeta_\x(\s_0)$ has bias
\bse
b_\x^{\rm semi}&=&-
\left[\frac{\partial  E\{\phi_\x(\O_i, \zeta_\x(\s_0), m_\x,
  f_{\epsilon_\x\mid\x})\mid \s_0\}
}{\partial \zeta_\x(\s_0)}f_\S(\s_0)\right]^{-1}\\
    &&\times h^{t}\frac{\partial^t E[\phi_\x\{\O_i,\zeta_\x(\s),m_\x,f_{\epsilon_\x\mid\x}\}\mid\s]f_{\S}(\s)}{\partial \s^t}\mid_{\s=\s_0}\int \u^tK(\u)d\u+o(h^t),
\ese
Combine with \eqref{equ:asylinear-semi}, we have
\be
\label{equ:asylinear-semi2}
&&(n_1h^{d_s})^{1/2}\{\wh\zeta_\x(\s_0)-\zeta_\x(\s_0)-b_\x^{\rm semi}\}\\
&=&-\frac{(n_1h^{-d_s})^{-1/2}\sumia\bigl\{\phi_\x(\O_i,\zeta_\x(\s_0),m_\x,f_{\epsilon_\x\mid\x})K_h(\S_i-\s_0)-E(\phi_\x K_h)\bigr\}}{ g_{\x,0}f_\S(\s_0)}+o_p(1),\n
\ee
where $E(\phi_\v K_h)=E\{\phi_\v(\O_i,\zeta_\v(\s_0),m_\v,f_{\epsilon_\v\mid\v})K_h(\S_i-\s_0)\}$ and 
\bse
g_{\x,0}\equiv -\frac{\partial  E\{\phi_\x(\O_i, \zeta_\x(\s_0), m_\x,
  f_{\epsilon_\x\mid\x})\mid \s_0\}
}{\partial \zeta_\x(\s_0)}=[f_{\epsilon_\x\mid\s}\{\zeta_\x(\s_0),\s_0\}+f_{\epsilon_\x\mid\s}\{-\zeta_\x(\s_0) ,\s_0\}].
\ese
Further
\bse
(n_1h^{d_s})^{1/2}\{\wh\zeta_\x(\s_0)-\zeta_\x(\s_0)-b_\x^{\rm semi}\}
\to N(0, \sigma_{\x\x}),
\ese
where
\bse
\sigma_{\x\x}=
\left\{g_{\x,0}f_\S(\s_0)\right\}^{-2}
E[\phi_\x^2\{\O_i, \zeta_\x(\s_0), m_\x, f_{\epsilon_\x\mid\x}\}\mid\s_0] f_\S(\s_0)\int
K^2(\u)d\u.
\ese
\qed 

\subsection{Proof of Theorem \ref{th:semi}}
Let $n_j=|\mathcal I_j|$, $j=1,2$.  Apply
Lemma~\ref{lemm:semi-onefold} to each evaluation fold, using the
nuisance estimators trained on the other fold.  For $j=1,2$,
\begin{align*}
&(n_jh^{d_s})^{1/2}
\{\wh\zeta_\v^{(j)}(\s_0)-\zeta_\v(\s_0)-b_\v^{\rm semi}\}\\
&=-\frac{(n_jh^{-d_s})^{-1/2}
\sum_{i\in\mathcal I_j}\bigl\{\phi_\v(\O_i,\zeta_\v(\s_0),m_\v,
f_{\epsilon_\v\mid\v})K_h(\S_i-\s_0)-E(\phi_\v K_h)\bigr\}}
{g_{\v,0}f_\S(\s_0)}+o_p(1).
\end{align*}
Since
$\wh\zeta_\v(\s_0)=\{\wh\zeta_\v^{(1)}(\s_0)
+\wh\zeta_\v^{(2)}(\s_0)\}/2$, averaging the two expansions gives
\begin{align*}
&(nh^{d_s})^{1/2}
\{\wh\zeta_\v(\s_0)-\zeta_\v(\s_0)-b_\v^{\rm semi}\}\\
&=-\frac{(nh^{-d_s})^{-1/2}
\sumi\bigl\{\phi_\v(\O_i,\zeta_\v(\s_0),m_\v,
f_{\epsilon_\v\mid\v})K_h(\S_i-\s_0)-E(\phi_\v K_h)\bigr\}}
{g_{\v,0}f_\S(\s_0)}+o_p(1).
\end{align*}
As in the proof of Theorem~\ref{th:semi0}, the discrepancy caused by
an odd $n$ is $o_p(1)$.  This proves
\eqref{equ:theo1-linear}.  The asymptotic normality follows from the
same kernel-weighted central limit theorem used in
Lemma~\ref{lemm:semi-onefold}.
\qed

\subsection{Proof of Lemma~\ref{lemm:conformal-onefold}}
We prove the result for $\v=\x$; the proof for $\v=\w$ is identical.
The conformal prediction based estimator $\wt\zeta_\x(\s_0)$ satisfies
\bse
\wt\zeta_\x(\s_0)=\argmin_{\zeta\in\mR}\sumia \rho_{1-\alpha}\{\left|y_i-\wh m_\x(\x_i)\right|-\zeta\}K_h(\s_i-\s_0),
\ese
where $\rho_{1-\alpha}(t)\equiv
t\{1-\alpha-I(t<0)\}$. Define  $\psi_{1-\alpha}(t)\equiv 1-\alpha-I(t<0)$ and
\bse
\Psi_n(\zeta,m)\equiv\frac{1}{n_1}\sumia\psi_{1-\alpha}\{\left|y_i-m(\x_i)\right|-\zeta\}K_h(\s_i-\s_0).
\ese

\medskip\noindent\textbf{Step 1 (Consistency).}
Let
\[
Q_n(\zeta)=\sumia \rho_{1-\alpha}\{|y_i-\wh m_\x(\x_i)|-\zeta\}K_h(\s_i-\s_0).
\]
Its left derivative satisfies
\[
Q_{n,-}'(\zeta)
=
-\sumia \psi_{1-\alpha}\{|y_i-\wh m_\x(\x_i)|-\zeta\}K_h(\s_i-\s_0)
=
-n_1\Psi_n(\zeta,\wh m_\x).
\]
Since $K_h\ge0$ and $\rho_{1-\alpha}$ is convex, $Q_n(\zeta)$ is convex in
$\zeta$. Thus, every minimizer $\hat\zeta$ satisfies $Q_{n,-}'(\hat\zeta)\le 0\le Q_{n,+}'(\hat\zeta)$.
Moreover, if $Q_{n,-}'(a)<0$, then no minimizer can lie to the left of
$a$; if $Q_{n,-}'(b)>0$, then no minimizer can lie to the right of $b$.
It is therefore sufficient to show that, for any $0<\varepsilon<c_3$,
\be\label{eq:quantile-bracket}
\Psi_n\{\zeta_\x(\s_0)-\varepsilon,\wh m_\x\}>0,\qquad
\Psi_n\{\zeta_\x(\s_0)+\varepsilon,\wh m_\x\}<0
\ee
with probability approaching 1.

For $\zeta\in(-\zeta_\x(\s_0)-c_3,\,-\zeta_\x(\s_0)+c_3)\cup(\zeta_\x(\s_0)-c_3,\,\zeta_\x(\s_0)+c_3)$, denote $\Delta_n\equiv\sup_\x|\wh m_\x{(\x)}-m_\x{(\x)}|$. 
\bse
&&E\!\left(\left[I\{|Y_i-\wh m_\x(\X_i)|<\zeta\}-I\{|Y_i-m_\x(\X_i)|<\zeta\}\right]^2\,\Big|\, \S_i, D_2\right)\\
&\le& \pr\!\left[\zeta-\Delta_n\le |\epsilon_\x|\le \zeta+\Delta_n\,\Big|\, \S_i, D_2\right]\\
&\le& 2\sup_{\s}\{f_{\epsilon_\x\mid\s}(\xi,\s)+f_{\epsilon_\x\mid\s}(-\xi,\s)\}\Delta_n
\ese
for some $\xi\in[\zeta-\Delta_n,\zeta+\Delta_n]$.
Since $\Delta_n=o_p(1)$,
$\xi$ lies in the $c_3$-neighborhood of $\zeta$ with probability
approaching 1. By the boundedness of $f_{\epsilon_\x\mid\s}$ in Assumption~\ref{ass:lip2},
\be
&&E[\{\Psi_n(\zeta,\wh m_\x)-\Psi_n(\zeta,m_\x)\}^2\mid D_2]\notag\\
&=&n_1^{-1}E\!\left\{\left(\left[I\{|Y_i-\wh m_\x(\X_i)|<\zeta\}-I\{|Y_i-m_\x(\X_i)|<\zeta\}\right]K_h(\S_i-\s_0)\right)^2\,\Big|\, D_2\right\}\notag\\
&+& \frac{n_1(n_1-1)}{n_1^2}E\!\left(\left[I\{|Y_i-\wh m_\x(\X_i)|<\zeta\}-I\{|Y_i-m_\x(\X_i)|<\zeta\}\right]K_h(\S_i-\s_0)\,\Big|\, D_2\right)^2\notag\\
&=&O_p\{\Delta_n(n_1h^{d_s})^{-1}\}+O_p(\Delta_n^2)=o_p(1)\label{eq:11}.
\ee
Then by Lemma~\ref{lemm:conditional},
\be \label{equ:12} \Psi_n(\zeta,\wh
m_\x)-\Psi_n(\zeta,m_\x)\xrightarrow{p}0.
\ee
Moreover, 
\bse
\lim_{h\to 0}E\{\Psi_n(\zeta,m_\x)\}
&=&\lim_{h\to 0}\int\{1-\alpha-\pr(|\epsilon_\x|<\zeta\mid\s_0+h\u)\}
f_\S(\s_0+h\u)K(\u)d\u\\
&=& \{1-\alpha-\pr(|\epsilon_\x|<\zeta\mid\s_0)\}f_\S(\s_0).
\ese 
Also,
\bse
\var\{\Psi_n(\zeta,m_\x)\}
\le \frac{1}{n_1}E\{K_h^2(\S-\s_0)\}
=O\{(n_1h^{d_s})^{-1}\}=o(1).
\ese
Thus,
\be\label{equ:13}\Psi_n(\zeta,m_\x)\xrightarrow{p}\{1-\alpha-\pr(|\epsilon_\x|<\zeta\mid\s_0)\}f_\S(\s_0).
\ee
By the local positivity of
$f_{\epsilon_\x\mid\s_0}$ in Assumption~\ref{ass:lip2},
\bse
\pr\{|\epsilon_\x|<\zeta_\x(\s_0)-\varepsilon\mid\s_0\}<1-\alpha<
\pr\{|\epsilon_\x|<\zeta_\x(\s_0)+\varepsilon\mid\s_0\}.
\ese
Combining \eqref{equ:12} and \eqref{equ:13}, we obtain
\eqref{eq:quantile-bracket}.

The sign change in \eqref{eq:quantile-bracket}, together with
convexity of 
the weighted check-loss objective, implies
$\pr\{\zeta_\x(\s_0)-\varepsilon\le\wt\zeta_\x(\s_0)\le
\zeta_\x(\s_0)+\varepsilon\}\to1$. Since $\varepsilon>0$ is arbitrary,
\be\label{eq:tilde-cons}
\wt\zeta_\x(\s_0)\xrightarrow{p}\zeta_\x(\s_0).
\ee

\medskip\noindent\textbf{Step 2 (Asymptotic normality).}
We use the identity (\citealp{knight1998limiting})
\be\label{eq:knight}
\rho_{1-\alpha}(u-v)-\rho_{1-\alpha}(u)=-v\,\psi_{1-\alpha}(u)+\int_0^v\{I(u\le s)-I(u\le 0)\}ds,
\ee
which holds for all $u,v\in\mR$. 
Let $\wh\epsilon_{\x,i}=y_i-\wh m_\x(\x_i)$. Then applying \eqref{eq:knight} with $u=|\wh\epsilon_{\x,i}|-\zeta_\x(\s_0)$ and $v=\zeta-\zeta_\x(\s_0)$,
\bse
Z_n(\zeta)&\equiv& h^{d_s}\sumia\left[\rho_{1-\alpha}\left\{|y_i-\wh m_\x(\x_i)|-\zeta\right\}-\rho_{1-\alpha}\{|y_i-\wh m_\x(\x_i)|-\zeta_\x(\s_0)\}\right]K_h(\s_i-\s_0)\\
&=&h^{d_s}\sumia\left[\rho_{1-\alpha}\left(|\wh \epsilon_{\x,i}|-\zeta\right)-\rho_{1-\alpha}\{|\wh\epsilon_{\x,i}|-\zeta_\x(\s_0)\}\right]K_h(\s_i-\s_0)
\\
&=&-(n_1h^{d_s})\Psi_n\{\zeta_\x(\s_0),\wh m_\x\}\{\zeta-\zeta_\x(\s_0)\}\\
&&+h^{d_s}\sumia K_h(\s_i-\s_0)\int_0^{\zeta-\zeta_\x(\s_0)}I\{|\wh \epsilon_{\x,i}|\le \zeta_\x(\s_0)+c\}-I\{|\wh \epsilon_{\x,i}|\le \zeta_\x(\s_0)\}dc.
\ese
First, we prove for fixed $(n_1h^{d_s})^{1/2}\{\zeta-\zeta_\x(\s_0)\}$,
\be
\label{eq:bi}
&& B_n\equiv h^{d_s}\sumia K_h(\s_i-\s_0)\int_0^{\zeta-\zeta_\x(\s_0)}I\{|\wh \epsilon_{\x,i}|\le \zeta_\x(\s_0)+c\}-I\{|\wh \epsilon_{\x,i}|\le \zeta_\x(\s_0)\}dc\notag\\
&&\xrightarrow{p}\frac{1}{2}g_{\x,0} f_\S(\s_0)(n_1h^{d_s})\{\zeta-\zeta_\x(\s_0)\}^2,
\ee
where $g_{\x,0}=f_{\epsilon_\x\mid\s}\{\zeta_\x(\s_0),\s_0\}+f_{\epsilon_\x\mid\s}\{-\zeta_\x(\s_0),\s_0\}$.
We exchange sum and integral. Then,
\bse
E\{B_n\mid D_2\}=n_1h^{d_s}\int\int_0^{\zeta-\zeta_\x(\s_0)}\pr\{\zeta_\x(\s_0)\le|\wh\epsilon_{\x,i}|<\zeta_\x(\s_0)+c\mid \s,D_2\}dc K_h(\s-\s_0)f_\S(\s)d\s,
\ese
By the boundedness of $f_{\epsilon_\x\mid\s}$ in Assumption~\ref{ass:lip2}, uniformly for $|c| \le |\zeta-\zeta_\x(\s_0)|< c_3$,
\bse
&&\left|\pr\{|Y_i-\wh m_\x(\X_i)|<\zeta_\x(\s_0)+c\mid \s,D_2\}-\pr\{|Y_i- m_\x(\X_i)|<\zeta_\x(\s_0)+c\mid \s,D_2\}\right|\\
&\leq& 2L'\sup_\x|\wh m_\x-m_\x|=o_p(1).
\ese
Thus, 
uniformly in $|c|\le |\zeta-\zeta_\x(\s_0)|$,
\bse
E\{B_n\mid D_2\}=n_1h^{d_s}\int\int_0^{\zeta-\zeta_\x(\s_0)}[\pr\{\zeta_\x(\s_0)\le|\epsilon_{\x,i}|<\zeta_\x(\s_0)+c\mid \s\}+o_p(1)] dc K_h(\s-\s_0)f_\S(\s)d\s,
\ese
By Taylor expansion and $|\zeta-\zeta_\x(\s_0)|=o_p(1)$ for fixed $(n_1h^{d_s})^{1/2}\{\zeta-\zeta_\x(\s_0)\}$,
\bse
&&\pr\{\zeta_\x(\s_0)\le|\epsilon_{\x,i}|<\zeta_\x(\s_0)+c\mid \s\}
=[f_{\epsilon_\x\mid\s}\{\zeta_\x(\s_0),\s\}+f_{\epsilon_\x\mid\s}\{-\zeta_\x(\s_0),\s\}]c+o(c).
\ese
It follows that
\bse
E\{B_i\mid D_2\}=\frac{1}{2}g_{\x,0} f_\S(\s_0)(n_1h^{d_s})\{\zeta-\zeta_\x(\s_0)\}^2+o(1).
\ese
By $\var(X\mid D_2)\le E(X^2\mid D_2)$ and Cauchy--Schwarz inequality,
\bse
&&(n_1h^{2d_s})^{-1}\var\{B_n\mid D_2\}\\&=&\var\!\left\{K_h(\S_i-\s_0)\!\int_0^{\zeta-\zeta_\x(\s_0)}I\{|\wh \epsilon_{\x,i}|\le \zeta_\x(\s_0)+c\}-I\{|\wh \epsilon_{\x,i}|\le \zeta_\x(\s_0)\}\,dc\,\bigg|\,D_2\right\}\\
&\le&E\!\left[ K_h^2(\S_i-\s_0)\left\{\!\int_0^{\zeta-\zeta_\x(\s_0)}I\{|\wh \epsilon_{\x,i}|\le \zeta_\x(\s_0)+c\}-I\{|\wh \epsilon_{\x,i}|\le \zeta_\x(\s_0)\}\,dc\right\}^{\!2}\,\bigg|\,D_2\right]\\
&\le& |\zeta-\zeta_\x(\s_0)|\\&&\times E\!\left[ K_h^2(\S_i-\s_0)\!\int_0^{|\zeta-\zeta_\x(\s_0)|}\!\!\big[I\{|\wh \epsilon_{\x,i}|\le \zeta_\x(\s_0)+c\}-I\{|\wh \epsilon_{\x,i}|\le \zeta_\x(\s_0)\}\big]^2\,dc\,\bigg|\,D_2\right].
\ese
For any $c\ge 0$, $\{I(t<c)-I(t<0)\}^2=I(0\le t<c)\le I(|t|\le c)$, and analogously for $-c$, so $$\{I\{|\wh \epsilon_{\x,i}|\le \zeta_\x(\s_0)\pm c\}-I\{|\wh \epsilon_{\x,i}|\le \zeta_\x(\s_0)\}\}^2\le I\{|\wh \epsilon_{\x,i}|-\zeta_\x(\s_0)|\le c\}.$$ Hence,
\bse
&&E\!\left[\!\int_0^{|\zeta-\zeta_\x(\s_0)|}\!\!\big[I\{|\wh \epsilon_{\x,i}|\le \zeta_\x(\s_0)+c\}-I\{|\wh \epsilon_{\x,i}|\le \zeta_\x(\s_0)\}\big]^2\,dc\,\bigg|\,\S_i,D_2\right]\\
&\le&\int_0^{|\zeta-\zeta_\x(\s_0)|}\!\!\pr\!\left\{\bigl||\wh\epsilon_{\x,i}|-\zeta_\x(\s_0)\bigr|\le c\,\bigg|\,\S_i,D_2\right\}\,dc\\
&=&\int_0^{|\zeta-\zeta_\x(\s_0)|}\!\!\pr\!\left\{\zeta_\x(\s_0)-c\le|\wh \epsilon_{\x,i}|\le \zeta_\x(\s_0)+c\,\bigg|\,\S_i,D_2\right\}\,dc\\
&\le& \int_0^{|\zeta-\zeta_\x(\s_0)|}\!\!\pr\!\left\{\zeta_\x(\s_0)-c-\Delta_n\le|\epsilon_\x|\le \zeta_\x(\s_0)+c+\Delta_n\,\bigg|\,\S_i\right\}\,dc \\
&=&\int_0^{|\zeta-\zeta_\x(\s_0)|} 2\{f_{\epsilon_\x\mid\s}(\xi',\S_i)+f_{\epsilon_\x\mid\s}(\xi'',\S_i)\}(c+\Delta_n)\,dc\\
&\le&\int_0^{|\zeta-\zeta_\x(\s_0)|}\!\!4L'(c+\Delta_n)\,dc\;\le\; 2L'\{\zeta-\zeta_\x(\s_0)\}^2+4L'\Delta_n|\zeta-\zeta_\x(\s_0)|,
\ese
 where $\Delta_n=\sup_\x|\wh m_\x(\x)-m_\x(\x)|=o_p\{(nh^{d_s})^{-1/2}\}$ and
 $|\xi'-\zeta_\x(\s_0)|\vee|\xi''+\zeta_\x(\s_0)|<|\zeta-\zeta_\x(\s_0)|+\Delta_n=o_p(1)$. 
   In  the above, we used the fact that $\xi'$ and $\xi''$ lies in the
 $c_3$-neighborhood of $\zeta_\x(\s_0)$ when $n\to\infty$ with
 probability approaching 1, hence by Assumption~\ref{ass:lip2},
 $f_{\epsilon_\x\mid\s}(\xi',\S_i)+f_{\epsilon_\x\mid\s}(\xi'',\S_i)<2L'$.
Thus, 
\bse
\var\{B_n\mid D_2\}
=O_p\left[(n_1h^{d_s})\left\{|\zeta-\zeta_\x(\s_0)|^3 +\{\zeta-\zeta_\x(\s_0)\}^2\Delta_n\right\}\right]=o_p(1).
\ese
Hence, \eqref{eq:bi} has been proved. 
And
\bse
Z_n(\zeta)=-(n_1h^{d_s})\Psi_n\{\zeta_\x(\s_0),\wh m_\x\}\{\wt\zeta_\x(\s_0)-\zeta_\x(\s_0)\}+(n_1h^{d_s}) \frac{1}{2}g_{\x,0} f_\S(\s_0)\{\wt\zeta_\x(\s_0)-\zeta_\x(\s_0)\}^2+o_p(1) .
\ese
Nota that $\wt\zeta_\x(\s_0)=\argmin_{\zeta\in\mR} Z_n(\zeta)$.
Applying the Basic Corollary of \cite{hjort2011asymptotics} with $s=(n_1h^{d_s})^{1/2}\{\zeta-\zeta_\x(\s_0)\}$ to $Z_n(\zeta)$,
\bse\label{eq:wttau-rep}
&&(n_1h^{d_s})^{1/2}\{\wt\zeta_\x(\s_0)-\zeta_\x(\s_0)\}=\frac{(n_1h^{d_s})^{1/2}\Psi_n\{\zeta_\x(\s_0),\wh m_\x\}}{g_{\x,0} f_\S(\s_0)}+o_p(1).
\ese
Note that
\bse
&&E[\Psi_n\{\zeta_\x(\s_0),m_\x\}] \\
&=&E\left[\psi_{1-\alpha}\{|Y-m_\x(\X)|-\zeta_\x(\s_0)\}\mid\s_0\right] f_\S(\s_0)\\&&+h^{t}\frac{\partial^t E[\psi_{1-\alpha}\{\left|Y_i-m_\x(\X_i)\right|-\zeta_{\x}(\s_0)\}\mid\s]f_{\S}(\s)}{\partial \s^t}\mid_{\s=\s_0}\int \u^tK(\u)d\u+o(h^t)\\
  &=&h^{t}\frac{\partial^t E[\psi_{1-\alpha}\{\left|Y_i-m_\x(\X_i)\right|-\zeta_{\x}(\s_0)\}\mid\s]f_{\S}(\s)}{\partial \s^t}\mid_{\s=\s_0}\int \u^tK(\u)d\u+o(h^t),
\ese
and 
\bse
&&\var[\Psi_n\{\zeta_\x(\s_0),m_\x\}] \\
&=&n_1^{-1}\var\left[\psi_{1-\alpha}\{\left|Y_i-m_\x(\X_i)\right|-\zeta_{\x}(\s_0)\}K_h(\S_i-\s_0)\right] \\
&=&n_1^{-1} E\left[\psi_{1-\alpha}^2\{\left|Y_i-m_\x(\X_i)\right|-\zeta_\x(\s_0)\}K_h^2(\S_i-\s_0)\right]+O(n_1^{-1}h^{2t})\\
&=& (n_1 h^{d_s})^{-1}\alpha(1-\alpha)f_{\S}(\s_0)\int
K^2(\u)d\u+O\{n_1^{-1}(h^{2t}+h^{-d_s+2})\}.
\ese
Thus,
\bse\label{eq:linear-CLT-m}
(n_1h^{d_s})^{1/2}[\Psi_n\{\zeta_\x(\s_0),m_\x\}-b_n]\xrightarrow{d}N(0,\sigma_0^2),
\ese
with
\bse
\sigma_0^2&=&\alpha(1-\alpha)\,f_\S(\s_0)\!\int K^2(\u)d\u,\\
b_n&=&h^{t}\frac{\partial^t \left(E[\psi_{1-\alpha}\{|Y_i-m_\x(\X_i)|-\zeta_{\x}(\s_0)\}\mid\s]f_{\S}(\s)\right)}{\partial \s^t}\bigg|_{\s=\s_0}\int \u^tK(\u)d\u+o(h^t).
\ese
By \eqref{eq:11} and $ \Delta_n=o_p\{(nh^{d_s})^{-1/2}\}$, $(n_1h^{d_s})^{1/2}[\Psi_n\{\zeta_\x(\s_0),\wh m_\x\}-\Psi_n\{\zeta_\x(\s_0),m_\x\}]=o_p(1)$, so by Slutsky's theorem,
\be\label{eq:linear-CLT}
(n_1h^{d_s})^{1/2}[\Psi_n\{\zeta_\x(\s_0),\wh m_\x\}-b_n]\xrightarrow{d}N(0,\sigma_0^2).
\ee
Thus, $\wt\zeta_\x(\s_0)$ has bias
\bse
b_\x^{\rm conf}\equiv \{g_{\x,0} f_\S(\s_0)\}^{-1} h^{t}\frac{\partial^t\bigl[E\{\psi_{1-\alpha}(\left|\epsilon_\x\right|-\zeta_\x(\s_0))\mid\s\}f_\S(\s)\bigr]}{\partial \s^t}\bigg|_{\s=\s_0}\!\!\int \u^t K(\u)d\u+o(h^t).
\ese
And, 
\be
\label{eq:conf-linear2}
&&(n_1h^{d_s})^{1/2}\{\wt\zeta_\x(\s_0)-\zeta_\x(\s_0)-b_\x^{\rm conf}\}\notag\\
&=&\frac{(n_1h^{d_s})^{1/2}[\Psi_n\{\zeta_\x(\s_0),\wh m_\x\}-b_n]}{g_{\x,0} f_\S(\s_0)}+o_p(1)\notag\\
&=&\frac{(n_1h^{-d_s})^{-1/2}\sumia\bigl[\psi_{1-\alpha}\{|Y_i-m_\x(\X_i)|-\zeta_\x(\s_0)\} K_h(\S_i-\s_0)-b_n\bigr]}{g_{\x,0}f_\S(\s_0)}+o_p(1),
\ee
because of $b_n=E\bigl[\psi_{1-\alpha}\{|Y_i-m_\x(\X_i)|-\zeta_\x(\s_0)\} K_h(\S_i-\s_0)\bigr]$.

Moreover,
\be\label{eq:tilde-AN}
(n_1h^{d_s})^{1/2}\{\wt\zeta_\x(\s_0)-\zeta_\x(\s_0)-b_\x^{\rm conf}\}\xrightarrow{d}N\left(0,\;\{g_{\x,0} f_\S(\s_0)\}^{-2}\,\alpha(1-\alpha)f_\S(\s_0)\!\int K^2(\u)d\u\right).
\ee
\qed

\subsection{Proof of Theorem~\ref{th:conformal}}
Let $n_j=|\mathcal I_j|$, $j=1,2$.  Apply
Lemma~\ref{lemm:conformal-onefold} to each evaluation fold, using
$\wh m_\v^{(-j)}$ trained on the other fold.  For $j=1,2$,
\begin{align*}
&(n_jh^{d_s})^{1/2}
\{\wt\zeta_\v^{(j)}(\s_0)-\zeta_\v(\s_0)-b_\v^{\rm conf}\}\\
&=\frac{(n_jh^{-d_s})^{-1/2}
\sum_{i\in\mathcal I_j}\bigl[\psi_{1-\alpha}\{|\epsilon_{\v,i}|-\zeta_\v(\s_0)\}
K_h(\S_i-\s_0)-E(\psi_{\v,1-\alpha}K_h)\bigr]}
{g_{\v,0}f_\S(\s_0)}+o_p(1).
\end{align*}
Since
$\wt\zeta_\v(\s_0)=\{\wt\zeta_\v^{(1)}(\s_0)
+\wt\zeta_\v^{(2)}(\s_0)\}/2$, averaging the two expansions gives
\begin{align*}
&(nh^{d_s})^{1/2}
\{\wt\zeta_\v(\s_0)-\zeta_\v(\s_0)-b_\v^{\rm conf}\}\\
&=\frac{(nh^{-d_s})^{-1/2}
\sumi\bigl[\psi_{1-\alpha}\{|\epsilon_{\v,i}|-\zeta_\v(\s_0)\}
K_h(\S_i-\s_0)-E(\psi_{\v,1-\alpha}K_h)\bigr]}
{g_{\v,0}f_\S(\s_0)}+o_p(1).
\end{align*}
As in the proof of Theorem~\ref{th:semi0}, the discrepancy caused by
an odd $n$ is $o_p(1)$.  This proves
\eqref{equ:theo2-linear}.  The asymptotic normality follows from the
same kernel-weighted central limit theorem used in
Lemma~\ref{lemm:conformal-onefold}.
\qed

\subsection{Proofs of Corollaries~\ref{th:semi-delta-uncondi}, \ref{th:conf-delta-uncondi}, \ref{th:semi-delta}, and \ref{th:conformal-delta}}\label{sec:proof-delta}

\paragraph{Proof of Corollary~\ref{th:semi-delta-uncondi}.}
By \eqref{equ:theo1-linear0} in Theorem~\ref{th:semi0}, applied jointly to
$\v=\x$ and $\v=\w$,
\bse
n^{1/2}\begin{pmatrix}\wh\zeta_\x-\zeta_\x\\
\wh\zeta_\w-\zeta_\w\end{pmatrix}
=-n^{-1/2}\sumi
\begin{pmatrix}
g_\x^{-1}\phi_\x(\O_i,\zeta_\x,m_\x,f_{\epsilon_\x\mid\x})\\
g_\w^{-1}\phi_\w(\O_i,\zeta_\w,m_\w,f_{\epsilon_\w\mid\w})
\end{pmatrix}+o_p(1).
\ese
The multivariate central limit theorem therefore gives a centered normal
limit whose $(\v_1,\v_2)$ entry is $\sigma_{\v_1\v_2}^0$.
Taking the contrast $(1,-1)$ and using
$\wh\Delta=\wh\zeta_\x-\wh\zeta_\w$ proves the result.
\qed

\paragraph{Proof of Corollary~\ref{th:conf-delta-uncondi}.}
By \eqref{equ:theo2-linear0} in Theorem~\ref{th:conformal0}, for each
$\v\in\{\x,\w\}$,
\bse
n^{1/2}(\wt\zeta_\v-\zeta_\v)
=-g_\v^{-1}n^{1/2}\overline B_\v
+g_\v^{-1}n^{-1/2}\sumi
\psi_{1-\alpha}(|\epsilon_{\v,i}|-\zeta_\v)
+O_p(\overline a_{\v,n}^2n^{1/2})+o_p(1).
\ese
Subtracting the representation for $\v=\w$ from that for $\v=\x$ and
using $\wt\Delta=\wt\zeta_\x-\wt\zeta_\w$ and
$\Delta=\zeta_\x-\zeta_\w$, we obtain
\begin{align*}
n^{1/2}(\wt\Delta-\Delta)
={}&-n^{1/2}\{g_\x^{-1}\overline B_\x-g_\w^{-1}\overline B_\w\}\\
&+n^{-1/2}\sumi\left[
g_\x^{-1}\psi_{1-\alpha}(|\epsilon_{\x,i}|-\zeta_\x)
-g_\w^{-1}\psi_{1-\alpha}(|\epsilon_{\w,i}|-\zeta_\w)
\right]\\
&+O_p\left[n^{1/2}\{\overline a_{\x,n}^2+\overline a_{\w,n}^2\}\right]+o_p(1),
\end{align*}
which is \eqref{equ:theo-conf-delta-linear0}.
\qed

\paragraph{Proof of Corollary~\ref{th:semi-delta}.}
First, we prove
\be
\label{eq:joint-semi}
(nh^{d_s})^{1/2}\!\left\{\!\begin{pmatrix}\wh\zeta_\x(\s_0)\\\wh\zeta_\w(\s_0)\end{pmatrix}-\begin{pmatrix}\zeta_\x(\s_0)\\\zeta_\w(\s_0)\end{pmatrix}-\begin{pmatrix}b_\x^{\rm semi}\\ b_\w^{\rm semi}\end{pmatrix}\!\right\}\xrightarrow{d}N\!\left(\!\0,\;\bSigma^{\rm semi}\right),
\ee
with
\bse
\bSigma^{\rm semi}=\frac{\int K^2(\u)d\u}{f_\S(\s_0)}\begin{pmatrix}\sigma_{\x\x}&\sigma_{\x\w}\\\sigma_{\x\w}&\sigma_{\w\w}\end{pmatrix},
\ese
where for $\v_i\in\{\x,\w\}$, $i=1,2$, 
\bse \sigma_{\v_1\v_2}\equiv E[\phi_{\v_1}\{\O_i, \zeta_{\v_1}(\s_0),m_{\v_1},f_{\epsilon_{\v_1}\mid\v_1}\}\phi_{\v_2}\{\O_i, \zeta_{\v_2}(\s_0),m_{\v_2},f_{\epsilon_{\v_2}\mid\v_2}\}\mid\s_0]/(g_{\v_1,0}g_{\v_2,0}).\ese

Recall \eqref{equ:theo1-linear} in Theorem~\ref{th:semi},
\bse
&&(nh^{d_s})^{1/2}\{\wh\zeta_\v(\s_0)-\zeta_\v(\s_0)-b_\v^{\rm semi}\}\\
&=&-\frac{(nh^{-d_s})^{-1/2}\sumi\bigl\{\phi_\v(\O_i,\zeta_\v(\s_0),m_\v,f_{\epsilon_\v\mid\v})K_h(\S_i-\s_0)-E(\phi_\v K_h)\bigr\}}{ g_{\v,0}f_\S(\s_0)}+o_p(1),
\ese
where $E(\phi_\v K_h)=E\{\phi_\v(\O_i,\zeta_\v(\s_0),m_\v,f_{\epsilon_\v\mid\v})K_h(\S_i-\s_0)\}$.

Stack the two representations for $\v=\x$ and $\v=\w$ into a $2$-vector. The leading
term is a kernel-weighted i.i.d.\ sum over $i\in\{1,\ldots,n\}$,
\bse
\label{eq:semi-linear}
&&(nh^{d_s})^{1/2}\!\left\{\!\begin{pmatrix}\wh\zeta_\x(\s_0)\\\wh\zeta_\w(\s_0)\end{pmatrix}-\begin{pmatrix}\zeta_\x(\s_0)\\\zeta_\w(\s_0)\end{pmatrix}-\begin{pmatrix}b_\x^{\rm semi}\\ b_\w^{\rm semi}\end{pmatrix}\!\right\}\notag\\
&=&-(nh^{-d_s})^{-1/2}\sumi \!\left[\!\begin{matrix}\{g_{\x,0}f_\S(\s_0)\}^{-1}\{\phi_\x(\O_i,\zeta_\x(\s_0),m_\x,f_{\epsilon_\x\mid\x})K_h(\S_i-\s_0)-E(\phi_\x K_h)\}\\\{g_{\w,0}f_\S(\s_0)\}^{-1}\{\phi_\w(\O_i,\zeta_\w(\s_0),m_\w,f_{\epsilon_\w\mid\w})K_h(\S_i-\s_0)-E(\phi_\w K_h)\}\end{matrix}\!\right]+o_p(1).
\ese
Note that, for $\v_1,\v_2\in\{\x,\w\}$,
\be
\label{eq:cov-semi}
&&\cov\!\left\{\frac{\phi_{\v_1}}{g_{\v_1,0}f_\S(\s_0)}K_h(\S_i\!-\!\s_0),\;\frac{\phi_{\v_2}}{ g_{\v_2,0}f_\S(\s_0)}K_h(\S_i\!-\!\s_0)\right\}\notag\\
&=&\frac{E\!\left\{\phi_{\v_1}\phi_{\v_2}K_h^2(\S-\s_0)\right\}}{g_{\v_1,0}g_{\v_2,0}f_{\S}^2(\s_0)}+O(h^{2t})\notag\\
&=&\frac{h^{-d_s}\int K^2(\u)d\u}{g_{\v_1,0}g_{\v_2,0}f_\S(\s_0)}E\{\phi_{\v_1}\phi_{\v_2}\mid\s_0\}+O(h^{-d_s+2}+h^{2t})
\ee
where we abbreviate $\phi_{\v}\{\O_i,\zeta_\v(\s_0),m_\v,f_{\epsilon_\v\mid\v}\}$ as $\phi_{\v}$.
Thus, 
\bse
&&\var\bigl((nh^{-d_s})^{-1/2}\sumi \!\left[\!\begin{matrix}\{g_{\x,0}f_\S(\s_0)\}^{-1}\{\phi_\x(\O_i,\zeta_\x(\s_0),m_\x,f_{\epsilon_\x\mid\x})K_h(\S_i-\s_0)-E(\phi_\x K_h)\}\\\{g_{\w,0}f_\S(\s_0)\}^{-1}\{\phi_\w(\O_i,\zeta_\w(\s_0),m_\w,f_{\epsilon_\w\mid\w})K_h(\S_i-\s_0)-E(\phi_\w K_h)\}\end{matrix}\!\right] \bigr)\\&=&\bSigma^{\rm semi}+O(h^2+h^{2t+d_s}).
\ese
Then \eqref{eq:joint-semi} has been proved. From the joint distribution, we can directly get limit for
$\wh\Delta(\s_0)=\wh \zeta_\x(\s_0)-\wh \zeta_\w(\s_0)$:
\bse
(nh^{d_s})^{1/2}\{\wh\Delta(\s_0)-\Delta(\s_0)-(b_\x^{\rm semi}-b_\w^{\rm semi})\}\xrightarrow{d}N(0,v_\Delta^{\rm semi}),
\ese
with \bse v_\Delta^{\rm semi}=(\sigma_{\x\x}+\sigma_{\w\w}-2\sigma_{\x\w})\int K^2(\u)d\u/f_\S(\s_0).\ese
\qed

\paragraph{Proof of Corollary~\ref{th:conformal-delta}.}
First, we prove
\be
\label{eq:joint-conf}
(nh^{d_s})^{1/2}\!\left\{\!\begin{pmatrix}\wt\zeta_\x(\s_0)\\\wt\zeta_\w(\s_0)\end{pmatrix}-\begin{pmatrix}\zeta_\x(\s_0)\\\zeta_\w(\s_0)\end{pmatrix}-\begin{pmatrix}b_\x^{\rm conf}\\ b_\w^{\rm conf}\end{pmatrix}\!\right\}\xrightarrow{d}N\!\left(\!\0,\;\bSigma^{\rm conf}\right),
\ee
with
\bse
\bSigma^{\rm conf}=\frac{\int K^2(\u)d\u}{f_\S(\s_0)}\begin{pmatrix}\dfrac{\alpha(1-\alpha)}{g_{\x,0}^{2}}&\dfrac{\rho_0}{g_{\x,0}g_{\w,0}}\\[6pt]\dfrac{\rho_0}{g_{\x,0}g_{\w,0}}&\dfrac{\alpha(1-\alpha)}{g_{\w,0}^{2}}\end{pmatrix},
\ese
and
\bse
\rho_0\equiv\cov\bigl(I\{|\epsilon_\x|<\zeta_\x(\s_0)\},\;I\{|\epsilon_\w|<\zeta_\w(\s_0)\}\bigm|\s_0\bigr).
\ese
Recall \eqref{equ:theo2-linear} in Theorem~\ref{th:conformal},
\bse
&&(nh^{d_s})^{1/2}\{\wt\zeta_\v(\s_0)-\zeta_\v(\s_0)-b_\v^{\rm conf}\}\notag\\
&=&\frac{(nh^{-d_s})^{-1/2}\sumi\bigl[\psi_{1-\alpha}\{|\epsilon_{\v,i}|-\zeta_\v(\s_0)\} K_h(\S_i-\s_0)-E(\psi_{\v,1-\alpha}K_h)\bigr]}{g_{\v,0}f_\S(\s_0)}+o_p(1),
\ese
where $E(\psi_{\v,1-\alpha}K_h)=E\bigl[\psi_{1-\alpha}\{|\epsilon_{\v,i}|-\zeta_\v(\s_0)\} K_h(\S_i-\s_0)\bigr]$ and $\epsilon_{\v,i}=Y_i-m_\v(\V_i)$. Stack the two representations for $\v=\x$ and $\v=\w$ into a $2$-vector.
The leading term is a kernel-weighted i.i.d.\ sum over $i\in\{1,\ldots,n\}$,
\bse
\label{eq:conf-linear}
&&(nh^{d_s})^{1/2}\!\left\{\!\begin{pmatrix}\wt\zeta_\x(\s_0)\\\wt\zeta_\w(\s_0)\end{pmatrix}-\begin{pmatrix}\zeta_\x(\s_0)\\\zeta_\w(\s_0)\end{pmatrix}-\begin{pmatrix}b_\x^{\rm conf}\\ b_\w^{\rm conf}\end{pmatrix}\!\right\}\notag\\
&=&(nh^{-d_s})^{-1/2}\sumi \!\left[\!\begin{matrix}\{g_{\x,0}f_\S(\s_0)\}^{-1}\{\psi_{1-\alpha}\{|\epsilon_{\x,i}|-\zeta_\x(\s_0)\}K_h(\S_i-\s_0)-E(\psi_{\x,1-\alpha} K_h)\}\\\{g_{\w,0}f_\S(\s_0)\}^{-1}\{\psi_{1-\alpha}\{|\epsilon_{\w,i}|-\zeta_\w(\s_0)\}K_h(\S_i-\s_0)-E(\psi_{\w,1-\alpha} K_h)\}\end{matrix}\!\right]+o_p(1).
\ese
Note that, 
\bse
\label{eq:cov-conf}
&&\cov\!\left[\frac{\psi_{1-\alpha}\{|\epsilon_{\x}|-\zeta_\x(\s_0)\}}{g_{\x,0}(\s_0)f_\S(\s_0)}K_h(\S_i\!-\!\s_0),\frac{\psi_{1-\alpha}\{|\epsilon_{\w}|-\zeta_\w(\s_0)\}}{g_{\w,0}f_\S(\s_0)}K_h(\S_i\!-\!\s_0)\right]\notag\\
&=&\frac {E\!\left[\psi_{1-\alpha}\{|\epsilon_{\x}|-\zeta_\x(\s_0)\}\psi_{1-\alpha}\{|\epsilon_{\w}|-\zeta_\w(\s_0)\}K_h^2(\S-\s_0)\right]}{g_{\x,0}g_{\w,0}f_{\S}^2(\s_0)}+O(h^{2t})\notag\\
&=&\frac{h^{-d_s}E[\psi_{1-\alpha}\{|\epsilon_{\x}|-\zeta_\x(\s_0)\}\psi_{1-\alpha}\{|\epsilon_{\w}|-\zeta_\w(\s_0)\}\mid\s_0]\int K^2(\u)d\u}{g_{\x,0}g_{\w,0}f_\S(\s_0)}+O(h^{-d_s+2}+h^{2t})\notag\\
&=&\frac{\rho_0h^{-d_s}\int K^2(\u)d\u}{g_{\x,0}g_{\w,0}f_\S(\s_0)}+O(h^{-d_s+2}+h^{2t});
\ese
and for $\v\in\{\x,\w\}$,
\bse
\var\!\left[\frac{\psi_{1-\alpha}\{|\epsilon_{\v}|-\zeta_\v(\s_0)\}}{g_{\v,0}(\s_0)f_\S(\s_0)}K_h(\S_i\!-\!\s_0)\right]=\frac{h^{-d_s}\alpha(1-\alpha)\int K^2(\u)d\u}{g_{\v,0}^2f_\S(\s_0)}+O(h^{-d_s+2}+h^{2t}),
\ese
where $\epsilon_\v=Y-m_\v(\V)$.
Thus,
\bse
&&\var\bigl((nh^{-d_s})^{-1/2}\sumi \!\left[\!\begin{matrix}\{g_{\x,0}f_\S(\s_0)\}^{-1}\{\psi_{1-\alpha}\{|\epsilon_{\x,i}|-\zeta_\x(\s_0)\}K_h(\S_i-\s_0)-E(\psi_{\x,1-\alpha} K_h)\}\\\{g_{\w,0}f_\S(\s_0)\}^{-1}\{\psi_{1-\alpha}\{|\epsilon_{\w,i}|-\zeta_\w(\s_0)\}K_h(\S_i-\s_0)-E(\psi_{\w,1-\alpha} K_h)\}\end{matrix}\!\right] \bigr)\\&=&\bSigma^{\rm conf}+O(h^2+h^{2t+d_s}).
\ese
Then \eqref{eq:joint-conf}
has been proved. From the joint distribution, we can directly get limit for
$\wt\Delta(\s_0)=\wt \zeta_\x(\s_0)-\wt \zeta_\w(\s_0)$:
\bse
(nh^{d_s})^{1/2}\{\wt\Delta(\s_0)-\Delta(\s_0)-(b_\x^{\rm conf}-b_\w^{\rm conf})\}\xrightarrow{d}N(0,v_\Delta^{\rm conf}),
\ese
with
\bse
v_\Delta^{\rm conf}=\frac{\int K^2(\u)d\u}{f_\S(\s_0)}\!\left\{\frac{\alpha(1-\alpha)}{g_{\x,0}^{2}}+\frac{\alpha(1-\alpha)}{g_{\w,0}^{2}}-\frac{2\rho_0}{g_{\x,0}g_{\w,0}}\right\}.
\ese
\qed

\subsection{Proof of Proposition~\ref{pro:var-cons}}
Based on the construction of $\wh v_\Delta^{\rm semi}$ and
$\wh v_\Delta^{\rm conf}$, we just need to prove
\bse
\wh f_{\S}(\s_0)\xrightarrow{p}f_{\S}(\s_0),\quad
(\wh g_{\v,0})^{-1}\xrightarrow{p}g_{\v,0}^{-1},\quad
\wh\sigma_{\v_1\v_2}\xrightarrow{p}\sigma_{\v_1\v_2},\quad
\wh\rho_0\xrightarrow{p}\rho_0,
\ese
where $\v,\v_1,\v_2\in\{\x,\w\}$.

First, for $j=1,2$,
\bse
E\left\{\frac1{n_j}\sum_{i\in\mathcal I_j}K_h(\S_i-\s_0)\right\}
&=&\int K_h(\s-\s_0)f_\S(\s)d\s
=f_\S(\s_0)+o(1),
\ese
and
\bse
\var\left\{\frac1{n_j}\sum_{i\in\mathcal I_j}K_h(\S_i-\s_0)\right\}
&=&O\{(n_jh^{d_s})^{-1}\}=o(1).
\ese
Averaging the two folds, we have
\be\label{cons-f}
\wh f_\S(\s_0)\xrightarrow{p}f_\S(\s_0).
\ee

Second,
\begin{align*}
(\wh g_{\v,0})^{-1}
= \frac12\sum_{j=1}^2\left\{
\frac{\zeta_\v^{+}-\zeta_\v^{-}}{2c_n}
+\frac{\wt\zeta_\v^{ (j),+}-\zeta_\v^{+}}{2c_n}
-\frac{\wt\zeta_\v^{(j),-}-\zeta_\v^{-}}{2c_n}
\right\},
\end{align*}
where
$\pr\{|\epsilon_\v|<\zeta_\v^{+}\mid\S=\s_0\}=1-\alpha+c_n$
and
$\pr\{|\epsilon_\v|<\zeta_\v^{-}\mid\S=\s_0\}=1-\alpha-c_n$.
Moreover,
\bse
\lim_{c_n\to0}\frac{\zeta_\v^{+}-\zeta_\v^{-}}{2c_n}
=(g_{\v,0})^{-1}.
\ese
Based on Theorem~\ref{th:conformal}, applied to each fold,
\bse
\frac{\wt\zeta_\v^{(j),+}-\zeta_\v^{+}}{2c_n}
-\frac{\wt\zeta_\v^{(j),-}-\zeta_\v^{-}}{2c_n}
=O_p[c_n^{-1}\{h^t+(n_jh^{d_s})^{-1/2}\}].
\ese
Since $n_j/n\to1/2$, when $c_n^{-1}h^t\to0$ and
$c_n^2nh^{d_s}\to\infty$,
\be
(\wh g_{\v,0})^{-1}\xrightarrow{p}(g_{\v,0})^{-1}.
\ee

Third, by $\wh f_\S(\s_0)\xrightarrow{p}f_\S(\s_0)$ and
$\wh g_{\v,0}\xrightarrow{p}g_{\v,0}$,
\begin{align*}
\wh\sigma_{\v_1\v_2}
=\frac{\displaystyle\frac12\sum_{j=1}^2\frac1{n_j}
\sum_{i\in\mathcal I_j}\wh\phi_{\v_1,i}^{(-j)}
\wh\phi_{\v_2,i}^{(-j)}K_h(\s_i-\s_0)}
{\wh f_\S(\s_0)\wh g_{\v_1,0}\wh g_{\v_2,0}}
=
\frac{\displaystyle\frac12\sum_{j=1}^2\frac1{n_j}
\sum_{i\in\mathcal I_j}\wh\phi_{\v_1,i}^{(-j)}
\wh\phi_{\v_2,i}^{(-j)}K_h(\s_i-\s_0)}
{f_\S(\s_0)g_{\v_1,0}g_{\v_2,0}}+o_p(1).
\end{align*}
For  $i\in\mathcal I_j$, let
$\phi_{\v,i}=\phi_\v\{\o_i,\zeta_\v(\s_0),m_\v,
f_{\epsilon_\v\mid\v}\}$. By the uniform consistency of the nuisance
estimators and $\wh\zeta_\v^{(j)}(\s_0)-\zeta_\v(\s_0)=o_p(1)$,
\bse
E[\{\wh\phi_{\v_1,i}^{(-j)}\wh\phi_{\v_2,i}^{(-j)}
-\phi_{\v_1,i}\phi_{\v_2,i}\}K_h(\s_i-\s_0)\mid D_{-j}]
=o_p(1).
\ese
Also,
\begin{align*}
\var\left[\frac1{n_j}\sum_{i\in\mathcal I_j}
\{\wh\phi_{\v_1,i}^{(-j)}\wh\phi_{\v_2,i}^{(-j)}
-\phi_{\v_1,i}\phi_{\v_2,i}\}K_h(\s_i-\s_0)\mid D_{-j}\right]
=O_p\{(n_jh^{d_s})^{-1}\}=o_p(1).
\end{align*}
Then by Lemma~\ref{lemm:conditional},
\bse
\frac1{n_j}\sum_{i\in\mathcal I_j}
\{\wh\phi_{\v_1,i}^{(-j)}\wh\phi_{\v_2,i}^{(-j)}
-\phi_{\v_1,i}\phi_{\v_2,i}\}K_h(\s_i-\s_0)=o_p(1).
\ese
It follows that
\bse
\wh\sigma_{\v_1\v_2}=
\frac{\frac12\sum_{j=1}^2\frac1{n_j}
\sum_{i\in\mathcal I_j}\phi_{\v_1,i}\phi_{\v_2,i}K_h(\s_i-\s_0)}
{f_\S(\s_0)g_{\v_1,0}g_{\v_2,0}}+o_p(1).
\ese
For each $j=1,2$,
\begin{align*}
E\left\{\frac1{n_j}\sum_{i\in\mathcal I_j}
\phi_{\v_1,i}\phi_{\v_2,i}K_h(\s_i-\s_0)\right\}
&=E(\phi_{\v_1,i}\phi_{\v_2,i}\mid\s_0)f_\S(\s_0)+o(1),\\
\var\left\{\frac1{n_j}\sum_{i\in\mathcal I_j}
\phi_{\v_1,i}\phi_{\v_2,i}K_h(\s_i-\s_0)\right\}
&=O\{(n_jh^{d_s})^{-1}\}=o(1).
\end{align*}
Thus,
\be
\wh\sigma_{\v_1\v_2}\xrightarrow{p}
\frac{E(\phi_{\v_1,i}\phi_{\v_2,i}\mid\s_0)}
{g_{\v_1,0}g_{\v_2,0}}=\sigma_{\v_1\v_2}.
\ee
The continuous mapping theorem then gives
\bse
\wh v_\Delta^{\rm semi}
=\frac{\int K^2(\u)d\u}{\wh f_\S(\s_0)}
(\wh\sigma_{\x\x}+\wh\sigma_{\w\w}-2\wh\sigma_{\x\w})
\xrightarrow{p}v_\Delta^{\rm semi}.
\ese

Lastly, by $\wh f_\S(\s_0)\xrightarrow{p}f_\S(\s_0)$,
\begin{align*}
\wh\rho_0
=\frac1{f_\S(\s_0)}\frac12\sum_{j=1}^2\frac1{n_j}
\sum_{i\in\mathcal I_j}
I\{|\wh\epsilon_{\x,i}^{(-j)}|<\wt\zeta_\x^{(j)}(\s_0)\}
I\{|\wh\epsilon_{\w,i}^{(-j)}|<\wt\zeta_\w^{(j)}(\s_0)\}
K_h(\s_i-\s_0)-(1-\alpha)^2+o_p(1).
\end{align*}
For $i\in\mathcal I_j$, by the uniform consistency of
$\wh m_\v^{(-j)}$ and
$\wt\zeta_\v^{(j)}(\s_0)-\zeta_\v(\s_0)=o_p(1)$,
\begin{align*}
E\bigl(&[ I\{|\wh\epsilon_{\x,i}^{(-j)}|<\wt\zeta_\x^{(j)}(\s_0)\}
I\{|\wh\epsilon_{\w,i}^{(-j)}|<\wt\zeta_\w^{(j)}(\s_0)\}\\
&-I\{|\epsilon_{\x,i}|<\zeta_\x(\s_0)\}
I\{|\epsilon_{\w,i}|<\zeta_\w(\s_0)\}]
K_h(\s_i-\s_0)\mid D_{-j}\bigr)=o_p(1).
\end{align*}
Moreover, the conditional variance of the corresponding fold average is
$O_p\{(n_jh^{d_s})^{-1}\}=o_p(1)$. Then by
Lemma~\ref{lemm:conditional},
\begin{align*}
&\frac1{n_j}\sum_{i\in\mathcal I_j}
I\{|\wh\epsilon_{\x,i}^{(-j)}|<\wt\zeta_\x^{(j)}(\s_0)\}
I\{|\wh\epsilon_{\w,i}^{(-j)}|<\wt\zeta_\w^{(j)}(\s_0)\}
K_h(\s_i-\s_0)\\
&=\frac1{n_j}\sum_{i\in\mathcal I_j}
I\{|\epsilon_{\x,i}|<\zeta_\x(\s_0)\}
I\{|\epsilon_{\w,i}|<\zeta_\w(\s_0)\}K_h(\s_i-\s_0)+o_p(1).
\end{align*}
For each $j=1,2$,
\begin{align*}
&E\left\{\frac1{n_j}\sum_{i\in\mathcal I_j}
I\{|\epsilon_{\x,i}|<\zeta_\x(\s_0)\}
I\{|\epsilon_{\w,i}|<\zeta_\w(\s_0)\}K_h(\s_i-\s_0)\right\}\\
&\quad=E[I\{|\epsilon_\x|<\zeta_\x(\s_0)\}
I\{|\epsilon_\w|<\zeta_\w(\s_0)\}\mid\s_0]f_\S(\s_0)+o(1),\\
&\var\left\{\frac1{n_j}\sum_{i\in\mathcal I_j}
I\{|\epsilon_{\x,i}|<\zeta_\x(\s_0)\}
I\{|\epsilon_{\w,i}|<\zeta_\w(\s_0)\}K_h(\s_i-\s_0)\right\}
=O\{(n_jh^{d_s})^{-1}\}=o(1).
\end{align*}
Thus,
\be
\wh\rho_0\xrightarrow{p}
E[I\{|\epsilon_\x|<\zeta_\x(\s_0)\}
I\{|\epsilon_\w|<\zeta_\w(\s_0)\}\mid\s_0]-(1-\alpha)^2
=\rho_0.
\ee
The continuous mapping theorem gives
$\wh v_\Delta^{\rm conf}\xrightarrow{p}v_\Delta^{\rm conf}$.

For the marginal setting, under the assumptions of
Corollary~\ref{th:semi-delta-uncondi}, let $c_n\to0$ and
$nc_n^2\to\infty$. Taking $K_h=1$ in the preceding proof gives
$\wh f_\S=1$ and replaces each conditional expectation by its marginal
counterpart. Therefore, the same argument proves
\bse
\wh v_\Delta^{\rm semi}\xrightarrow{p}
\sigma_{\x\x}^0+\sigma_{\w\w}^0-2\sigma_{\x\w}^0.
\ese
\qed

\bibliographystyle{agsm}
\bibliography{extra}
\end{document}